\documentclass[12pt,a4paper]{article}
\usepackage[utf8]{inputenc}
\usepackage[T1]{fontenc}
\usepackage[bulgarian,english]{babel} 
\usepackage[margin=1.9cm]{geometry}
\usepackage[pdfstartview={FitH},bookmarks=false,linktoc=page,colorlinks=true,linkcolor=blue,citecolor=blue,urlcolor=blue]{hyperref}
\usepackage[numbered]{bookmark}
\usepackage{mathtools}
\usepackage{amssymb}
\usepackage{graphicx}
\usepackage[font=small,labelfont=bf]{caption}
\usepackage{authblk}
\usepackage{cite}
\usepackage{dsfont}
\usepackage[dvipsnames]{xcolor} 
\usepackage[titles]{tocloft}
\usepackage{float}
\usepackage{subcaption}
\usepackage{color}
\usepackage{tikz}
\usepackage{ifthen}
\usepackage[warn]{textcomp}
\numberwithin{equation}{section}
\allowdisplaybreaks
\usepackage{multirow}
\usepackage{cancel}

\usetikzlibrary{patterns}

\usepackage{comment}

\usepackage{caption}
\usepackage{subcaption}

\begin{document}
\title{\vspace{2cm}\textbf{ Optimal Control Theory of the (2+1)-Dimensional BTZ Black Hole}\vspace{1cm}}

\author[a]{M. Radomirov\thanks{Email: radomirov@phys.uni-sofia.bg}}
\author[a,c]{R. C. Rashkov\thanks{Email: rash@phys.uni-sofia.bg}}
\author[a]{G. S. Stoilov\thanks{Email: grigorstoilov234@gmail.com}}
\author[a]{T. Vetsov\thanks{Email: vetsov@phys.uni-sofia.bg}}

\affil[a]{\textit{Department of Physics, Sofia University,}\authorcr\textit{5 J. Bourchier Blvd., 1164 Sofia, Bulgaria}

\vspace{-10pt}\texttt{}\vspace{0.0cm}}

\affil[c]{\textit{Institute for Theoretical Physics, Vienna University of Technology,}
  \authorcr\textit{Wiedner Hauptstr. 8–10, 1040 Vienna, Austria}}

\date{}
\maketitle

\begin{abstract}

We apply a finite-time geometric optimization framework to investigate thermal fluctuations and (non)equilibrium optimal processes in the $(2+1)$-dimensional BTZ black hole. Employing Hessian thermodynamic information metrics, we construct geodesic trajectories that define optimal protocols connecting distinct thermodynamic configurations. Finite-time state transitions are described by paths that extremize entropy production or energy dissipation, depending on the chosen thermodynamic representation. { We compare our optimization framework with a non-optimal blackbody Hawking evaporation model, revealing substantial differences between the two descriptions. Finally, we quantify the intrinsic efficiency of both types of processes in terms of the extractable rotational energy stored in the black hole configurations.} This work presents the first formulation of a geometric optimal control theory for the BTZ black hole.

\end{abstract}

\thispagestyle{empty}
\tableofcontents

\section{Introduction}

A key problem in black hole physics is to characterize the evolution of their thermodynamic states under different physical constraints. Their macroscopic properties can change under processes such as matter accretion \cite{Luminet:1979nyg, Page:1974he, Thorne:1974ve, Gyulchev:2019tvk, Gyulchev:2021dvt}, mergers \cite{LIGOScientific:2016aoc, LIGOScientific:2016vlm}, gravitational-wave emission \cite{1982TaylorPulse}, and Hawking radiation \cite{Hawking:1975vcx}. In addition to their fundamental relevance, these mechanisms have practical implications, such as energy extraction in the Penrose process \cite{penrose1971extraction}. This naturally raises the question of whether other mechanisms can also generate changes in black holes' thermodynamics.

Thermodynamic Geometry (TG) provides the essential tools for addressing this problem \cite{Weinhold:1975a, Ruppeiner:1983zz, Ruppeiner:1995ss}. By encoding fluctuations in geometric terms, it captures intrinsic relations among macrostates of opened or closed systems. Two widely used constructions are Weinhold’s metric, defined as the Hessian of the internal energy, and Ruppeiner’s metric, defined as the Hessian of the entropy. These are conformally related via the temperature \cite{Ruppeiner:1995ss, salamon1984relation, mrugala1984equivalence}. 

However, the Hessian approach is not always sufficient to capture the full range of physical features in black hole thermodynamics. To address this, Geometrothermodynamics (GTD) was developed \cite{Quevedo:2007ws, Quevedo:2007mj, Quevedo:2017tgz, pineda2019physical}, extending the Hessian framework to construct Legendre-invariant metrics \cite{Quevedo:2023vip}. These geometric methods have since been widely applied to study thermodynamic properties of black holes and other gravitational systems \cite{Ruppeiner:2007hr, Mansoori:2016jer, Vetsov:2018dte}.

Equipped with appropriate metrics on the space of states, one can introduce geometric structures such as curvature, geodesic paths between states, and thermodynamic length. These quantities play a central role in analyzing fluctuations in (non)equilibrium systems and in formulating valid optimization protocols under certain conditions. Such geometric features are particularly important in the control theory of classical and quantum systems, where the task is to drive the system from one state to another while maximizing efficiency or minimizing losses \cite{Ramirez2025}.   

In thermodynamic geometry, equilibrium states form a Riemannian manifold whose metric reflects thermal fluctuations \cite{Ruppeiner:1983zz, Ruppeiner:1995ss}. The corresponding scalar curvature has direct physical significance: it is zero for noninteracting systems, negative for attractive interactions, and positive for repulsive ones. Its absolute value relates to the correlation volume and diverges at criticality, mirroring the behavior of statistical correlation lengths. These features make thermodynamic curvature a powerful tool for probing microscopic interactions \cite{Ruppeiner:2007hr, Mansoori:2016jer, Vetsov:2018dte}.

On the other hand, geodesics  on the state space correspond to trajectories that extremize the thermodynamic length \cite{andresen1996finite, Crooks2007, cafaro2022thermodynamic}. They represent the natural notion of shortest paths between points and admit a physical interpretation as statistically preferred fluctuations connecting neighboring states. Within finite-time thermodynamics (FTT) \cite{salamon1977finite, andresen1977extremals, salamon1985length, Avramov:2025tlh}, geodesics identify minimally dissipative processes, thus providing a geometric criterion for optimal evolution. Applied to black hole systems \cite{Bravetti:2015xsp, Gruber:2016mqb, avramov2023thermodynamic, Avramov:2025tlh, Avramov:2025zxc},  geodesics have proven effective in characterizing (non)equilibrium dynamics and phase transitions. 

Recently, a finite-time geometric optimization framework was introduced to study thermal fluctuations and optimal processes in the state space of the Kerr black hole solution \cite{Avramov:2025tlh}. It was shown that geodesic evolution of thermodynamic states can reproduce a Hawking-like evaporation of the Kerr black hole, albeit arising from entirely different physical mechanisms. These optimal evaporation processes manifest as fluctuations on the event horizon, which may occur spontaneously or be triggered by external accretion. The thermodynamic length determines both the likelihood of such optimal processes and the duration over which they proceed.

The aim of this work is to investigate thermal fluctuations and optimal processes on the event horizon of the Ba\~nados–Teitelboim–Zanelli (BTZ) black hole within the finite-time geometric optimization framework developed in \cite{Avramov:2025tlh}. The BTZ solution describes a black hole in $(2+1)$-dimensional Anti-de Sitter spacetime (AdS$_3$) \cite{Banados:1992wn}. First identified in 1992, it serves as a lower-dimensional analogue of more familiar $(3+1)$-dimensional black holes such as the Schwarzschild and Kerr solutions. Owing to its reduced dimensionality, the BTZ black hole is mathematically more tractable and thus widely employed as a simplified model in theoretical investigations. Moreover, its asymptotically AdS geometry, rather than flat spacetime, renders it particularly important in the context of the AdS/CFT correspondence and holography\footnote{The BTZ solution is holographically dual to a thermal state in a two-dimensional conformal field theory (CFT$_2$) living on the asymptotic boundary of AdS$_3$ \cite{Maldacena:1997re}. The black hole mass  and angular momentum are directly related to the energy and momentum of the dual CFT, while the Hawking temperature and entropy are reproduced by the thermal properties of the boundary theory \cite{Strominger:1997eq}. In particular, the microscopic entropy of the BTZ black hole can be derived from the Cardy formula for the asymptotic density of states in a unitary two-dimensional CFT with central charge\footnote{Here $\ell$ is the AdS$_3$ radius and $G_3$ is the Newton's gravitational constant in ($2+1$) dimensions.} $\tilde c = 3\ell/(2G_3)$, obtained from the Brown-Henneaux analysis of the AdS$_3$ asymptotic symmetry algebra \cite{Brown:1986nw}. This remarkable agreement established one of the earliest and most explicit demonstrations of holography in lower dimensions, and it continues to serve as a foundational example in the study of black hole thermodynamics and quantum aspects of gravity.}.

{ 
The paper is organized as follows. In Section \ref{secTDBTZBH}, we summarize the fundamental thermodynamic properties of the BTZ black hole in both SI and 3$d$ Planck units. Section \ref{secTGO} provides a brief review of the main aspects of the Thermogeometric Optimization (TGO) framework introduced in \cite{Avramov:2025tlh}. In Section \ref{secOPER}, we investigate thermal fluctuations and optimal processes with correlated initial conditions in the energy representation for both static and rotating BTZ black holes. Section \ref{secOPEntrR} is devoted to the corresponding analysis in the entropy representation. In Section \ref{secNLGRFL}, we explore whether a static BTZ black hole can dynamically evolve into a rotating configuration through spontaneous or externally induced optimal fluctuations originating from the nonlinear structure of the optimal process. In Section \ref{secHawkEvapModels}, we study Hawking evaporation models for both static and rotating BTZ black holes based on the Stefan--Boltzmann power law. Section \ref{OptimalvsHawking} compares the non-optimal Hawking evaporation scenario with the optimal TGO protocol, highlighting substantial differences between the two approaches. In Section \ref{IntrThdEffPr}, we analyze the decomposition of the total black hole energy $E$ into rotational, irreducible, and emitted contributions in order to determine the intrinsic thermodynamic (conversion) efficiency, $\eta(t)$, of the processes. This quantity is defined as the ratio of the extractable rotational energy to the initial energy input at a given time $t$, and therefore represents the theoretical upper bound on the fraction of energy that can be extracted and stored in the spin of the BTZ black hole through mechanisms such as the Penrose process. Finally, Section \ref{secConcl} contains a brief discussion and summary of our main results.
}

\section{Thermodynamics of the BTZ black hole}\label{secTDBTZBH}

We shortly review the fundamental thermodynamic relations for the BTZ black hole in both the energy and entropy ensembles, with all quantities expressed in SI units. The conversion to Planck and other units is detailed in Appendix \ref{appPlancktoSI}.
 
\subsection{The BTZ black hole solution}
The BTZ black hole metric in $D=2+1$ dimensions is given by \cite{Banados:1992wn, Carlip:1995qv}:
\begin{equation}\label{eqBTZMetric}
ds^2 =-\frac{(r^2-r_{+}^2) (r^2-r_{-}^2)}{\ell^2 r^2} c^2dt^2+\frac{\ell^2 r^2 }{(r^2-r_{+}^2) (r^2-r_{-}^2)} dr^2+r^2\bigg(d\varphi-\frac{r_{+} r_{-}}{\ell r^2} c dt\bigg)^{\!2},
\end{equation}
where $c$ is the speed of light,  $\ell$ is the AdS radius, $r_{+}$ is the event horizon, and $r_{-}$ is the inner Cauchy horizon. The  energy\footnote{The standard relation between mass and energy $E=Mc^2$ is assumed.} $E$, angular momentum $J$, entropy $S$, Hawking temperature $T$, and angular velocity $\Omega$ of the BTZ black hole solution are given by\footnote{Here $\hbar$ is the Dirac constant, $k$ is the Boltzmann constant, and $G_3$ is the 3-dimensional gravitational constant with units $\text{m}^2 \cdot \text{kg}^{-1} \cdot \text{s}^{-2}$.}:
\begin{align}\label{eqBTZTD}
S=\frac{\pi k c^3}{2 \hbar G_3} r_+,\quad E = \frac{r_+^2 + r_-^2}{8 G_3 \ell^2}  c^4,\quad J = \frac{c^3}{4 G_3 \ell} r_+ r_-,\quad T = \frac{\hbar c}{k} \frac{r_{+}^2-r_{-}^2}{2 \pi\ell^2 r_+},\quad 
\Omega = c \frac{r_-}{\ell r_+},
\end{align}
These determine the  thermodynamics of the BTZ system with first law and a Smarr relation:
\begin{equation}
    dE=T dS+\Omega dJ, \quad E = \frac{1}{2}TS + \Omega J.
\end{equation}

\subsection{Energy and entropy thermodynamic representations}

Solving Eqs. (\ref{eqBTZTD}) for $r_{\pm}$ in terms of the relevant thermodynamic parameters one can find the fundamental relations  for energy and entropy:
%
\begin{align}\label{eqESenergyRep}
E(S,J)= \frac{\zeta^2 S^4 +\lambda^2 J^2}{2\lambda S^2},
\quad
S(E,J) 
=\frac{1}{\zeta} \sqrt{\lambda\big(E +\sqrt{E^2-\zeta^2 J^2} \,\big)},
\end{align}
where following notations were introduce:
\begin{equation}
    \lambda= \frac{\pi ^2 c^4  k^2}{ G_3 \hbar ^2},\quad \zeta=\frac{c}{\ell}.
\end{equation}

The temperature and the angular velocity follow naturally in energy representation:
\begin{align}
T=\frac{\partial E}{\partial S}\bigg|_{J} = \frac{\zeta^2 S^4 -\lambda^2 J^2}{\lambda S^3} ,
\quad\Omega =\frac{\partial E}{\partial J}\bigg|_{S} = \frac{J \lambda }{S^2}.
\end{align}

In entropy representation one has:
\begin{align}\label{TOmega}     
&T= \bigg( \frac{\partial S}{\partial E} \bigg)^{\!\!-1}_{\!J} = \frac{2 \zeta  \sqrt{E^2-\zeta ^2 J^2}}{\sqrt{\lambda  \big(E+\sqrt{E^2-\zeta ^2 J^2}\big)}},
\quad\Omega= -T \frac{\partial S}{\partial J}\bigg|_{E} = \frac{\zeta ^2 J}{E+\sqrt{E^2-\zeta ^2 J^2}}.
\end{align}

Non-extremality requires the temperature of the BTZ black hole always be positive ($T>0$), which in the corresponding representations leads to\footnote{The case $T=0$ ($r_+=r_-$) violates classically the third law of thermodynamics.}:
\begin{equation}\label{eqExtremality}
\zeta  J<E\quad \text{or}\quad  \lambda J < \zeta S^2 .
\end{equation}
Note that the extremal case $\zeta  J = E$ is omitted since it violates classically the third law of thermodynamics. One can write (\ref{eqExtremality}) in terms of the BTZ dimensionless specific spin\footnote{Note that, assuming $G_3>0$, Eq. \eqref{eqBTZTD} imposes $J\geq 0$, thus $0\leq a<1$.} $a \in [0,1)$:
\begin{equation}\label{eqSpecificSpin}
   a=\frac{\zeta J}{E},\quad  0\leq a<1.
\end{equation}

The non-extremality condition $a<1$ imposes a constraint on the boundary quantum dynamics via the dimensionless central charge $\tilde{c}$ of the dual conformal field theory (CFT), i.e. \cite{Carlip:2005zn}:
\begin{equation}
\tilde{c} = \frac{3 c^3 \ell}{2 G_3 \hbar} < \frac{3 \hbar}{2 \pi^2 k^2} \frac{S^2}{J},
\end{equation}
thereby setting an upper bound on the effective number of degrees of freedom in the dual CFT.

\subsection{Thermodynamic stability}
The BTZ black hole is globally stable from classical thermodynamic point of view. This follows either from the eigenvalue criterion or the Sylvester criterion for positive-definiteness of the Hessian of the energy (energy is strictly convex) \cite{Avramov:2023eif, Avramov:2024hys}:
\begin{equation}\label{Sylvester}
    \frac{\partial^2 E}{\partial S^2}\bigg|_{J} = \frac{\zeta^2 S^4 +3\lambda^2 J^2}{\lambda S^4} > 0,\quad
     \frac{\partial^2 E}{\partial J^2}\bigg|_{S}=\frac{\lambda }{S^2}\!>0, \quad
     \det\hat{H}  =\frac{\zeta^2 S^4 -\lambda^2 J^2}{S^6} >0.
\end{equation}
The first two inequalities reflects the convexity of energy along its natural parameters, and the third one assures full convexity according to the Sylvester criterion.  Since the non-extremal BTZ solution is globally stable, it should also be locally stable, e.g. all heat capacities must be strictly positive \cite{Avramov:2023eif}:
\begin{align}\label{eqCE}
&C_E(S,J) =T\frac{\partial S}{\partial T}\bigg|_{E} = T\frac{\{S,E\}_{S,J}}{\{T,E\}_{S,J}}=\frac{S \big(\zeta^2 S^4 -\lambda^2 J^2 \big)}{3 \zeta ^2 S^4 +\lambda^2 J^2} >0,
\\[5pt]
&C_J (S,J) =T\frac{\partial S}{\partial T}\bigg|_{J} =  T\frac{\{S,J\}_{S,J}}{\{T,J\}_{S,J}}= \frac{S \big(\zeta^2 S^4 -\lambda^2 J^2 \big)}{\zeta^2 S^4 +3\lambda^2 J^2}> 0,
\\[5pt]\label{eqCOmega}
&C_\Omega(S,J) =T \frac{\partial S}{\partial T}\bigg|_\Omega= T\frac{\{S,\Omega\}_{S,J}}{\{T,\Omega\}_{S,J}}=S>0,
\end{align}
which is true for nonextremal BTZ black holes.
Here we use Nambu brackets (see \cite{Avramov:2023eif}) to find the corresponding heat capacities. The result further shows that there are no Davies phase transition curves for BTZ, besides the extremal one (\ref{eqExtremality}).

The global stability of the BTZ black hole implies that it cannot evaporate solely through standard Hawking radiation. However, like all physical systems, black holes are susceptible to spontaneous thermal or quantum fluctuations, as well as external perturbations at their event horizons. Such effects can induce changes in the black hole’s parameters, potentially driving an evaporation-like evolution that decreases its mass–energy content\footnote{The mass energy of the black hole can also increase under optimal  accretion.}.

In the following sections, we apply the thermogeometric optimization method  to the BTZ system to investigate this behavior. The TGO method is specifically designed to study optimal processes in thermal systems, including black holes \cite{Avramov:2025tlh}.

\section{Thermogeometric optimization method (TGO)}\label{secTGO}
The Thermogeometric Optimization (TGO) method, introduced in \cite{Avramov:2025tlh}, provides a systematic framework for analyzing thermal fluctuations and optimal processes on or near the event horizon of the  black hole. The procedure begins by specifying a natural thermodynamic potential $\Phi$, determined by the system’s control parameters, together with its associated thermodynamic metric $\hat{g}$. Within thermodynamic geometry, two  metric formulations are commonly employed,  namely  Weinhold’s metric, defined as the Hessian of the energy\footnote{In its original form, Weinhold’s metric is defined with $\epsilon=+1$.} ($\Phi \equiv E$) \cite{Weinhold:1975a}:
\begin{equation}\label{eqWeinhMetricdif}
ds^2_{W}=\epsilon\sum_{a,b=1}^n\frac{\partial^2 E}{\partial E^a\partial E^b} dE^a dE^b, \quad \epsilon\in\mathbb{R},
\end{equation}
and Ruppeiner’s metric\footnote{Importantly, these two metrics are conformally related through the temperature \cite{Ruppeiner:1995ss, salamon1984relation, mrugala1984equivalence}.}, defined as the Hessian of the entropy\footnote{In its original form, Ruppeiner’s metric is defined with  $\epsilon=-1$.} ($\Phi \equiv S$)\cite{Ruppeiner:1983zz, Ruppeiner:1995ss}:
\begin{equation}\label{eqRupMetGen}
ds^2_{R}=\epsilon\sum_{a,b=1}^n\frac{\partial^2 S}{\partial S^a\partial S^b} dS^a dS^b.
\end{equation}
Here, $E^a$ denote the natural control variables of the energy $E$, while $S^a$ are those associated to the entropy $S$. The parameter $\epsilon$, introduced in \cite{Avramov:2025tlh}, extends the equilibrium setup of TG to include also non-equilibrium situations.

The next step is to parametrize the natural coordinates $\Phi^a$ of the potential $\Phi$ with an affine time parameter $t$ and extermize the thermodynamic length in order to obtain the  thermodynamic geodesics with respect to the chosen metric. The later can be interpreted as fluctuation paths or optimal protocols. By definition the thermodynamic length functional is given by:
\begin{equation}\label{eqGeodAction2}
{\cal L}[\gamma(t)] = \int \nolimits_{0}^{\tau} \sqrt {{g_{ab}}\big(\vec \Phi(t)\big)\dot \Phi^a(t) \dot \Phi^b(t)}\, dt,
\end{equation}
and the corresponding nonlinear system of geodesic equations yields:
\begin{equation}\label{eqGeodesicGeneric}
\ddot \Phi^c(t)+\Gamma^c_{ab}\big(\hat g,\vec \Phi\big) \dot \Phi^a(t)\dot \Phi^b(t)=0,
\end{equation}
where $\Gamma^c_{ab}(\hat g,\vec \Phi)$ are the standard  Christoffel symbols, defined by:
\begin{equation}\label{eqChristofel}
\Gamma^{c}_{ab}=\frac{1}{2}g^{c d}(\partial_a g_{db}+\partial_b g_{da}-\partial_d g_{ab}).
\end{equation}
\section{Optimal processes in energy representation}\label{secOPER}

We employ the TGO framework to study the finite-time evolution of optimal processes on the event horizon of the BTZ black hole, formulated in the energy representation. Within this approach, we demonstrate how the BTZ system can either lose or gain energy through a ``least resistant'' paths on the space of its macrostates.

\subsection{Thermodynamic metric and curvature}
The natural thermodynamic metric in energy representation is the Weinhold metric:
\begin{equation}\label{Weinhold}
  ds^2 =\epsilon\, \frac{\zeta^2 S^4 +3\lambda^2 J^2 }{\lambda S^4} dS^2-\epsilon \frac{ \lambda J }{S^3} dS dJ+\epsilon\frac{\lambda}{S^2} dJ^2.
\end{equation}
Its thermodynamic curvature is given by:
\begin{equation}\label{eqCurvEn}
R=\frac{2 \lambda\zeta ^2   S^6}{\epsilon (\zeta S^2 -\lambda J)^2 (\zeta S^2 +\lambda J)^2}.
\end{equation}
It is divergent along the extremal curve $\zeta S^2 = J \lambda$, where interactions become infinitely strong. The nature of the underlying information geometry is governed by the sign of the scaling parameter $\epsilon$. For $\epsilon > 0$, the information space is elliptic with positive curvature $R > 0$, whereas for $\epsilon < 0$, a hyperbolic information geometry with $R < 0$ emerges. 

\begin{figure}[H]
    \centering
\includegraphics[width=7.0cm,height=6cm]{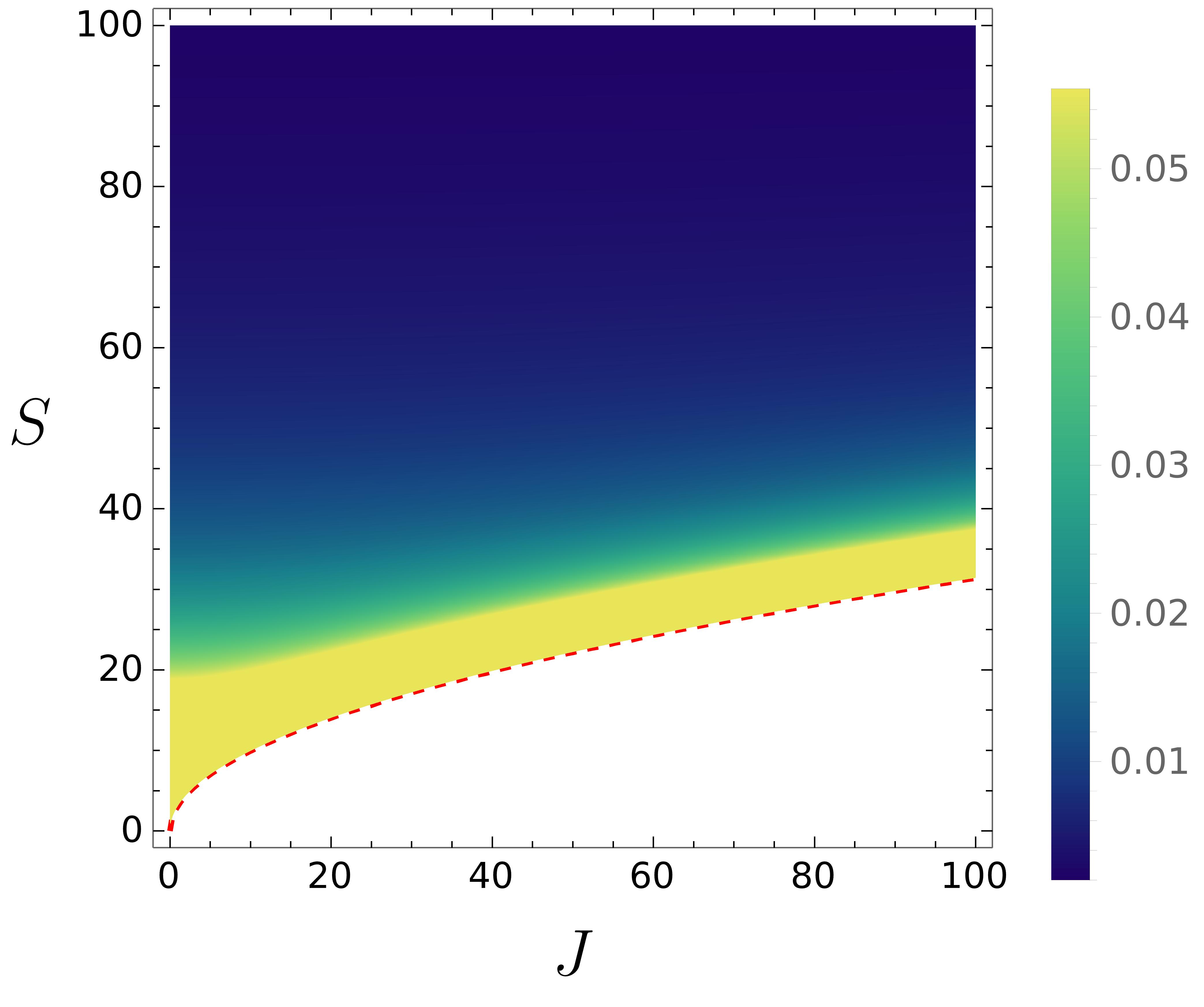}
\caption{The Weinhold thermodynamic curvature $R$ is presented as a function of $S$ and $J$ in Planck units with $\ell=1$. An increase in the interaction strength among the horizon degrees of freedom is observed near the extremal BTZ curve (dashed red), where a phase transition takes place. In the asymptotic regime, far from extremality, the thermodynamic geometry of the BTZ black hole approaches flatness, indicating weak or negligible interactions. Since $\epsilon > 0$ ensures a positive-definite thermodynamic length, the curvature $R$ remains positive, implying that the information geometry is elliptic.}
    \label{figTDRicciErep}
\end{figure}

The Weinhold thermodynamic curvature $R$ is shown in Fig. \ref{figTDRicciErep} as a function of $S$ and $J$. Since $R$ quantifies the effective interaction strength among horizon degrees of freedom, it is observed that the interactions intensify in the vicinity of the extremal BTZ curve (dashed red), where a phase transition occurs. As we show in Section \ref{secStaticBTZEnergyrep} in energy representation $\epsilon > 0$ ensures a positive-definite thermodynamic length, thus the curvature $R$ remains positive, implying that the information geometry is intrinsically elliptic.

\subsection{Geodesic equations in $(S,J)$ space}
We parametrize the $(S,J)$ coordinates by an affine time parameter $t$ and seek to obtain their optimal profiles by solving the geodesic equations on the space of macrostates in energy ensemble: 
\begin{align}\label{eqSener}
&\ddot S + \Gamma^S_{SS} \dot S^2 +2\Gamma^S_{S J}  \dot S\dot J +\Gamma^S_{JJ} \dot J^2=0, \\[5pt]\label{eqJener}
&\ddot J + \Gamma^J_{JJ}  \dot J^2 +2\Gamma^J_{S J} \dot S\dot J +\Gamma^J_{S S} \dot S^2=0,
\end{align}
where the Christoffel symbols are given by:
\begin{align}
&\Gamma^{S}_{SS}=0,\quad \Gamma^{S}_{SJ}=\frac{J \lambda ^2}{\zeta ^2 S^4-J^2 \lambda ^2},\quad \Gamma^{S}_{JJ}=\frac{\lambda ^2 S}{J^2 \lambda ^2-\zeta ^2 S^4},
\\[5pt]
&\Gamma^{J}_{SS}=\frac{3 J}{S^2},\quad \Gamma^{J}_{SJ}=\frac{3 J^2 \lambda ^2-\zeta ^2 S^4}{\zeta ^2 S^5-J^2 \lambda ^2 S},\quad \Gamma^{J}_{JJ}=\frac{2 J \lambda ^2}{J^2 \lambda ^2-\zeta ^2 S^4}.
\end{align}

The solutions to these nonlinear equations define the optimal profiles for entropy $S(t)$ and angular momentum $J(t)$ at various initial conditions:
\begin{equation}
S(0)=S_0,\quad \dot S(0)=\dot S_0,\quad 	J(0)=J_0,\quad \dot J(0)=\dot J_0,
\end{equation}
where $S_0$ and $J_0$ denote the initial entropy and angular momentum of the BTZ black hole, respectively, and $\dot{S}_0$ and $\dot{J}_0$ represent their initial rates of change. It is important to note that $S(t)$ and $J(t)$ represent valid geodesic profiles only when the thermodynamic length $\mathcal{L}$ along the path is real and positive. The profiles for energy $E(t)$ and specific spin $a(t)$ follow from their definitions (\ref{eqESenergyRep}) and (\ref{eqSpecificSpin}).

\subsection{Optimal evaporation of the static BTZ black hole}\label{secStaticBTZEnergyrep}

A simple analytical solution exists for the static ($J=0$) BTZ black hole. In this case, the second equation (\ref{eqJener}) is trivially satisfied, while the first equation (\ref{eqSener}) becomes $\ddot S(t)=0$ leading to a simple linear profile of the entropy:
\begin{equation}\label{eqEntropyOPstatic}
    S(t)=S_0-|\dot S_0| t.
\end{equation}
Here we already assumed an evaporation process with an initial rate: $\dot S_0=-|\dot S_0|< 0$. 

The thermodynamic length of the process from $S_0$ to $S_{\tau}$ at $t=\tau$ is given by:
\begin{equation}\label{eqTDLSJstatic}
    \mathcal{L}_{S_0\to S_\tau} = \int \nolimits_{0}^{\tau} \sqrt {{g_{ab}}\dot x^a(t) \dot x^b(t)}\, dt=\zeta \sqrt{\frac{\epsilon }{\lambda }}\,|\dot S_0| \,\tau=v \tau,
\end{equation}
where $x^a(t) = \big(S(t),J(t) = 0\big)$, and  $v$ is the thermodynamic speed of the process. The length is real and positive if $\epsilon>0$, hence it defines an elliptic information space ($R>0$). Additionally,  in the static case one can also find the direct thermodynamic length\footnote{Note that $\sqrt {{g_{ab}}\dot x^a \dot x^b}\propto \sqrt{dS^2}=|dS|=-dS>0$, since entropy decreases during an evaporation process.}:
\begin{equation}\label{ThermoLenghtBTZnorodEn}
    \mathcal{L}_{S_0\to S_\tau} = \int \sqrt {{g_{ab}} dx^a dx^b}=-\zeta  \sqrt{\frac{\epsilon }{\lambda }}\int_{S_0}^{S_\tau} dS=\zeta  \sqrt{\frac{\epsilon }{\lambda }} \,(S_0-S_\tau),
\end{equation}
where $x^a = (S,J = 0)$.
Insisting on both lengths be equal, we can make an estimation of the average relaxation time $\bar \tau$ of the process:
\begin{equation}\label{eqRelTE}
\bar \tau\approx \frac{S_0-S_\tau}{|\dot S_0|}.
\end{equation}
 This shows that the average relaxation time is proportional to the entropy change $\Delta S = S_0 - S_\tau$ produced by the BTZ black hole during the process. In the case of complete evaporation $(S_\tau = 0)$, the total evaporation time can be obtained from the condition $S(t_{\text{evap}})=0$, yielding:
\begin{equation}\label{eqOptevapTstatic}    
t_{\text{evap}} = \frac{S_0}{|\dot S_0|}.
\end{equation}
Hence, the evaporation thermodynamic length is:
\begin{equation}\label{eqLTDevapEnergyRep}
    \mathcal{L}_{evap}=\zeta  \sqrt{\frac{\epsilon }{\lambda }} S_0=\frac{\hbar  \sqrt{\epsilon G_3}}{\pi  c k \ell } S_0.
\end{equation}
Since the evaporation length is proportional solely to the initial entropy $S_0$ of the BTZ black hole, this indicates that larger black holes are less likely to undergo spontaneous evaporation. Conversely, smaller black holes may evaporate spontaneously with higher probability. 

The quantity $\mathcal{L}^2$ has units of energy, which aligns with its interpretation as the minimal energy required to drive the system from one thermodynamic state to another. The latter can be used to determine the scale factor $\epsilon$ for a complete evaporation process of the BTZ. Insisting on  $\mathcal{L}^2$ be at least the initial energy of the BTZ black hole $E_0=\zeta ^2 S_0^2/(2 \lambda )$,  we find $\epsilon=1/2$.

{ 

It is also easy to obtain the energy profile using \eqref{eqESenergyRep} at $J=0$:
\begin{equation}
E(t)= E_0 \Big(1-\frac{|\dot E_0|}{2E_0}t \Big)^2,
\end{equation}
 where the relation between the initial parameters of entropy and energy is given by:
 \begin{equation}
 E_0=\frac{\zeta^2S_0^2}{2\lambda}, \quad |\dot E_0|=\frac{\zeta^2}{\lambda} S_0 |\dot S_0|.
 \end{equation}

A comparison between these optimal profiles and the nonoptimal Hawking evaporation process is presented in Section \ref{secOvsHenergystatic}.
}

\subsection{Optimal processes of the rotating BTZ black hole}

We examine how optimal processes, initiated at different initial rates of the parameters, can transform the thermodynamic state of a rotating BTZ black hole. In all cases considered --- within the energy representation --- the system evolves toward a static (non-rotating) black hole with well-defined energy and entropy, 
and thus no complete evaporation of the rotating BTZ black hole occurs. However, the final energy and entropy may either increase or decrease, depending on the initial rates. We refer to processes that reduce both energy and entropy as optimal evaporation of the BTZ black hole. Conversely, when the final state exhibits higher energy and entropy, we have accretion-type process\footnote{An internal reconfiguration of the black hole states, driven by thermal or quantum fluctuations, may also lead to an increase in both energy and entropy at the expense of angular momentum. In such cases, no external driving (accretion) force is required to initiate or sustain the process.}. A comparison of optimal profiles with the nonoptimal Hawking evaporation process in the energy representation is presented in Section \ref{secOvsHenergyrot}.

\subsubsection{Setting up initial  conditions}

We will numerically integrate the geodesic equations (\ref{eqSener}) and (\ref{eqJener}) in energy representation. For consistency of numerical results, all quantities are rendered dimensionless and expressed in Planck units (see Appendix \ref{appPlancktoSI}). The resulting solutions represent valid geodesic trajectories --- interpreted as optimal thermodynamic processes in the $(S, J)$ parameter space --- provided that the thermodynamic length, defined by the integral (\ref{eqGeodAction2}), remains real and positive-definite. 

It is important to note that the algorithm does not encode or refer to any specific physical mechanism underlying the process. As a result, each geodesic represents a distinct optimal process, and multiple geodesics originating from the same initial state may intersect at a later point in the state space. That is, there may exist multiple optimal pathways to transition between the same two thermodynamic states\footnote{
Multiple optimal processes may correspond to different physical mechanisms --- such as conduction versus radiation --- or may arise under differing external constraints. A classic example of intersecting geodesics is provided by distinct great circles on a sphere, which always intersect at two antipodal points. Since, in our case, $(S, J)$ space exhibits positive curvature, similar to a sphere, the existence of intersecting geodesics is allowed.
}.

To demonstrate our optimal model, we chose the  initial state  be defined by the following initial parameters of the BTZ black hole:
\begin{equation}\label{eqIniCSJ}
   S_0 = 5,\quad J_0 = 1, \quad E_0 =\frac{\pi ^2 J_0^2}{2S_0^2}+\frac{S_0^2}{2\pi ^2\ell^2}= 1.46,\quad a_0 = \frac{J_0}{E_0\ell}= 0.68,
\end{equation}
where we set $\ell =1$. {  Since, in general, the initial rates $\dot S_0$ and $\dot J_0$ at $t=0$ can be chosen arbitrarily, it is convenient to introduce a new set of polar coordinates $(u,\phi)$ defined by}:   
\begin{equation}\label{eqPolarInitialCond}
\dot S_0=u\sin \phi, \quad  \dot J_0=u\cos \phi, \quad u=\sqrt{\dot S^2_0 +\dot J^2_0}, \quad \tan \phi=\frac{\dot S_0}{\dot J_0},
\end{equation}
where the angle $\phi$ determines the initial direction of the geodesic trajectory of states in $(S,J)$ space. The corresponding initial rates of energy $\dot{E}_0$ and specific spin $\dot a_0$ are then determined by evaluating the time derivatives of their numerical profiles at $t = 0$. Comparable results for other angles $\phi$ are summarized in Table \ref{tabicerep}. 

Furthermore, we can characterize the evolution of a given parameter $X$ by its relative change:
\begin{equation}
\delta X = \frac{X(\tau) - X(0)}{X(0)} \times 100\%,
\end{equation}
where $\tau$ typically denotes the duration of the process (the relaxation time). The relative change $\delta X$ is positive if $X$ increases over time and negative if it decreases.

In the following subsections, we illustrate the TGO method by fixing $u = 0.01$ and examining in detail  two representative cases: $\phi = 0^\circ$ (accretion-driven) and $\phi = 240^\circ$ (optimal evaporation process). 
The corresponding geodesic trajectories in the $(S, J)$ space for a wider range of initial angles, $\phi = \{0^\circ, 45^\circ, 60^\circ, 90^\circ, 135^\circ, 180^\circ, 225^\circ, 240^\circ, 270^\circ, 315^\circ\}$, along with the associated thermodynamic curvature, are displayed in Fig. \ref{figGeodSJ}.

\subsubsection{Optimal process at $\phi=0^\circ$}

\begin{figure}[H]
\centering \hspace{-1.5cm}
\begin{subfigure}{0.45\textwidth}
%
\includegraphics[width=8.3cm,height=5.5cm]{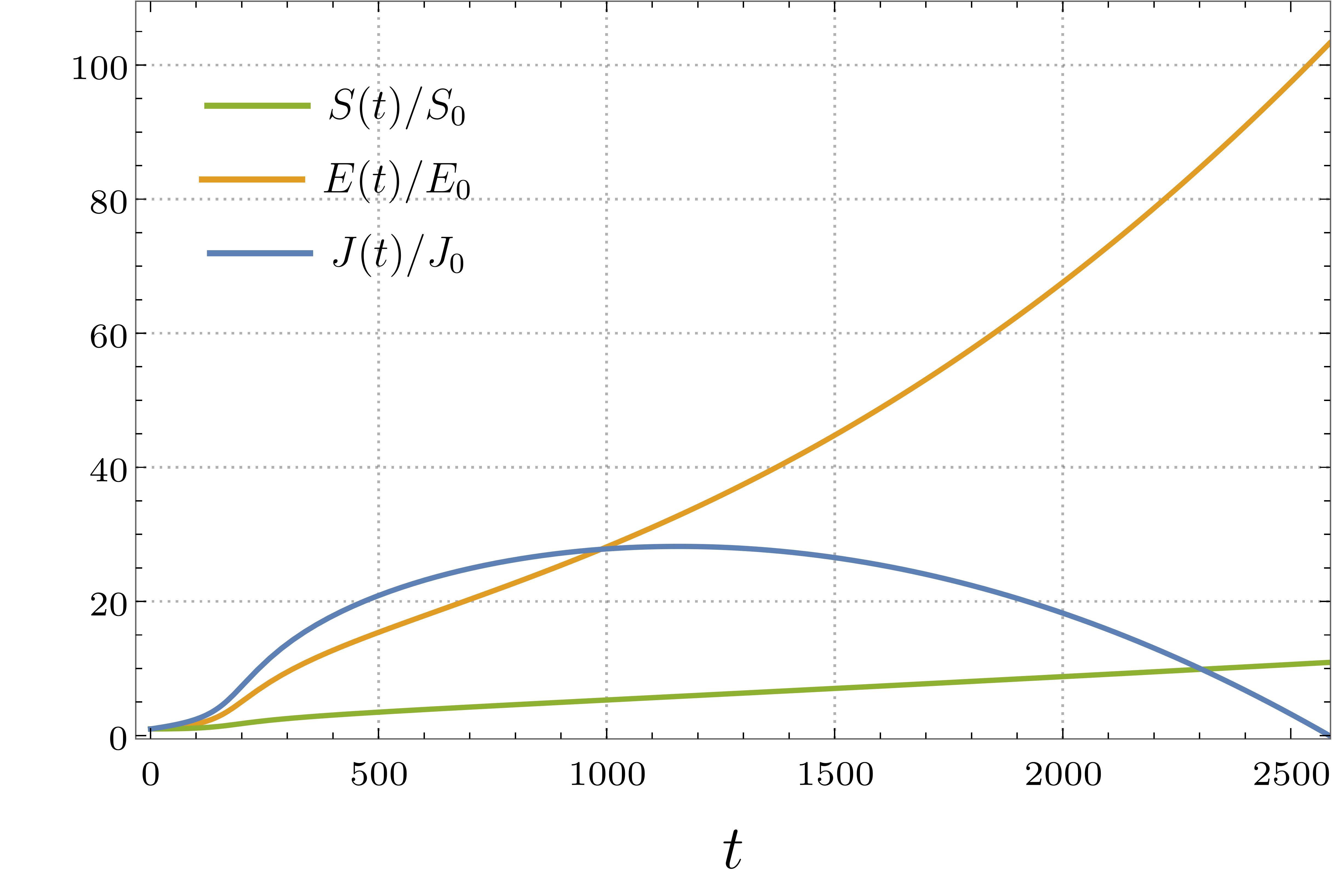}
\caption{\hspace*{-0cm} \scriptsize Optimal profiles: $E(t)/E_0$, $S(t)/S_0$, and $J(t)/J_0$.\vspace{0.3 cm}}\label{figESJt0}
\end{subfigure}
\hspace{0.45 cm}
\begin{subfigure}{0.45\textwidth}
\includegraphics[width=8.3cm,height=5.5cm]{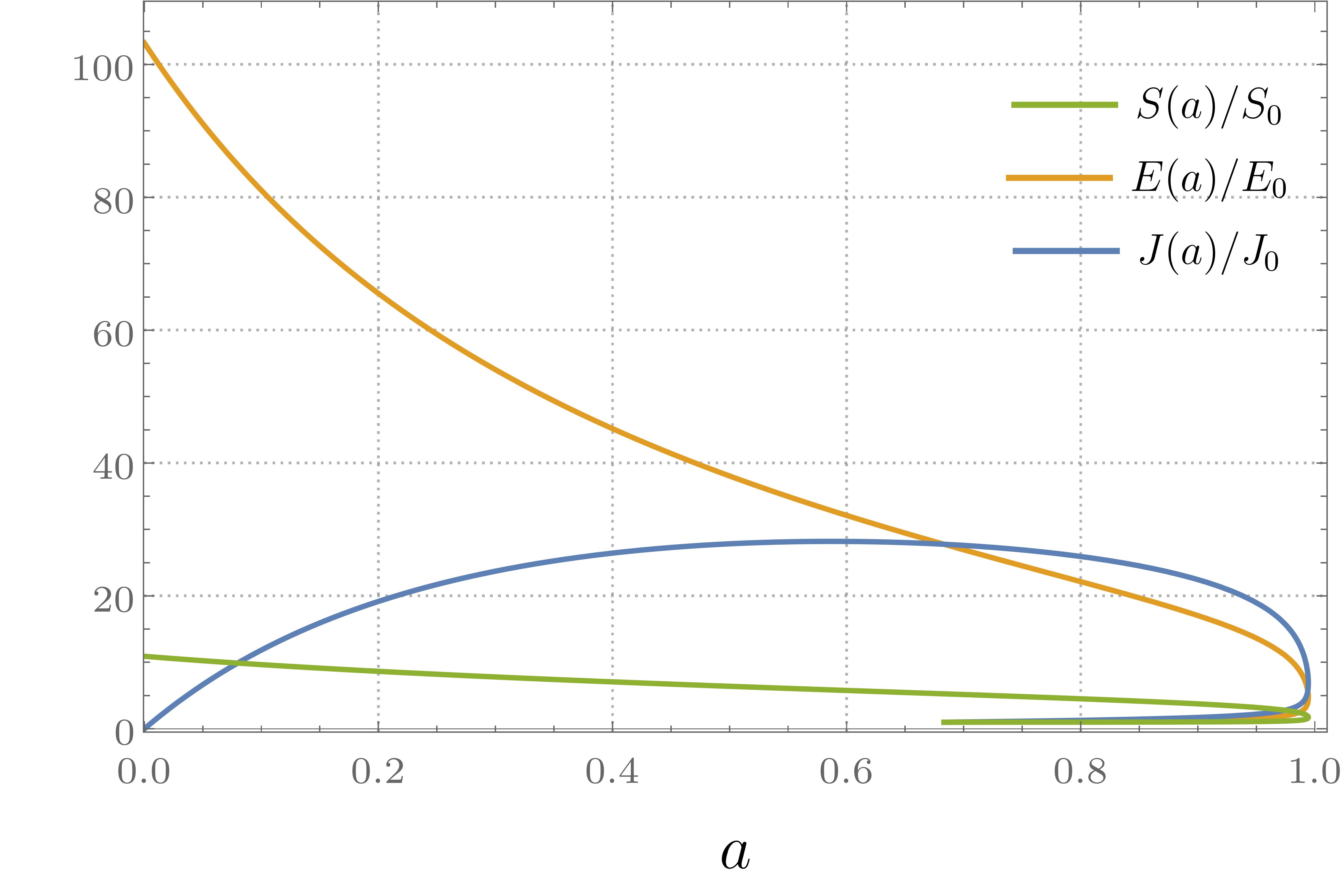}\vspace{-0.05cm}
\caption{\hspace*{0cm} \scriptsize Optimal profiles: $E(a)/E_0$, $S(a)/S_0$, and $J(a)/J_0$. \vspace{0.3 cm}}\label{figESJa0}
\end{subfigure}
\begin{subfigure}{0.43\textwidth}
\includegraphics[width=8.3cm,height=5.7cm]{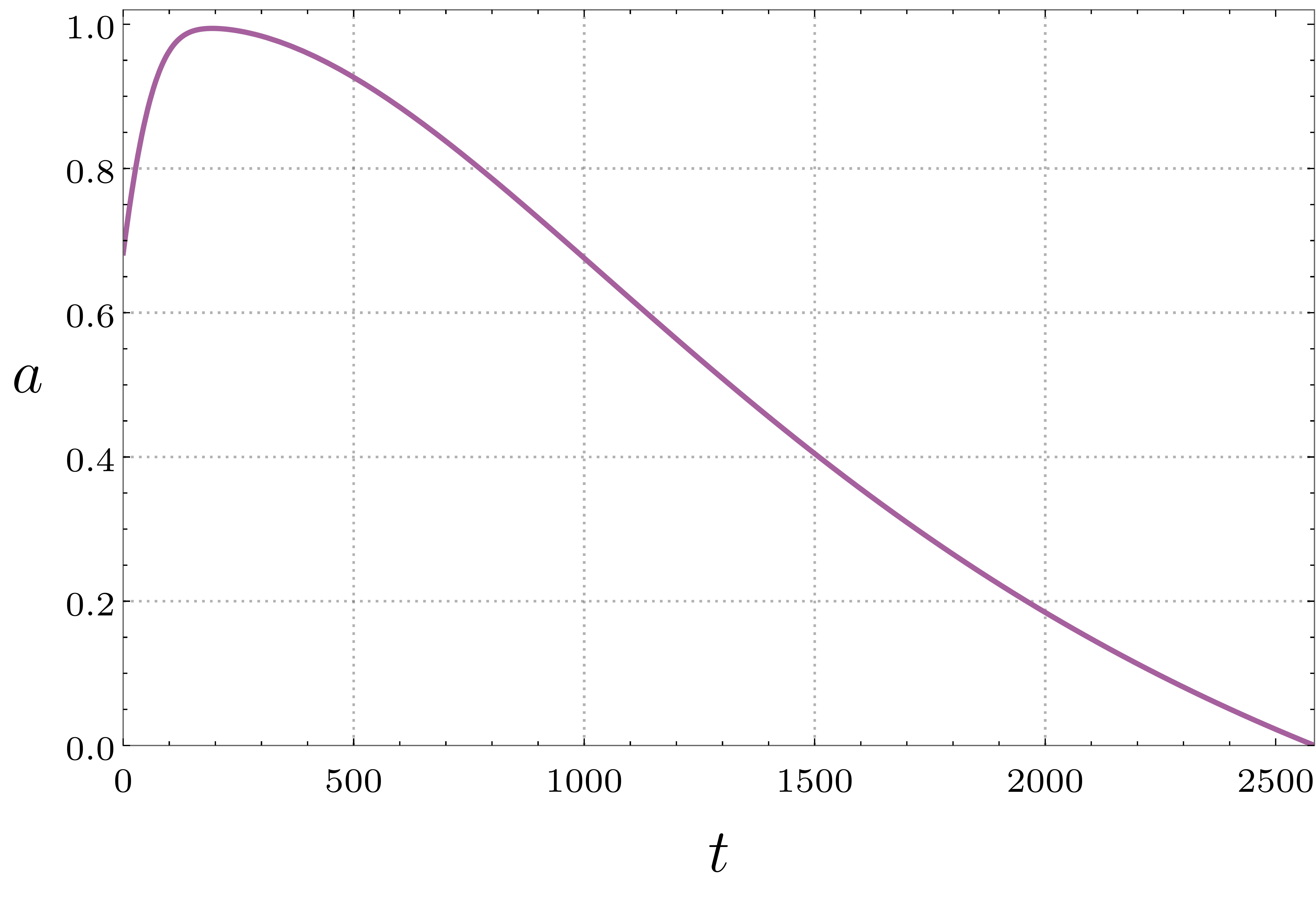}
\caption{\hspace*{-0cm} The specific spin $a(t)$.}\label{figat0}
\end{subfigure}
\hspace{0.9cm}
\begin{subfigure}{0.43\textwidth}
\includegraphics[width=8.3cm,height=5.6cm]{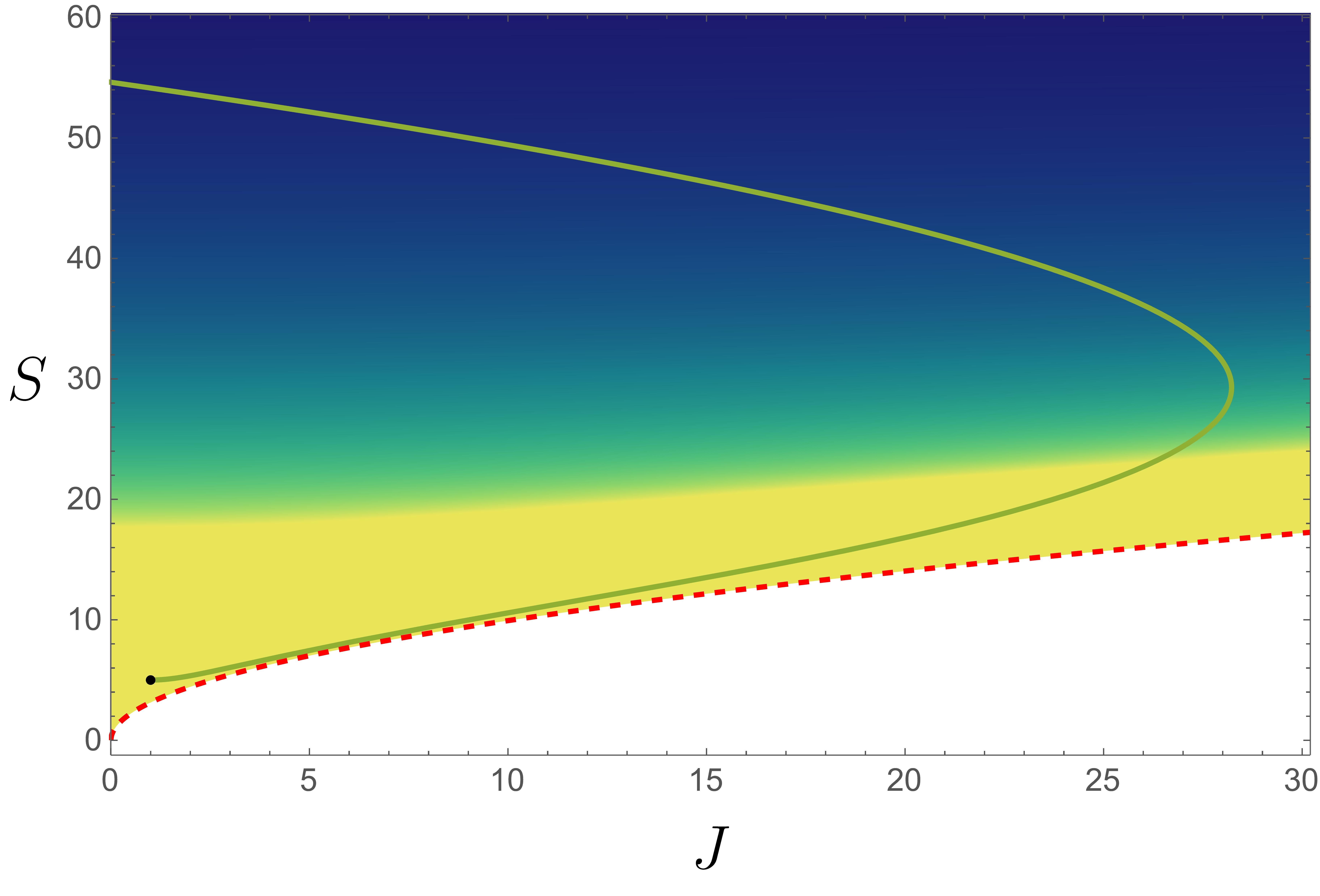}\vspace{0.05cm}
\caption{\hspace*{-0.0cm} The geodesic path of states in $(S, J)$ space.}\label{figSJ0}
\end{subfigure}
 \hspace{1cm} 
 \vspace{-0.0cm}
\caption{ Profiles for $\phi=0^\circ$. 
\textbf{(a)} Time evolution of $E$ (orange), $S$ (green), and $J$ (blue).
\textbf{(b)} The dependence of $E$, $S$, and $J$ on the specific spin $a$. The final configuration corresponds to a static BTZ black hole with $E_\tau= 151$ and $S_\tau= 55$ in Planck units. 
\textbf{(c) }Time evolution of the specific spin $a(t)$. It peaks at $a_{\text{peak}} \approx 0.994$ indicating the state closest to extremality, and then gradually decreases to zero.
\textbf{(d)} The geodesic trajectory of states (green curve) in $(S, J)$ space,  with a starting point at $S_0=5$ and $J_0=1$ (the black dot), terminates when the angular momentum vanishes. The extremal states are depicted by the dashed red curve.
}
\end{figure}

In this case, the final configuration  is a static ($a_\tau=0$) BTZ black hole with $E_\tau = 103\, E_0$ and $S_\tau = 11\, S_0$. It is characterized by a significant increase in energy and entropy, $\delta E = +10226\% $ and $\delta S = +992\%$, as shown on Figs. \ref{figESJt0} and \ref{figESJa0}. Note that the profiles on Fig. \ref{figESJa0}, with respect to $a$, are multi-valued functions, since the specific spin $a(t)$ has a local maximum (Fig. \ref{figat0}).

The results indicate that the optimal process is driven by accretion, whereby matter or energy is supplied to the system. This leads to an initial increase in the black hole’s energy, entropy, and angular momentum, with the specific spin peaking at $a_{\text{peak}} \approx 0.994$ (Fig. \ref{figat0}). Beyond this peak, $a(t)$ gradually decreases --- naturally avoiding any violation of the third law --- and ultimately vanishes in the final static  state. The latter corresponds to the endpoint of the associated geodesic trajectory in the $(S, J)$ space (Fig. \ref{figSJ0}). The subsequent evolution of the BTZ system can then be understood by analyzing the static case discussed in Sec. \ref{secStaticBTZEnergyrep}.

\subsubsection{Optimal process at $\phi=240^\circ$} 

At $\phi = 240^\circ$ the final state is a static BTZ black hole with reduced parameters: $E_\tau = 0.3\,E_0$ and $S_\tau = 0.6\,S_0$, corresponding to a decrease in energy $\delta E = -71\%$ and entropy $\delta S = -42\%$. Such an optimal evaporation may only occur through some physical emission of radiation (scalar, vector, or gravitational modes). 

Table \ref{tabicerep} demonstrates that as the initial angle $\phi$ increases from $240^\circ$, the extent of evaporation also grows, peaking around $\phi = 259^\circ$, where the system approaches its closest point to complete evaporation. Beyond this angle, the evaporation effect weakens, and accretion-driven processes begin to dominate the evolution. It is important to note, however, that complete direct evaporation of the rotating BTZ is never achieved within the energy representation.

\begin{figure}[H]
\centering \hspace{-1.5cm}
\begin{subfigure}{0.45\textwidth}
%
\includegraphics[width=8.3cm,height=5.5cm]{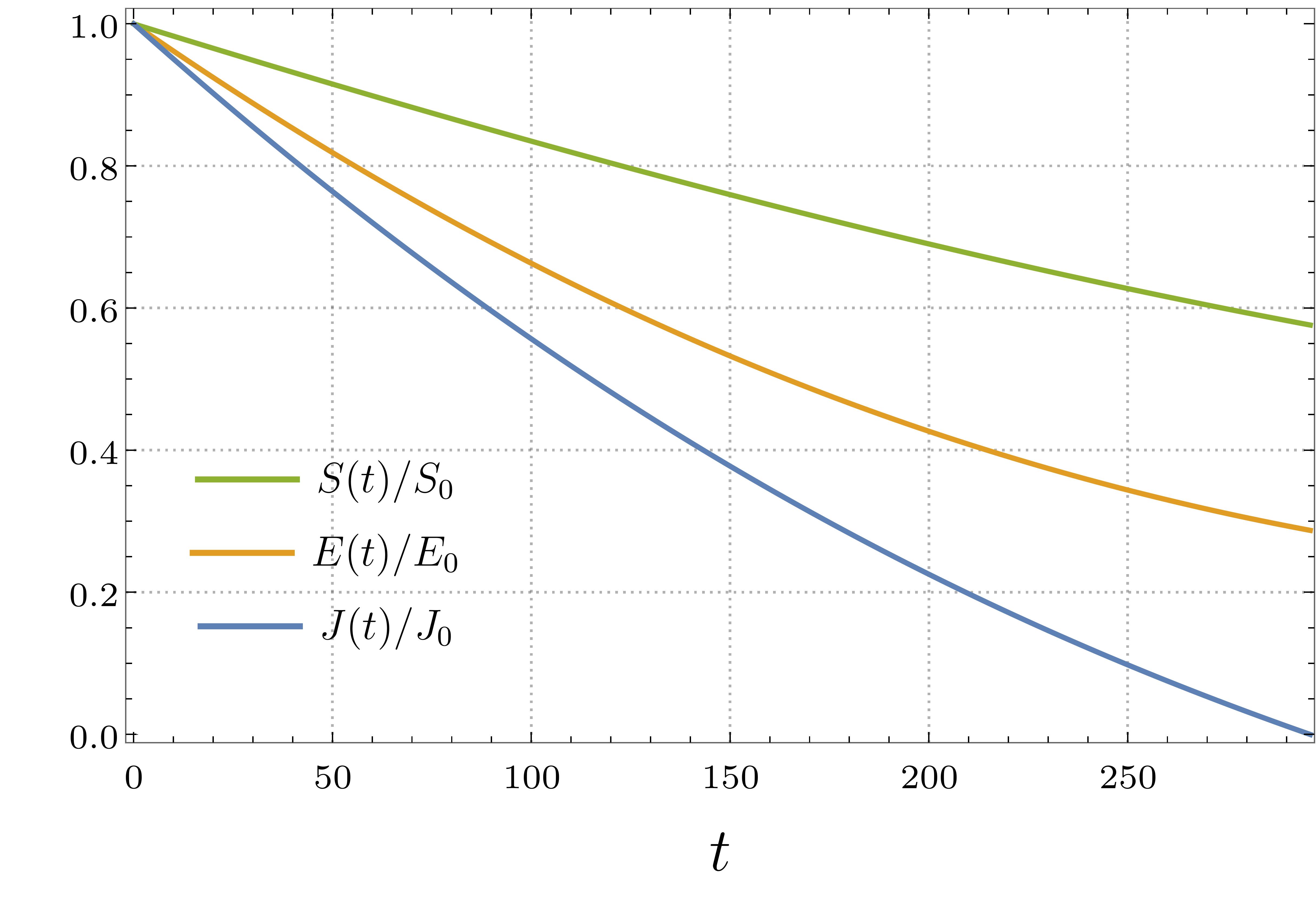}
\caption{\hspace*{-0cm} \scriptsize Optimal profiles: $E(t)/E_0$, $S(t)/S_0$, and $J(t)/J_0$.\vspace{0.3 cm}}\label{}
\end{subfigure}
\hspace{0.45 cm}
\begin{subfigure}{0.45\textwidth}
\includegraphics[width=8.3cm,height=5.5cm]{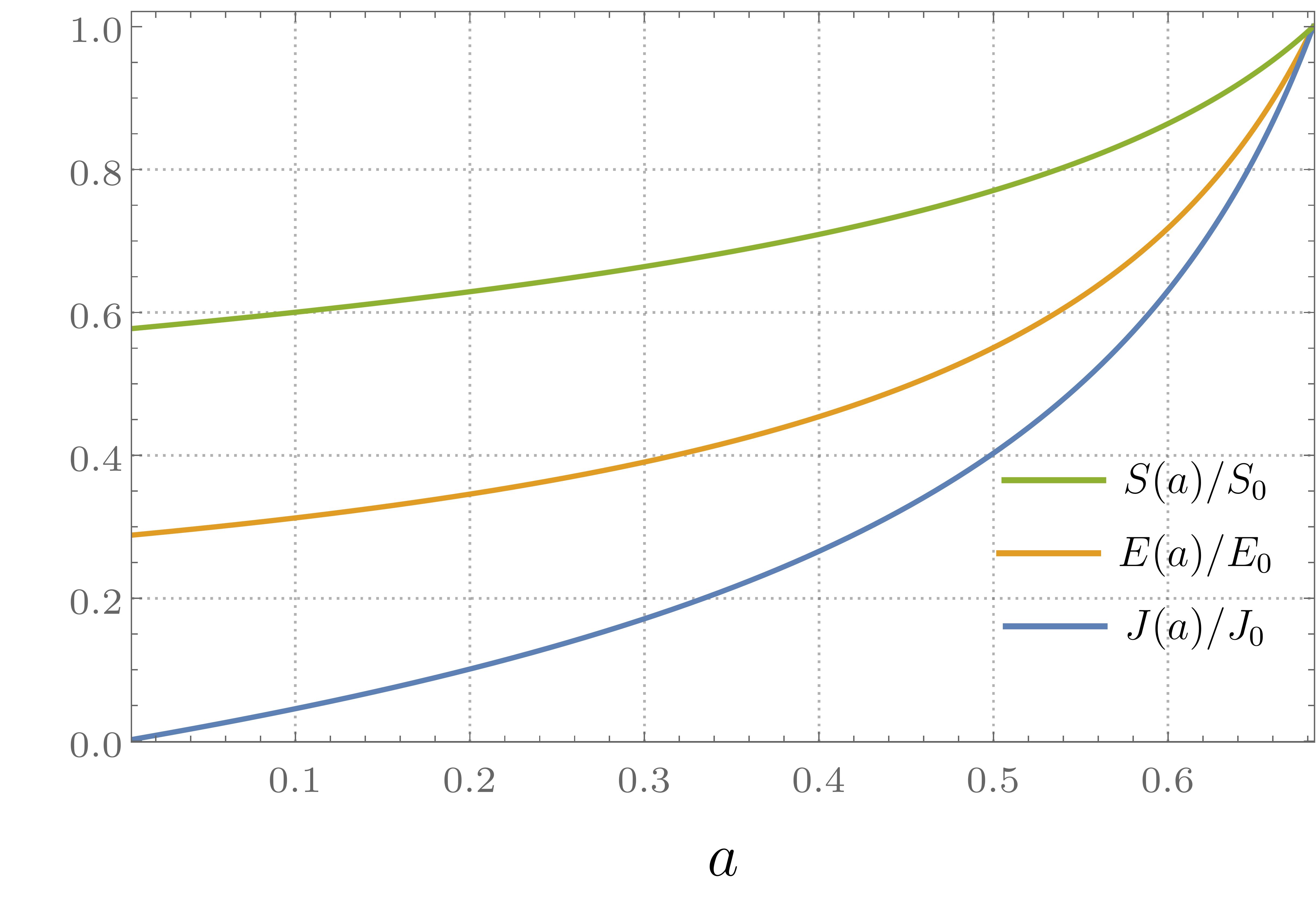}\vspace{-0.05cm}
\caption{\hspace*{-0cm} \scriptsize Optimal profiles: $E(a)/E_0$, $S(a)/S_0$, and $J(a)/J_0$. \vspace{0.3 cm}}\label{}
\end{subfigure}
%
\begin{subfigure}{0.45\textwidth}
\includegraphics[width=8.3cm,height=5.8cm]{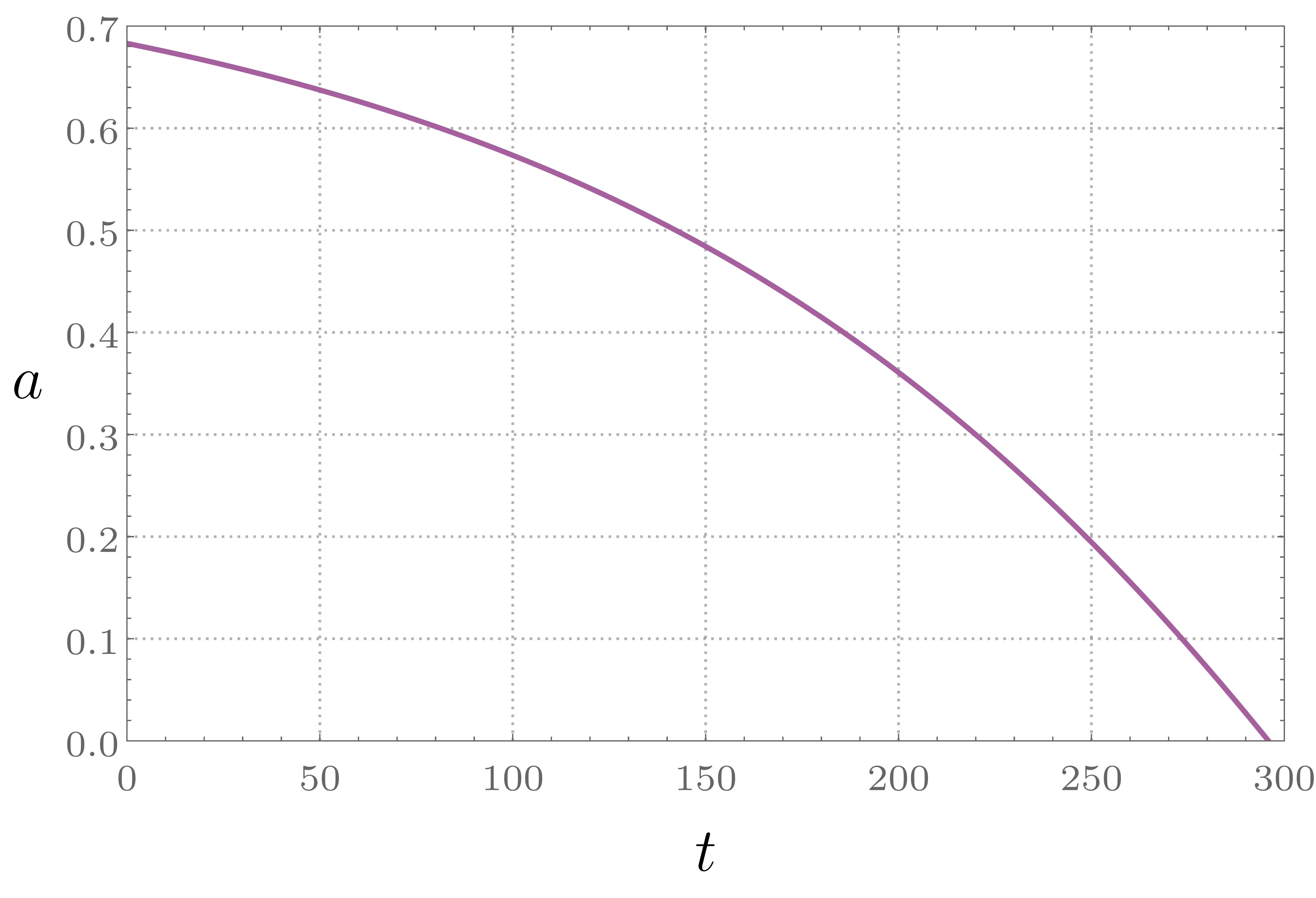}
\caption{\hspace*{-0cm} The specific spin $a(t)$.}\label{}
\end{subfigure}
\hspace{0.3 cm}
\begin{subfigure}{0.43\textwidth}
\includegraphics[width=8.3cm,height=5.6cm]{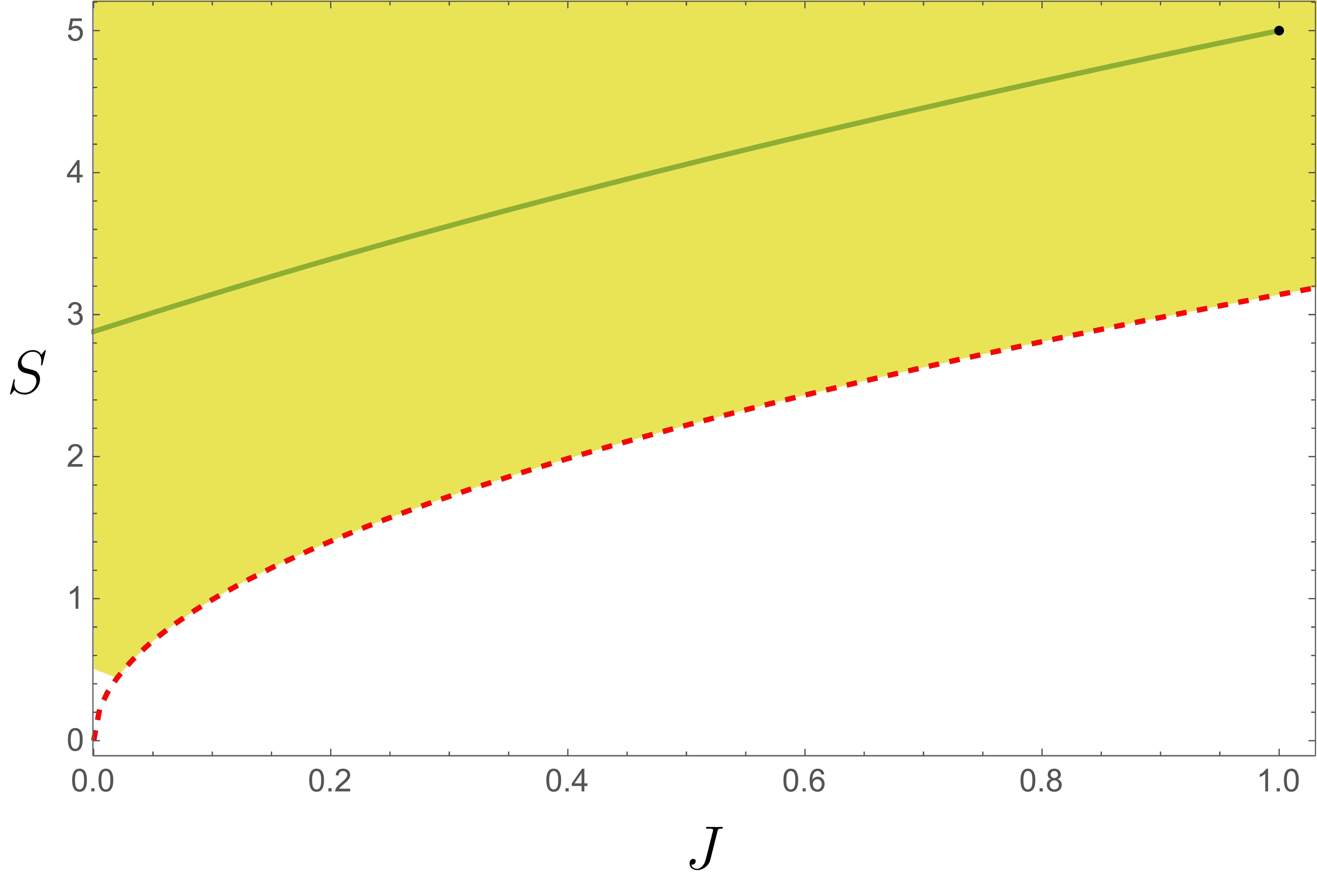}\vspace{0.05cm}
\caption{\hspace*{-0cm} The geodesic path of states in $(S, J)$ space.}\label{}
\end{subfigure}
 \hspace{1 cm} 
 \vspace{-0.0cm}
\caption{ Profiles for $\phi=240^\circ$. 
\textbf{(a)} Time evolution of $E$ (orange), $S$ (green), and $J$ (blue).
\textbf{(b)} The dependence of $E$, $S$, and $J$ on the specific spin $a$. The final configuration corresponds to a static BTZ black hole with $E_\tau= 0.4$ and $S_\tau= 2.9$ in Planck units. 
\textbf{(c) } The evolution of 
$a(t)$ is strictly monotonic, tending towards zero.
\textbf{(d)} The geodesic trajectory of states (green curve) in $(S, J)$ space,  with a starting point $S_0=5$ and $J_0=1$ (black dot), terminates when the angular momentum vanishes. Extremality is depicted by the dashed red curve.
}
\end{figure}	

\subsubsection{Summary for additional process}

\begin{figure}[H]
\centering \hspace{0.5cm}
\begin{subfigure}{0.4\textwidth}
\includegraphics[width=6.6cm,height=6cm]{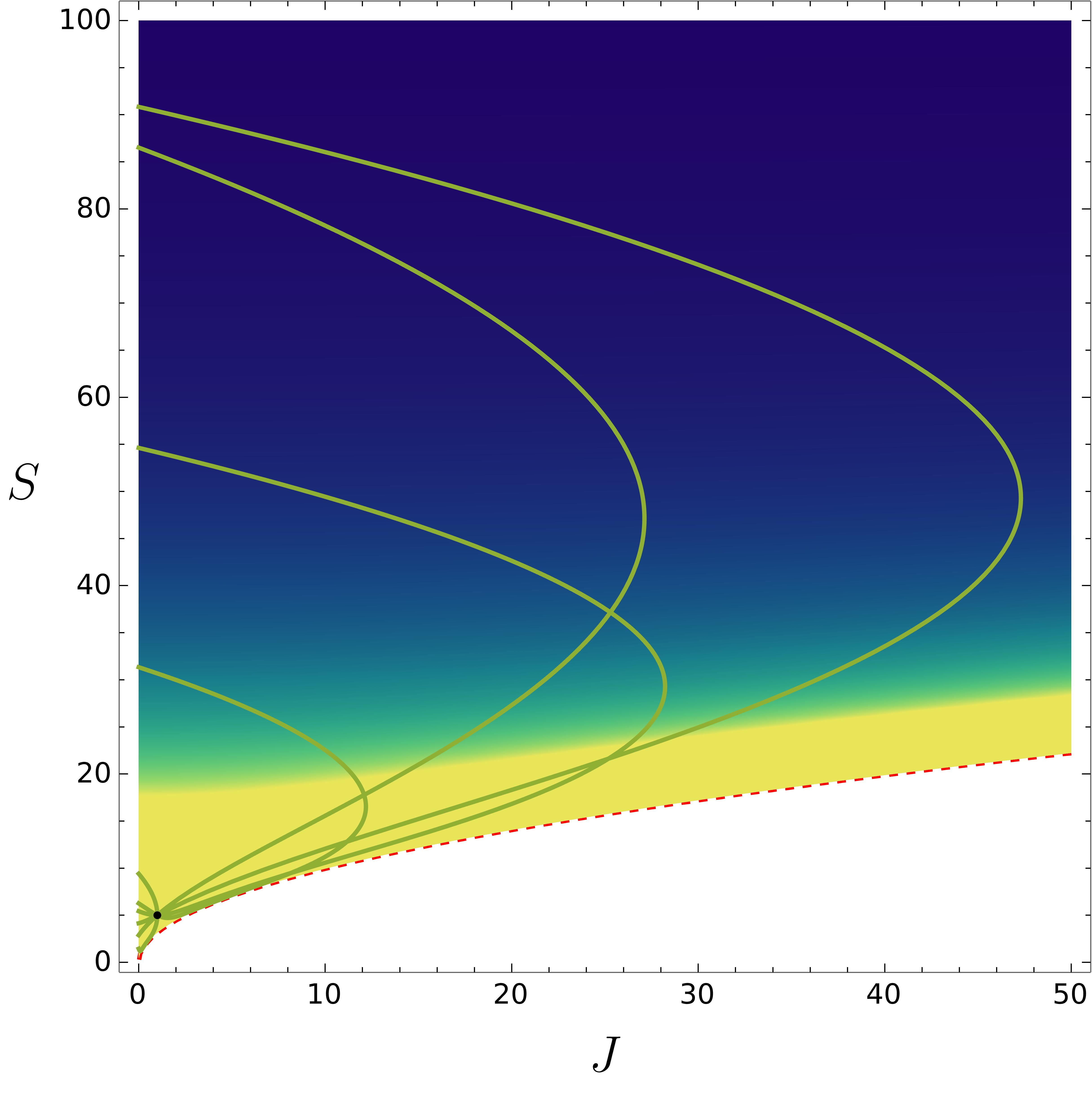}
\caption{\hspace*{-0.0cm} }\label{figGeodSJFull}
\end{subfigure}
\hspace{0.0 cm}
\begin{subfigure}{0.4\textwidth}
\includegraphics[width=6.2cm,height=6cm]{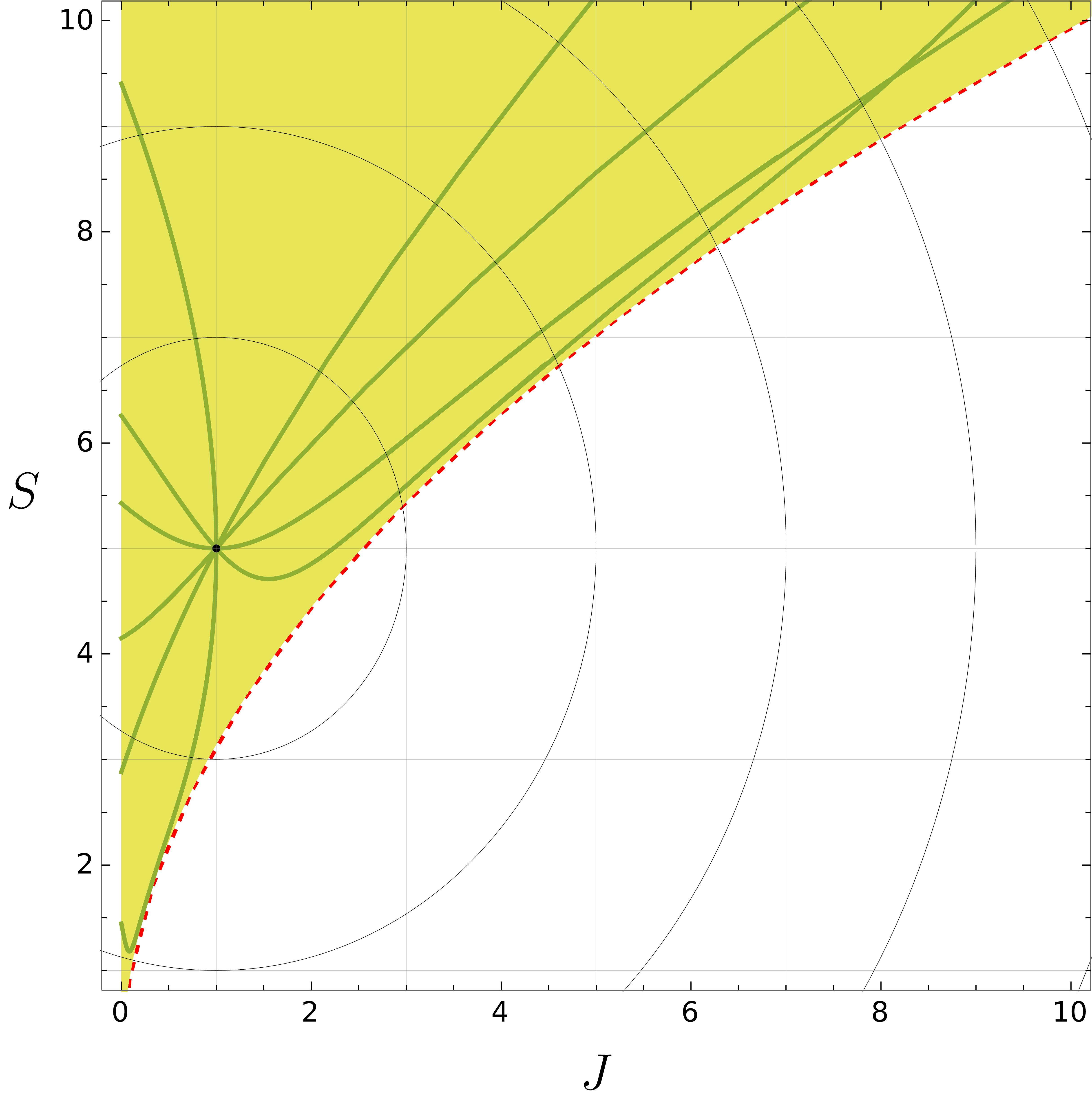}\vspace{-0.05cm}
\caption{\hspace*{-0.0cm} }\label{figGeodSJZoom}
\end{subfigure}
 \hspace{0.7 cm}
\caption{ Geodesics paths of states in $(S, J)$ space. \textbf{(a)} The thermodynamic curvature $R$ is represented as a colored background. The initial state $(S_0 = 5, J_0 = 1)$ lies in a relatively strong curved region (yellow) with high interactions, though it remains sufficiently distant from the extremal boundary (the dashed red curve).
\textbf{(b)} A magnified view near the initial point. Polar circles help visualize the initial angles $\phi$, which determine the direction of each geodesic. All paths naturally avoid extremality and drive the system to a static BTZ black hole configuration.
}\label{figGeodSJ}
\end{figure}

Figure \ref{figGeodSJ} illustrates several geodesic trajectories in the $(S, J)$ space, with the corresponding data\footnote{The data in the table is rounded up to two relevant digits.} summarized in Table \ref{tabicerep}. The results show that for initial angles $\phi$ between $0^\circ$ and $180^\circ$, the dynamics is dominated by accretion-driven optimal processes. As $\phi$ increases beyond $180^\circ$, evaporation processes become increasingly significant, reaching a maximum around $\phi = 259^\circ$, where the system comes closest to complete evaporation. Beyond this point, the influence of evaporation diminishes, and accretion-driven behavior once again dominates. 

Notably, full evaporation of the rotating BTZ black hole is never realized within the energy representation. The final configuration always corresponds to a static state ($a_\tau=0$) with non-zero energy and entropy.

\begin{table}[H]
\renewcommand{\arraystretch}{1.4} 
\centering
\footnotesize
\begin{tabular}{|r|r|r|r|r|r|r|r|r|r|r|r|}
\hline
 $\phi$ & $\dot E_0\, 10^{-3}$ & $\dot S_0\,10^{-3}$ & $\dot J_0\, 10^{-3}$ & $\dot a_0\, 10^{-3}$ & $a_\tau$ & $E_\tau$ & $\delta E$ \%   & $S_\tau$ & $\delta S$ \%  & $\tau$ & $\mathcal{L}^2$ \\ \hline\hline
 $0^\circ$ & 3.9 & 0.0 & 10 & 5.0 & 0.0 & 151 & $+10^4$  & 55 & $+10^3$  & 2584 & 264\\ \hline
 $45^\circ$ & 5.8 & 7.1 &  7.1 & 2.1  & 0.0 & 418 & $+10^4$  & 91 & $+10^3$ & 8070 & 741\\ \hline
 $60^\circ$ &  5.7 & 8.7 & 5.0 & 0.8 & 0.0 & 379 & $+10^4$  & 87 & $+10^3$ & 9539 & 668\\ \hline
 $90^\circ$ & 4.3 & 10 & 0.0 & $-2.0$ & 0.0 &  4.5 & $+200$  & 9.4 & $+88$ & 421 & 2.6\\ \hline
 $135^\circ$ & 0.2 & 7.1 & $-7.1$ & $-4.9$ & 0.0 & 2.0 & $+33$  &  6.3 & $+25$ & 122 & 0.6\\ \hline
 $180^\circ$ & $-3.9$ & 0.0 & $-10$ & $-5.0$ & 0.0  & 1.5 & $+0$  & 5.4 & $+9$ & 106 & 0.5\\ \hline
 $225^\circ$ &  $-5.8$ & $-7.1$ & $-7.1$  &  $-2.1$ & 0.0 & 0.9 & $-40$  & 4.1 & $-17$ & 185 & 0.4\\ \hline
 $240^\circ$ & $-5.7$ & $-8.7$ & $-5.0$ & $-0.8$  & 0.0 & 0.4 & $-73$  & 2.9 & $-42$  &  296 & 0.6\\ \hline
 $259^\circ$ &  $-4.9$ & $-9.8$ & $-1.9$ & 1.0 & 0.0 & 0.02 & $-99$  & 0.7 & $-86$ &  408 & 1.6\\ \hline
 $270^\circ$ &  $-4.3$ & $-10$ & 0.0 & 2.0 & 0.0 & 0.1 & $-93$ & 1.4 & $-72$ &  354 & 1.9\\ \hline
 $315^\circ$ & $-0.2$ & $-7.1$ & $7.1$  & 4.9 & 0.0  & 50 & $+10^3$  & 31 & $+527$ & 1439 &89 \\ \hline
\end{tabular}
\caption{
The table summarizes the initial rates of key parameters across various processes. It also indicates how the final state differs from the initial BTZ configuration after each process. Additionally, the duration $\tau$ of the process (in Planck units) and the thermodynamic length squared, $\mathcal{L}^2$, are provided. The thermodynamic length $\mathcal{L}^2$ quantifies the minimal energy required to perform the transformation. Smaller $\mathcal{L}^2$ indicates higher probability for the process to occur.}\label{tabicerep}
\end{table}
\normalsize

\section{Optimal process  in entropy representation}\label{secOPEntrR}

We now proceed by investigating  optimal processes on the space of macrostates of the BTZ black hole in entropy representation. A comparison of optimal profiles with the nonoptimal Hawking evaporation process  is presented in Section \ref{secOvsHentropyrot}.

\subsection{Thermodynamic metric and geodesic equations}

The Ruppeiner metric is given by the Hessian of entropy \eqref{eqESenergyRep} in $(E,J)$ space:
\begin{equation}
 \hat g=\epsilon \!\left(\!\!
\begin{array}{cc}
 \frac{\partial^2 S}{\partial E^2}\big |_J & \frac{\partial^2 S}{\partial E\partial J} \\[5pt]
 \frac{\partial^2 S}{\partial E\partial J} & \frac{\partial^2 S}{\partial J^2}\big |_E  \\
\end{array}
\!\! \right) 
=
 \epsilon \! \left(\!\!
\begin{array}{cc}
 \frac{(U-2 E) \sqrt{\lambda  (E+U)}}{4 \zeta  (E-\zeta  J)^{3/2} (E+\zeta  J)^{3/2}} & -\frac{\zeta  J \sqrt{\lambda } \left(\zeta ^2 J^2-3 E (E+U)\right)}{4 (E+U)^{3/2} (E-\zeta  J)^{3/2} (E+\zeta  J)^{3/2}} \\[5pt]
 -\frac{\zeta  J \sqrt{\lambda } \left(\zeta ^2 J^2-3 E (E+U)\right)}{4 (E+U)^{3/2} (E-\zeta  J)^{3/2} (E+\zeta  J)^{3/2}} & \frac{\zeta  (U-2 E) \sqrt{\lambda  (E+U)}}{4 (E-\zeta  J)^{3/2} (E+\zeta  J)^{3/2}} \\
\end{array}
\!\! \right)\!,
\end{equation}
where $U=\sqrt{E^2-\zeta ^2 J^2}$.  
Its thermodynamic curvature is zero, thus realizing a flat information space. We parametrize the $(E,J)$ coordinates by an affine time parameter $t$ and seek to obtain their optimal profiles by solving the geodesic equations on the space of macrostates:
\begin{align}\label{EqEenrep}
&\ddot E + \Gamma^E_{EE} \dot E^2 +2\Gamma^E_{E J}  \dot E\dot J +\Gamma^E_{JJ} \dot J^2=0, \\[5pt]\label{EqJenrep}
&\ddot J + \Gamma^J_{JJ} \dot J^2 +2\Gamma^J_{E J}\dot E\dot J +\Gamma^J_{E E} \dot E^2=0,
\end{align}
where the corresponding Christoffel symbols are given by:
\begin{align}
&\Gamma^{E}_{EE}=\frac{3 E}{4 \zeta ^2 J^2-4 E^2},\quad \Gamma^{E}_{EJ}=\frac{3 \zeta ^2 J}{4 E^2-4 \zeta ^2 J^2},\quad \Gamma^{E}_{JJ}=\frac{3 E \zeta ^2}{4 \zeta ^2 J^2-4 E^2},
\\[5pt]
&\Gamma^{J}_{EE}=\frac{3 J}{4 E^2-4 \zeta ^2 J^2},\quad \Gamma^{J}_{EJ}=\frac{3 E}{4 \zeta ^2 J^2-4 E^2},\quad \Gamma^{J}_{JJ}=\frac{3 \zeta ^2 J}{4 E^2-4 \zeta ^2 J^2}.
\end{align}

The solutions to these nonlinear equations define the optimal paths for energy $E(t)$ and angular momentum $J(t)$ at various initial conditions:
\begin{equation}
E(0)=E_0,\quad \dot E(0)=\dot E_0,\quad 	J(0)=J_0,\quad \dot J(0)=\dot J_0,
\end{equation}
where $E_0$ and $J_0$ denote the initial energy and angular momentum of the BTZ black hole, respectively, and $\dot{E}_0$ and $\dot{J}_0$ represent their initial rates of change. The profiles for $S(t)$ and $a(t)$ follow from their corresponding definitions (\ref{eqESenergyRep}) and (\ref{eqSpecificSpin}).

\subsection{Optimal evaporation of the static BTZ black hole}\label{OptimalEvap}

{  
	
In the static case $(J = 0)$, Eq. \eqref{EqJenrep} is trivially satisfied, while Eq. \eqref{EqEenrep} leads to the equation for the energy profile: 
\begin{align}
	\ddot E -\frac{3}{4 E} \dot E^2 =0, \quad E_0=E(0), \quad \dot E_0=\dot E(0).
\end{align}
The solution to this equation is nonlinear in time:
\begin{equation}
   E(t) = E_0 \bigg(\!1 - \frac{|\dot E_0| }{4 E_0}\, t\bigg)^{\!4}.
\end{equation}
Here we already assumed an evaporation process with an initial rate: $\dot E_0=-|\dot E_0| < 0 $. The optimal evaporation time is determined by $E(t)=0$:
\begin{equation}
 \tau_{evap} = \frac{4 E_0}{|\dot E_0|}.
\end{equation}
In contrast to the Hawking evaporation time (which is infinite, as shown in Sec. \ref{secHESBTZBHa}), here the optimal evaporation time is finite.
The thermodynamic length of the fluctuation, from $E_0$ at $t=0$ to $E_{\tau}$ at $t=\tau$, is given by:
\begin{equation}\label{eqTDLSJstatica}
	\mathcal{L}_{E_0\to E_\tau} = \int \nolimits_{0}^{\tau} \sqrt {{g_{ab}}\dot x^a \dot x^b}\, dt=\sqrt{-\epsilon } \,\frac{|\dot E_0| \sqrt[4]{\lambda } }{2^{3/4} E_0^{3/4} \sqrt{\zeta }}\,\tau=v \tau,
\end{equation}
where $x^a(t) = \big( E(t), J(t) = 0\big)$, and $v$ is the thermodynamic speed of the process. The length is real and positive if $\epsilon<0$. On the other hand, directly computing the thermodynamic length by inserting $x^a = (E,J = 0)$, one finds:
\begin{equation}\label{eqTDlengthentr}
	\mathcal{L}_{E_0\to E_\tau} = \int \sqrt {{g_{ab}} dx^a dx^b}=2  \sqrt[4]{ 2\lambda } \sqrt{\frac{-\epsilon }{\zeta }}\big(\sqrt[4]{E_0}-\sqrt[4]{E_
		\tau}\,\big).
\end{equation}
Comparing  both lengths we can estimate the average relaxation time $\bar \tau$ of the process:
\begin{equation}\label{eqTDtimeentr}
	\bar \tau\approx \frac{4 E_0^{3/4} }{|\dot E_0|}\big(\sqrt[4]{E_0}-\sqrt[4]{E_
		\tau} \,\big).
\end{equation}
The result for the average relaxation time is non-trivial. In the previous representation it was proportional to the entropy difference between the initial and the final state \eqref{eqRelTE}, here it involves the difference of the fourth root of the initial and the final energy of the BTZ system. 

For a full optimal evaporation the  time is $t_{evap}={4E_0}/{|\dot E_0|}$. Thus, the evaporation length is:
\begin{equation}\label{eqLTDevapEntropyRep}
	\mathcal{L}_{evap}=\frac{2 \sqrt[4]{2\lambda } \sqrt{-\epsilon }}{\sqrt{\zeta }} \,\sqrt[4]{E_0}
	=2 \sqrt{\frac{-\epsilon\pi c k \ell   }{\hbar }} \sqrt[4]{\frac{2 E_0}{G_3}} .
\end{equation}

The square of the thermodynamic length, $\mathcal{L}^2$, carries units of entropy (Joules per Kelvin), consistent with its interpretation as the minimal entropy produced during the process. This interpretation allows one to determine the metric scale factor $\epsilon$ for the complete evaporation of the BTZ. In this case, it is reasonable to require that $\mathcal{L}^2$ be at least equal to the initial entropy of the BTZ black hole, given by $S_0 = \sqrt{2\lambda E_0} / \zeta$. Imposing this condition yields $\epsilon = -1/4$. 

Finally, using Eq. \eqref{eqESenergyRep} we can also write the profile of the entropy:
\begin{equation}
S(t) =S_0 \Big(1 - \frac{|\dot S_0|}{2S_0}t \Big)^2, \quad S_0= \frac{\sqrt{2\lambda E_0}}{\zeta}, \quad |\dot S_0|=\frac{|\dot E_0|}{\zeta} \sqrt{\frac{\lambda}{2E_0}}. 
\end{equation} 

A comparison between these optimal profiles and the nonoptimal Hawking evaporation process is presented in Section \ref{secOvsHentropystatic}.
}

\subsection{Optimal processes of the rotating BTZ black hole}

In contrast to the results obtained in the energy representation, the entropy ensemble admits a broader variety of final configurations. For example, the analysis shows that certain optimal trajectories asymptotically drive the system toward near-extremal BTZ states. Yet this limit cannot be reached within finite classical time, in agreement with the third law of thermodynamics. Other trajectories instead drive the BTZ black hole toward a fixed, non-vanishing specific spin (far enough from extremality), while a third class of geodesics inevitably terminates in a static configuration within finite time.

\subsubsection{Initial conditions}

The initial state will be defined by the following initial parameters of the BTZ black hole:
\begin{equation}\label{eqIniCSJa}
   E_0 = 5,\quad J_0 = 1, \quad S_0 =\pi \sqrt{\ell\left(E_0 \ell +\sqrt{E_0^2 \ell^2-J_0^2} \,\right)}= 10,\quad a_0 = \frac{J_0}{E_0\ell}= 0.2,
\end{equation}
where $\ell =1$. The initial rates of change will be presented in polar coordinates $(u,\phi)$ by:  
\begin{equation}\label{eqInitCondRelsPolar}
\dot E_0=u\sin \phi, \quad  \dot J_0=u\cos \phi, \quad u=\sqrt{\dot E^2_0 +\dot J^2_0}, \quad \tan \phi=\frac{\dot E_0}{\dot J_0}.
\end{equation}
The initial rates for several angles $\phi$ and $u=0.01$ are presented in Table \ref{tabicsrep}. Their corresponding geodesic paths in $(E,J)$ space are displayed on Figure \ref{figRicciFlatGeodP}. In this case, the thermodynamic curvature $R$ is zero, thus the $(E,J)$ space is Ricci flat.

\subsubsection{Profiles at $\phi=0^\circ$}

\begin{figure}[H]
\centering \hspace{-1.5cm}
\begin{subfigure}{0.45\textwidth}
%
\includegraphics[width=8.3cm,height=5.5cm]{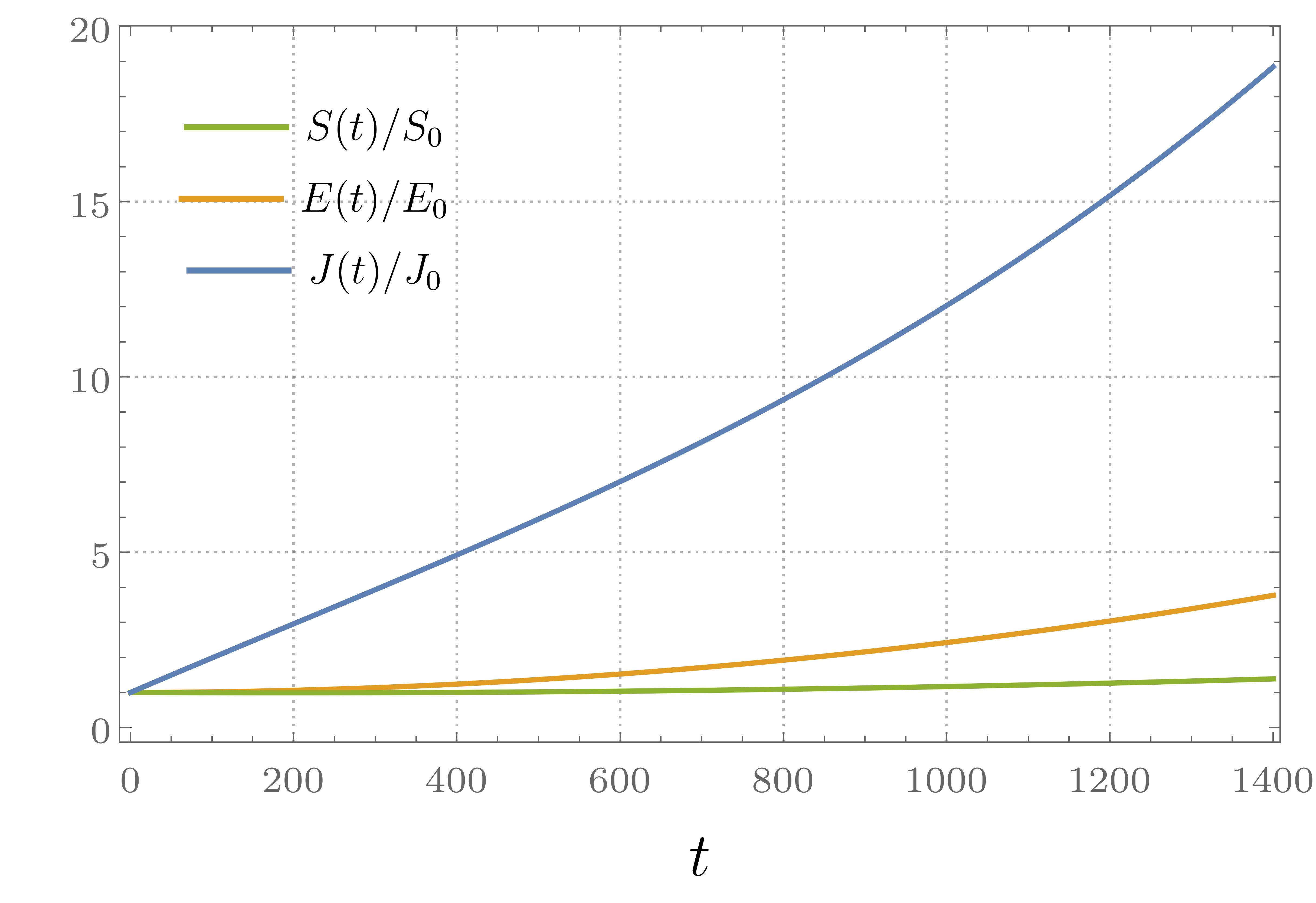}
\caption{\hspace*{-0cm} \scriptsize Optimal profiles: $E(t)/E_0$, $S(t)/S_0$, and $J(t)/J_0$.\vspace{0.3 cm}}\label{}
\end{subfigure}
\hspace{0.45 cm}
\begin{subfigure}{0.45\textwidth}
\includegraphics[width=8.3cm,height=5.5cm]{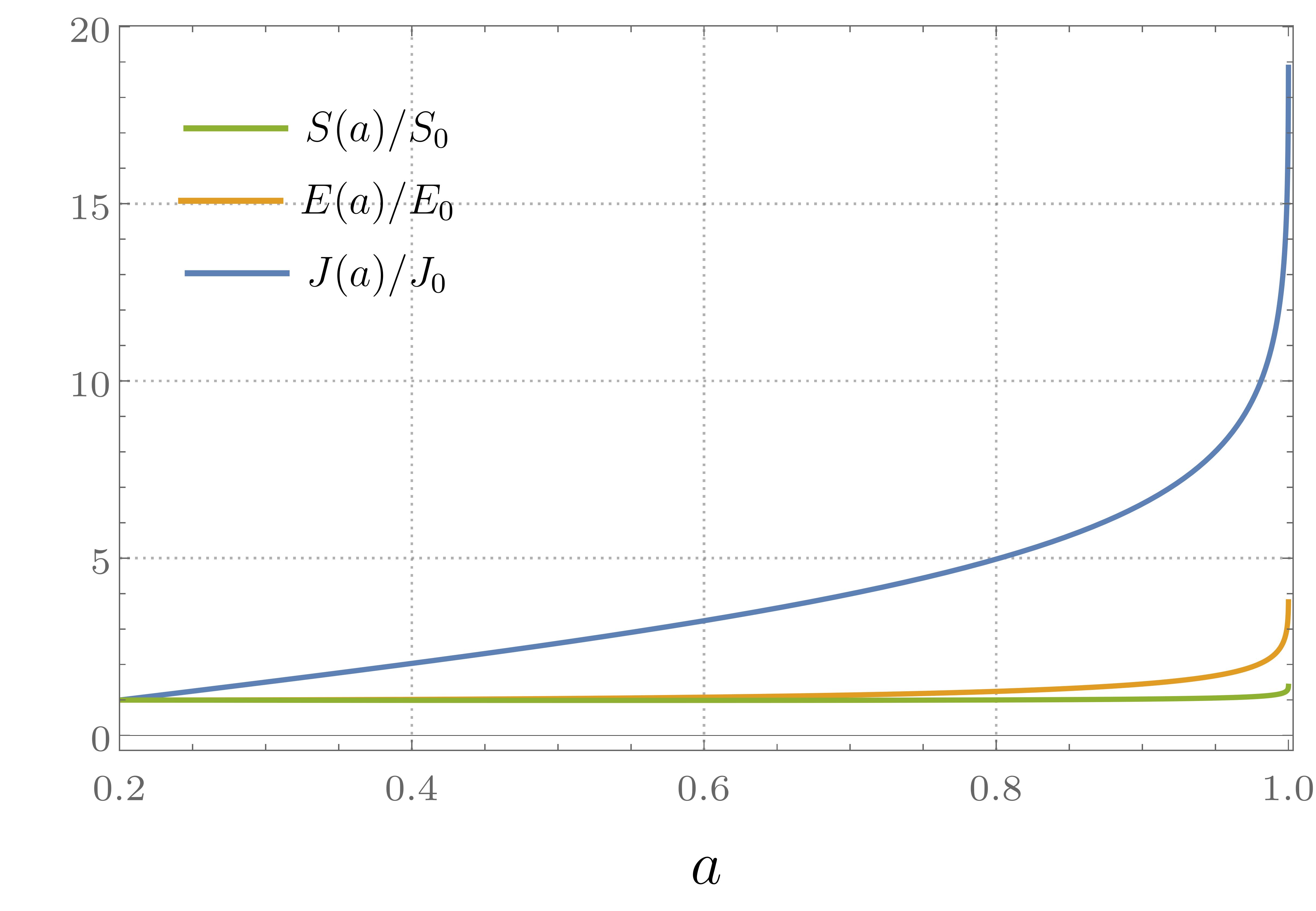}\vspace{-0.05cm}
\caption{\hspace*{-0cm} \scriptsize Optimal profiles: $E(a)/E_0$, $S(a)/S_0$, and $J(a)/J_0$. \vspace{0.3 cm}}\label{}
\end{subfigure}
%
\begin{subfigure}{0.45\textwidth}
\includegraphics[width=8.3cm,height=5.5cm]{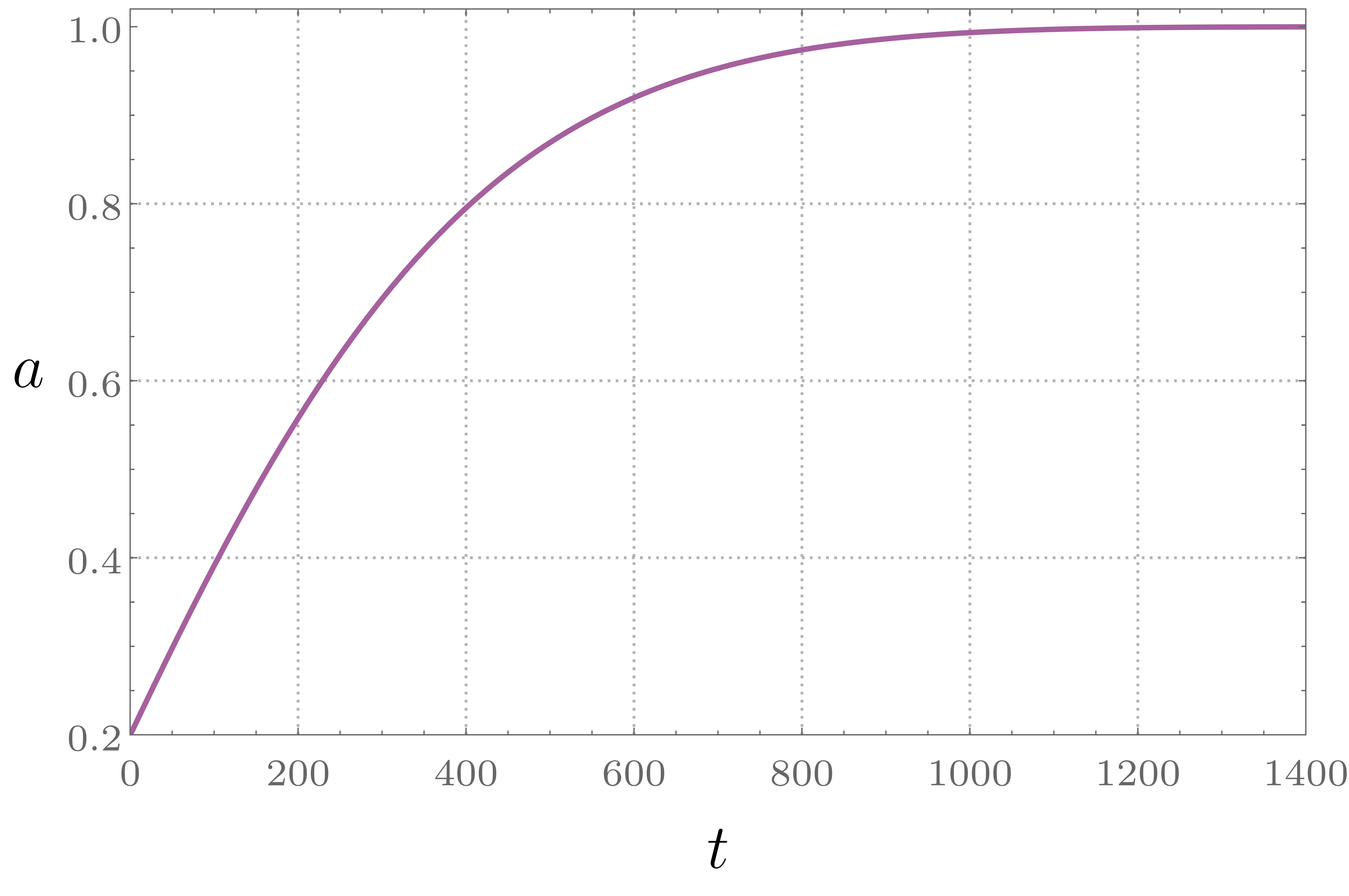}
\caption{\hspace*{-0cm} The specific spin $a(t)$.}\label{figsat0}
\end{subfigure}
\hspace{0.8cm}
\hspace{-0.7cm}
\begin{subfigure}{0.43\textwidth}
\includegraphics[width=8.3cm,height=5.5cm]{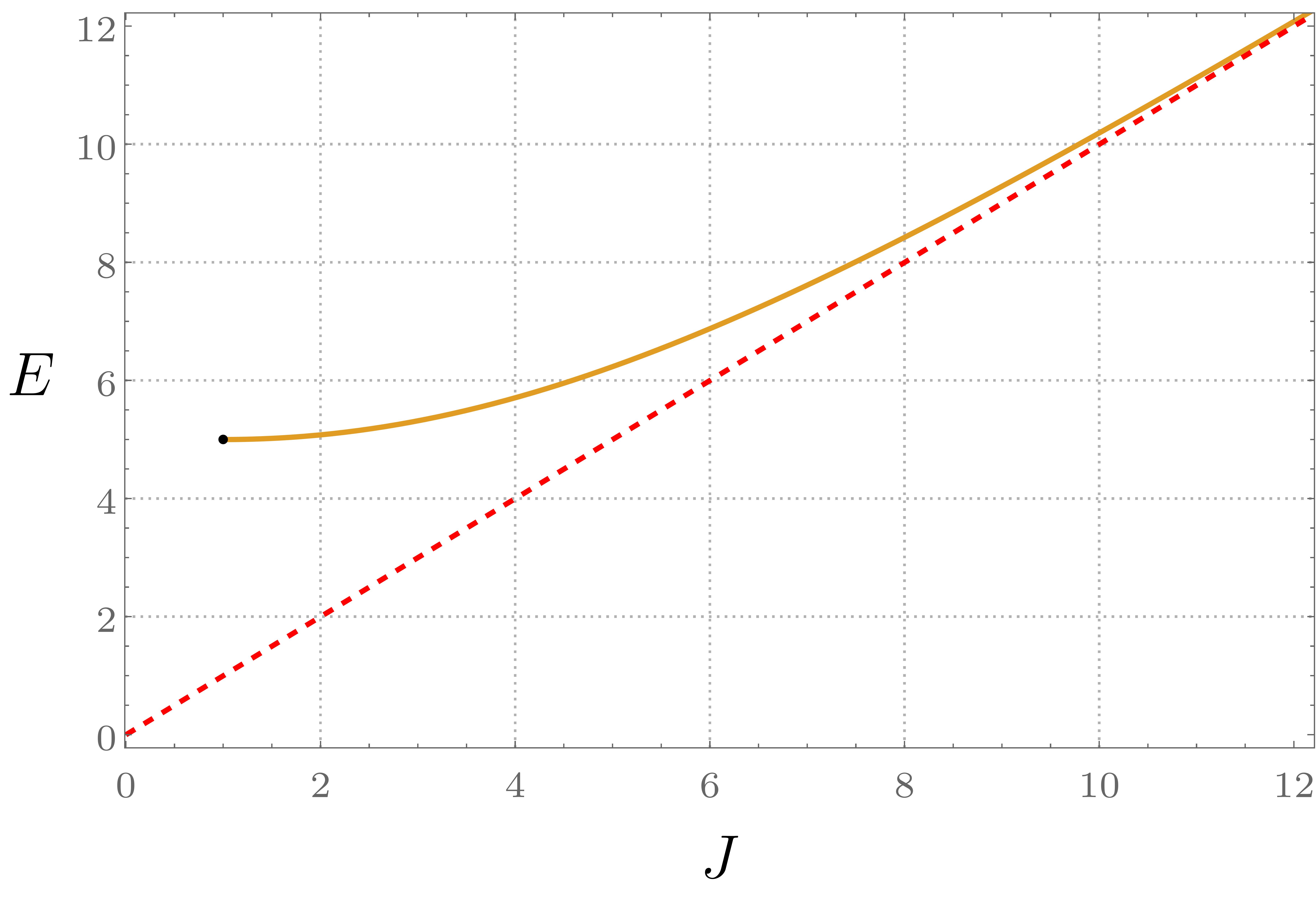}\vspace{0.05cm}
\caption{\hspace*{-0cm} The geodesic path of states in $(E, J)$ space.}\label{}
\end{subfigure}
 \hspace{1 cm} 
 \vspace{-0.0cm}
\caption{ Profiles for $\phi=0^\circ$. 
\textbf{(a)} Time evolution of $E$ (orange), $S$ (green), and $J$ (blue).
\textbf{(b)} The dependence of $E$, $S$, and $J$ on the specific spin $a$. The BTZ black hole does not settle to a final state. It fluctuates between near-extremal states forever asymptotically approaching extremality.  
\textbf{(c) }Time evolution of the specific spin $a(t)$. The spin asymptotically  increases towards the extremal value.
\textbf{(d)} The geodesic trajectory of states (orange curve) in the $(E, J)$ space starts at $E_0 = 5$ and $J_0 = 1$ (black dot), and asymptotically approaches the extremal curve (dashed red) over time.
}
\end{figure}

In this case, the specific spin $a(t)$ of the BTZ black hole asymptotically approaches its extremal value (Fig. \ref{figsat0}). Since this limit cannot be reached within finite time, the system continues to fluctuate indefinitely between nearly extremal states. The process leads to a monotonic growth of all black hole parameters, characterizing it as an accretion-type evolution.

\subsubsection{Profiles at $\phi=240^\circ$}

\begin{figure}[H]
\centering \hspace{-1.5cm}
\begin{subfigure}{0.45\textwidth}
%
\includegraphics[width=8.3cm,height=5.5cm]{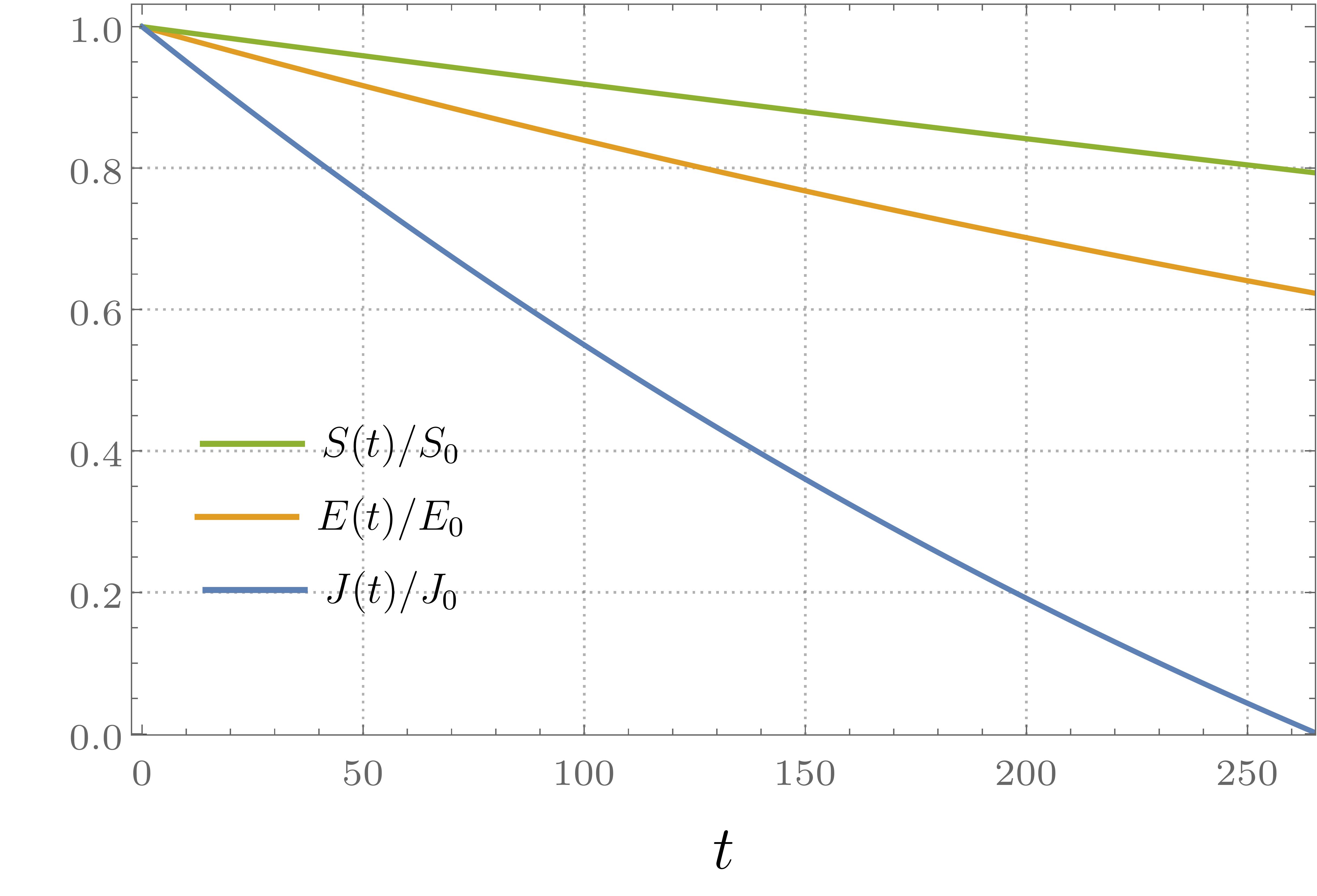}
\caption{\hspace*{-0cm} \scriptsize Optimal profiles: $E(t)/E_0$, $S(t)/S_0$, and $J(t)/J_0$.\vspace{0.3 cm}}\label{}
\end{subfigure}
\hspace{0.55 cm}
\begin{subfigure}{0.45\textwidth}
\includegraphics[width=8.3cm,height=5.5cm]{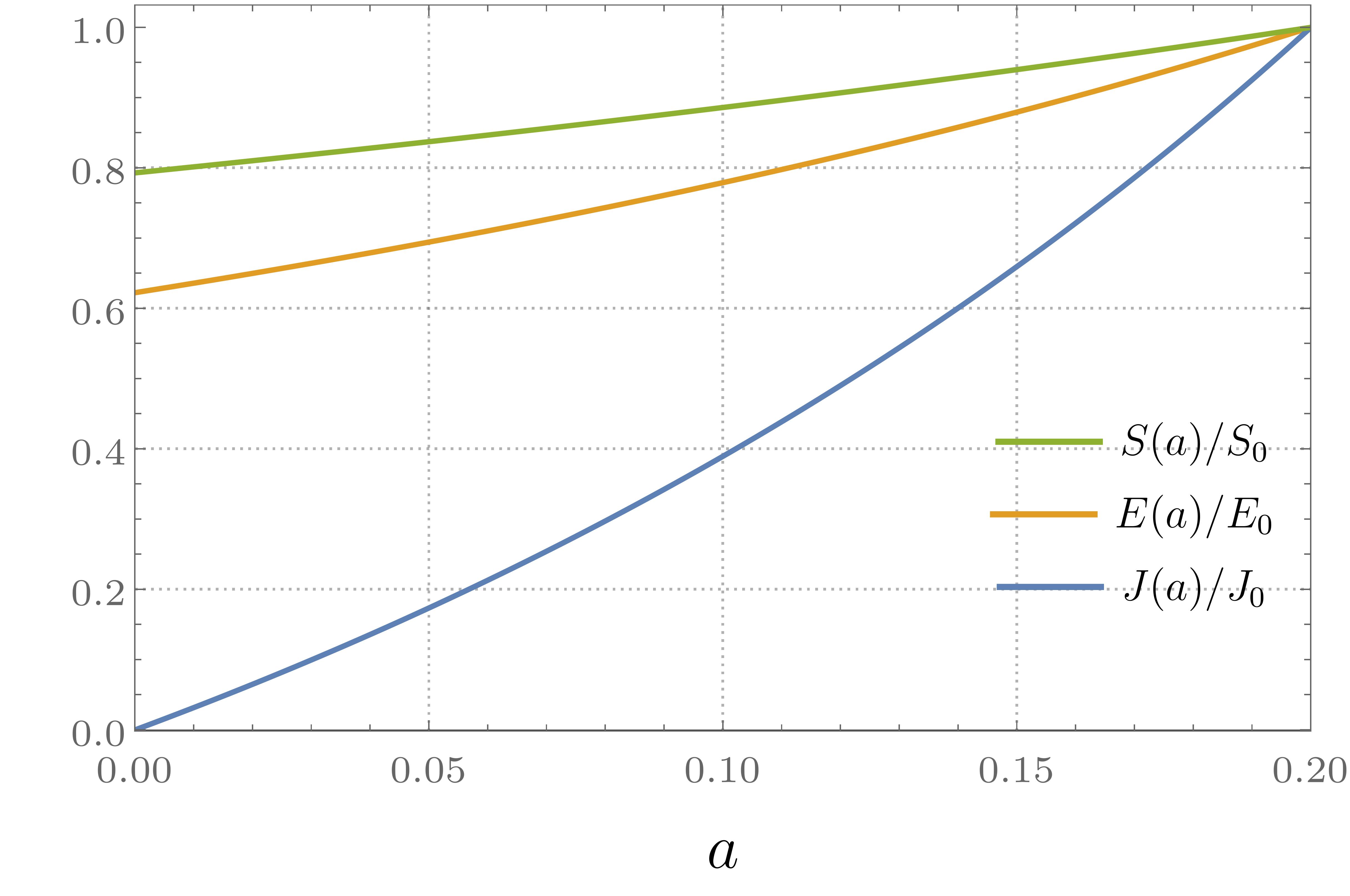}\vspace{-0.05cm}
\caption{\hspace*{-0cm} \scriptsize Optimal profiles: $E(t)/E_0$, $S(t)/S_0$, and $J(t)/J_0$. \vspace{0.3 cm}}\label{}
\end{subfigure}
%
\begin{subfigure}{0.45\textwidth}
\includegraphics[width=8.3cm,height=5.5cm]{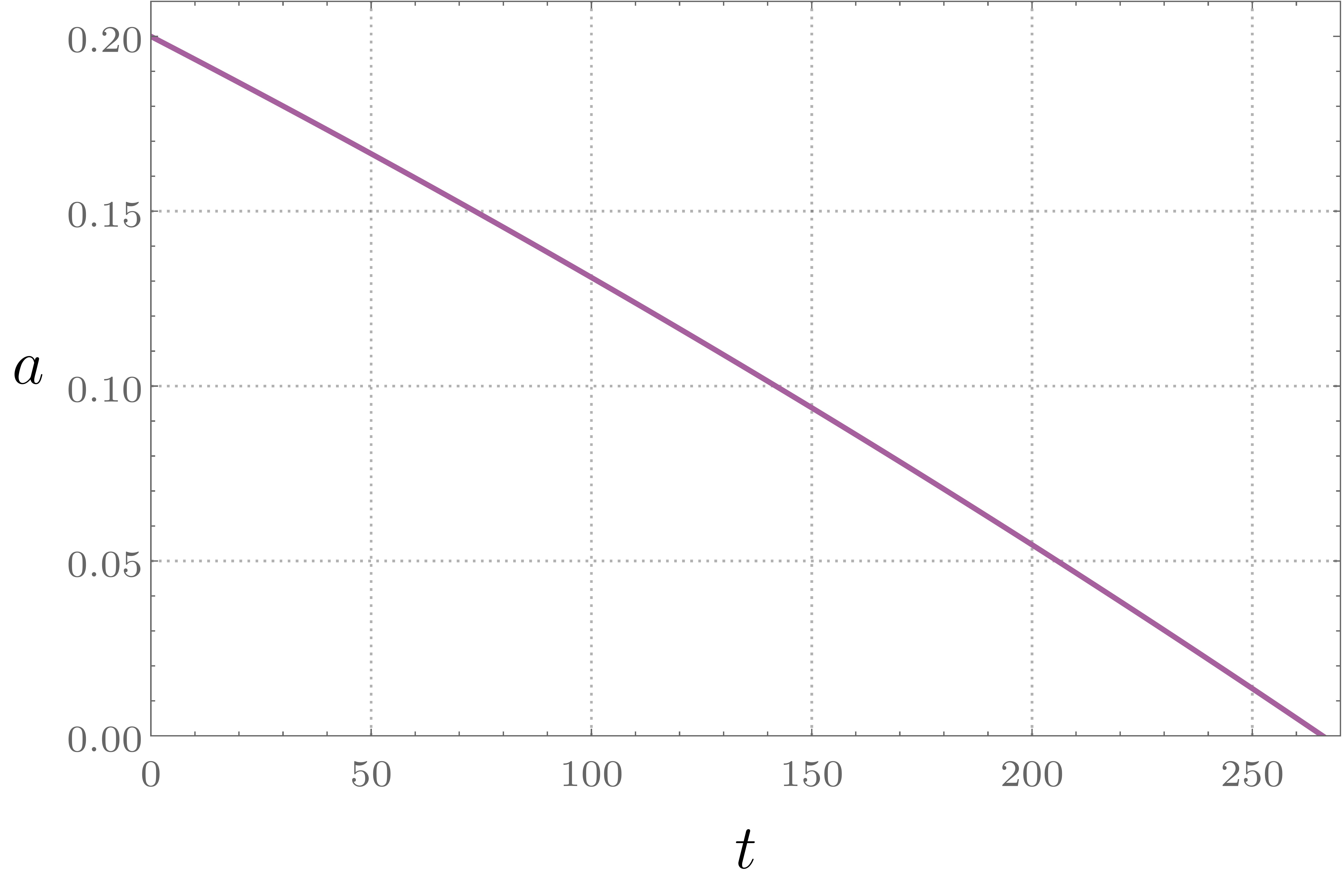}
\caption{\hspace*{-0cm} The specific spin $a(t)$.}\label{figsat240}
\end{subfigure}
\hspace{0.5cm}
\begin{subfigure}{0.43\textwidth}
\includegraphics[width=8.3cm,height=5.5cm]{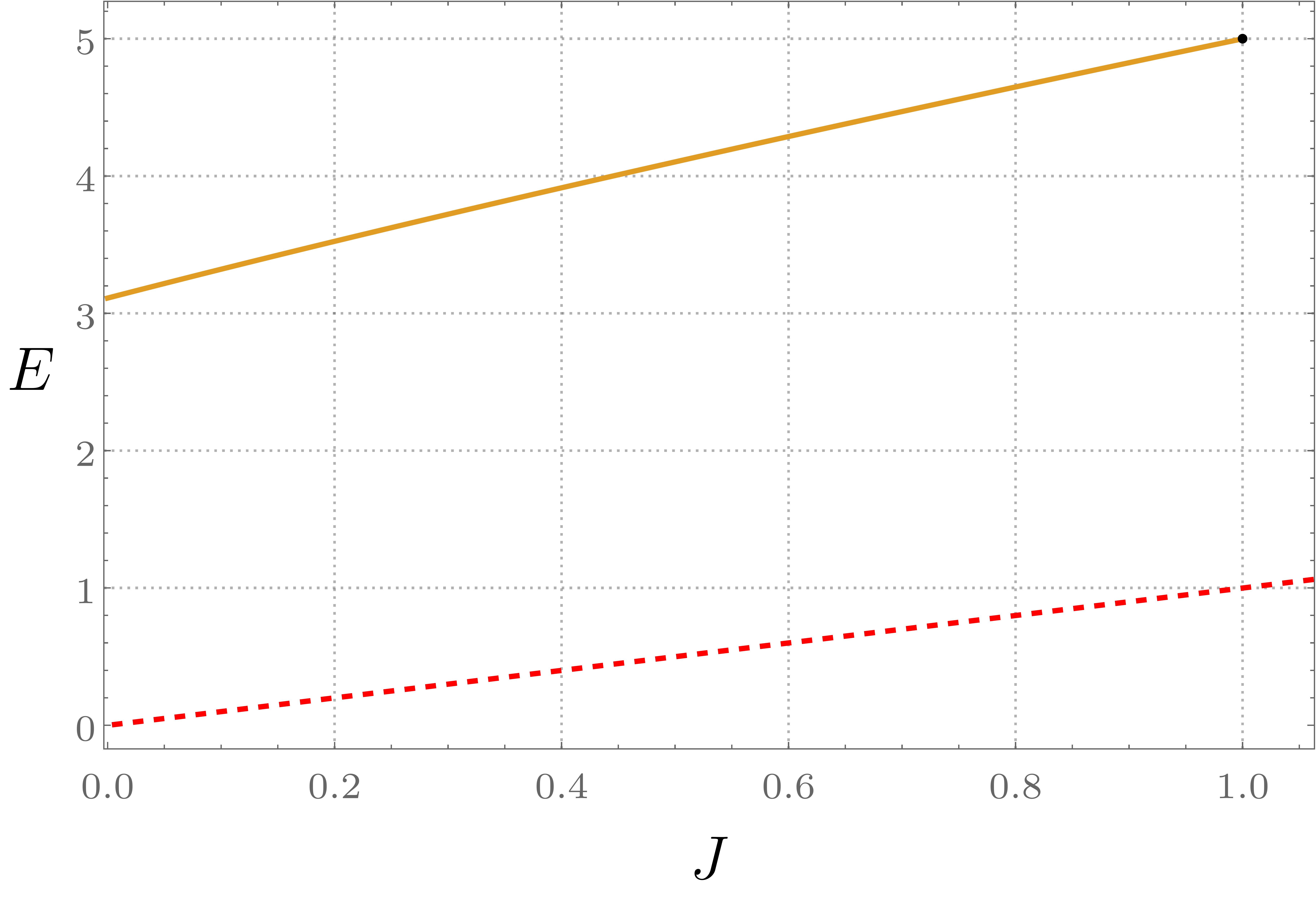}\vspace{0.05cm}
\caption{\hspace*{-0cm} The geodesic path of states in $(E, J)$ space.}\label{}
\end{subfigure}
 \hspace{1 cm} 
 \vspace{-0.0cm}
\caption{ Profiles for $\phi=240^\circ$. 
\textbf{(a)} Time evolution of $E$ (orange), $S$ (green), and $J$ (blue).
\textbf{(b)} The dependence of $E$, $S$, and $J$ on the specific spin $a$. The BTZ black hole  settles to a final static state with energy $E_\tau=3$ and $S_\tau=8$ in Planck units.  
\textbf{(c) }Time evolution of the specific spin $a(t)$. The spin monotonically decreases to zero.
\textbf{(d)} The geodesic trajectory of states (orange curve) in the $(E, J)$ space starts at $E_0 = 5$ and $J_0 = 1$ (black dot), and terminates when the angular momentum vanishes.
}
\end{figure}

In this case, the specific spin $a(t)$ of the BTZ black hole decreases monotonically until it vanishes (Fig. \ref{figsat240}). All black hole parameters decrease during the evolution, indicating an evaporation-type process in which the BTZ black hole loses energy. The final state is a static configuration with energy $E_\tau = 0.6\,E_0$ and entropy $S_\tau = 0.8\,S_0$, corresponding to a reduction of about $40\%$ in energy and $20\%$ in entropy.

\subsubsection{Summary for additional processes}

\begin{figure}[H]
    \centering
\includegraphics[width=6.0cm,height=6cm]{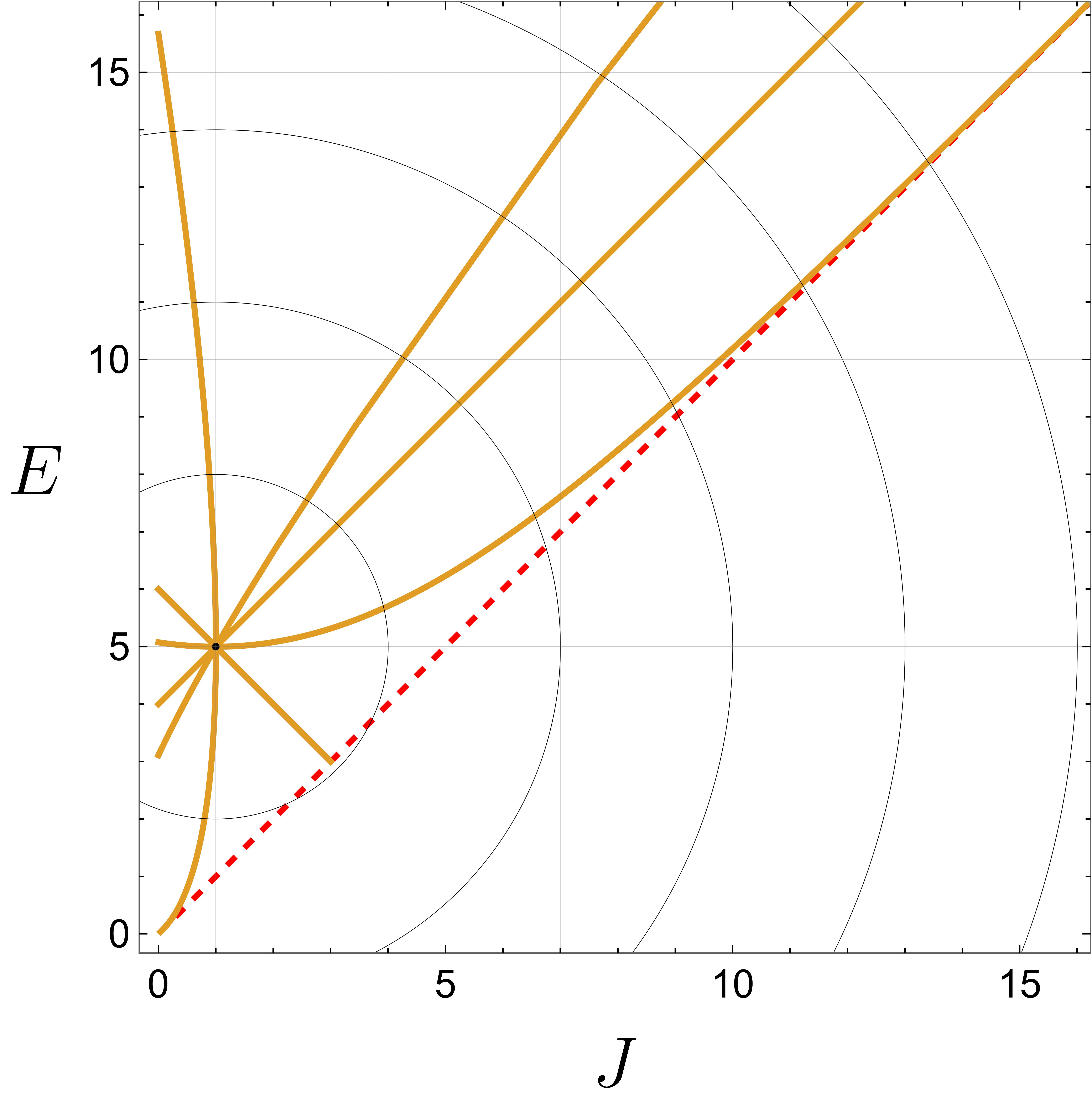}
\caption{Geodesic trajectories of states in the $(E, J)$ space. Since the thermodynamic curvature is $R = 0$, the manifold is Ricci-flat. The initial state $(E_0 = 5, J_0 = 1)$ is far  from the extremal boundary (dashed red curve). The figure presents a magnified view around this initial point. Concentric polar circles indicate the initial angles $\phi$, which set the direction of the corresponding geodesics. Each trajectory evolves toward a distinct final configuration, corresponding either to a near-extremal ($a_\tau\to 1$), finite spinning state $(0<a_\tau<1)$, or a static BTZ black hole $a_\tau=0$.
}
    \label{figRicciFlatGeodP}
\end{figure}

Figure \ref{figRicciFlatGeodP} displays representative geodesic trajectories in the $(E,J)$ space, with the corresponding initial conditions summarized in Table \ref{tabicsrep}. For initial angles $\phi$ in the range $0^\circ \leq \phi \leq 180^\circ$, the dynamics is predominantly governed by accretion-type optimal processes. As $\phi$ increases beyond $180^\circ$, evaporation channels become progressively more relevant. Strong evaporation of the BTZ black hole can be attained exclusively for $240^\circ<\phi <270^\circ$. Beyond $\phi >270^\circ$ the system is driven asymptotically toward extremality, but via evaporation.

The late-time behavior of the system separates into distinct regimes determined by the angle $\phi$. For $0^\circ \leq \phi \leq 45^\circ$, the evolution drives the system toward near-extremal configurations with $a_\tau \to 1$. In the range $45^\circ < \phi \leq 90^\circ$, the trajectories instead approach non-extremal states with spin parameter $a < 1$. For intermediate angles, $90^\circ < \phi \leq 240^\circ$, the dynamics settle into static final configurations within finite time.  Strong evaporation of the BTZ black hole occurs for $240^\circ<\phi <270^\circ$. Beyond this window, at large angles $\phi > 270^\circ$, the system is driven asymptotically toward extremality via evaporation.

\begin{table}[H]
\renewcommand{\arraystretch}{1.4} 
\centering
\footnotesize
\begin{tabular}{|r|r|r|r|r|r|r|r|r|r|r|r|}
\hline
 $\phi$ & $\dot E_0\, 10^{-3}$ & $\dot S_0\,10^{-3}$ & $\dot J_0\, 10^{-3}$ & $\dot a_0\, 10^{-3}$ & $a_\tau$ & $E_\tau$ & $\delta E$ \%  & $S_\tau$ & $\delta S$ \%  & $\tau$ & $\mathcal{L}^2$ \\ \hline\hline
 $0^\circ$ & 0.0 & $-1.0$ & 10 & 2.0  & $\to 1$ & $\to\infty$ & -- & $\to\infty$ & --  & $\to\infty$ & $\to\infty$\\ \hline
 $45^\circ$ & 7.1 & 6.4 &  7.1 & 1.1 & $\to1$  & $\to\infty$ & -- & $\to\infty$ & -- & $\to\infty$ & $\to\infty$\\ \hline
 $60^\circ$ &  8.7 & 8.2 & 5.0 & 0.7 & $\to 0.966$ & $\to\infty$ & --  & $\to\infty$ & -- & $\to\infty$ & $\to\infty$\\ \hline
 $70^\circ$ &  9.4 & 9.1 & 3.4 & 3.1 & $\to 0.725$ & $\to\infty$ & --  & $\to\infty$ & -- & $\to\infty$ & $\to\infty$\\ \hline
 $90^\circ$ & 10 & 10 & 0.0 & $-0.4$ & 0.0 &  16 & $+213$  & 18 & $+78$ & 651 & 4.5\\ \hline
 $135^\circ$ & 7.1 & 7.9 & $-7.1$ & $-1.7$ &0.0 & 6.0 & $+20$  &  11 & $+10$ & 121 & 0.2\\ \hline
 $180^\circ$ & $0.0$ & 1.0& $-10$ & $-2.0$& 0.0 & 5.1 & $+2$  & 10 & $+0$ & 98 & 0.1\\ \hline
 $225^\circ$ &  $-7.1$ & $-6.4$ & $-7.1$  &  $-1.1$ & 0.0 & 4.0 & $-20$  & 9.0 & $-10$ & 164 & 0.2\\ \hline
 $240^\circ$ & $-8.7$ & $-8.2$ & $-5.0$ & $-0.7$ & 0.0 & 3.0 & $-38$ & 8.0 & $-20$  &  266 & 0.6\\ \hline
 $270^\circ$ &  $-10$ & $-10$ & 0.0 & 0.4 & $\to 1$ & $\to 0$ & --  & $\to 0$ & -- &  $\to\infty$ & $\to\infty$ \\ \hline
 $315^\circ$ & $-7.1$ & $-7.9$ & $7.1$  & 1.7 & $\to 1$ & $\to 0$ & --  & $\to 0$ & -- & $\to\infty$ & $\to\infty$ \\ \hline
\end{tabular}
\caption{
The table summarizes the initial rates of key parameters across various processes in entropy representation. It also indicates how the final state differs from the initial BTZ configuration after each process. Additionally, the duration $\tau$ of the process (in Planck units) and the thermodynamic length squared, $\mathcal{L}^2$, are provided. The quantity  $\mathcal{L}^2$ is the minimal entropy produced during  the transformation. Smaller $\mathcal{L}^2$ indicates higher probability for the process to occur.}\label{tabicsrep}
\end{table}
\normalsize

{ 
\section{Nonlinear generation of rotations via fluctuations}\label{secNLGRFL}
We investigate the intriguing possibility that a static BTZ black hole may dynamically transition into a rotating configuration through spontaneous or externally induced optimal fluctuations generated by the nonlinear structure of the optimal processes. To this end, we numerically solve the thermodynamic geodesic equations in both the energy and entropy ensembles, imposing the natural initial condition $J(0)=0$ corresponding to a static BTZ black hole. In particular, we examine whether a nonvanishing initial angular-momentum rate, $\dot{J}(0)\neq0$, can drive the system from an initially static configuration, $a(0)=0$, to a rotating state with $a(t)\neq0$. This phenomenon was first identified by Dr. Vasil Avramov for the Kerr black hole in his 2026 PhD dissertation \cite{Avramov:2026}, and is therefore referred to as the Avramov effect.

\subsection{Generation of rotations in energy representation }
The numerical profiles of the optimal parameters $E(t)$, $S(t)$, and $a(t)=\zeta J(t)/E(t)$, describing the evaporation process in the energy representation, are obtained from Eqs. \eqref{eqSener} and \eqref{eqJener} subject to the initial conditions $J(0)=0$ and $\dot{J}(0)\neq0$. They are presented in Figure \ref{figAvrE}. We find that even small fluctuations in the initial angular momentum rate $\dot{J}(0) \neq 0$ are sufficient to induce an optimal rotation with $a(t)\neq0$ for the BTZ black hole. In the energy representation, the resulting optimal rotation corresponds to a relaxation process connecting two static BTZ configurations. Complete evaporation is not observed unless an Avramov's effect initiates.
 
 One possible interpretation of the observed phenomenon is the following. An initial (spontaneous) fluctuation in the angular momentum induces a non-zero rotational state of the BTZ black hole. The energy required for this rotation appears to originate from a reorganization of the black hole microstates due to evaporation, driving the system toward increasingly rapid rotation while its size simultaneously decreases\footnote{ During this stage of the evolution, not all of the energy necessarily is   converted into rotations --- a fraction of the energy may also be emitted from the system (see Section \ref{IntrThdEffPr}).}. At a certain stage, the specific spin $a(t)$ reaches a local maximum and subsequently decays rapidly toward zero. During this decay phase, the black hole continues to shrink, suggesting that energy is released throughout the relaxation process\footnote{To obey angular momentum conservation, the emitted radiation must carry an angular momentum opposite to that of the black hole.}.
\begin{figure}[H]
	\centering \hspace{-2cm}
	\begin{subfigure}{0.4\textwidth}
		\includegraphics[width=8.3cm,height=5.7cm]{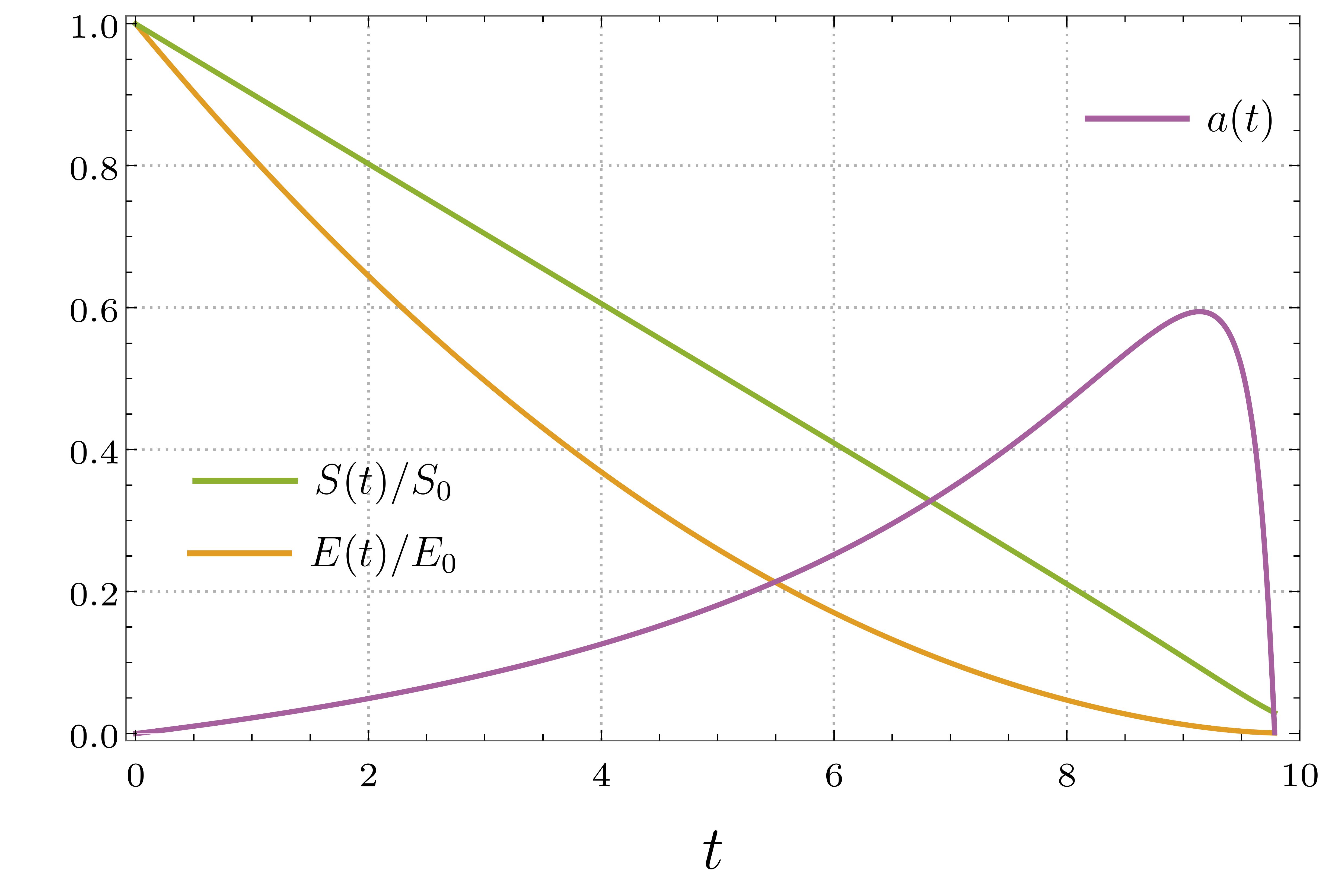}
		\caption{\hspace*{-0.0cm} The Avramov effect with $\dot{J}(0) = 0.02$. }\label{figAvr1E}
	\end{subfigure}
	\hspace{1.5 cm}
	\begin{subfigure}{0.4\textwidth}
		\includegraphics[width=8.3cm,height=5.7cm]{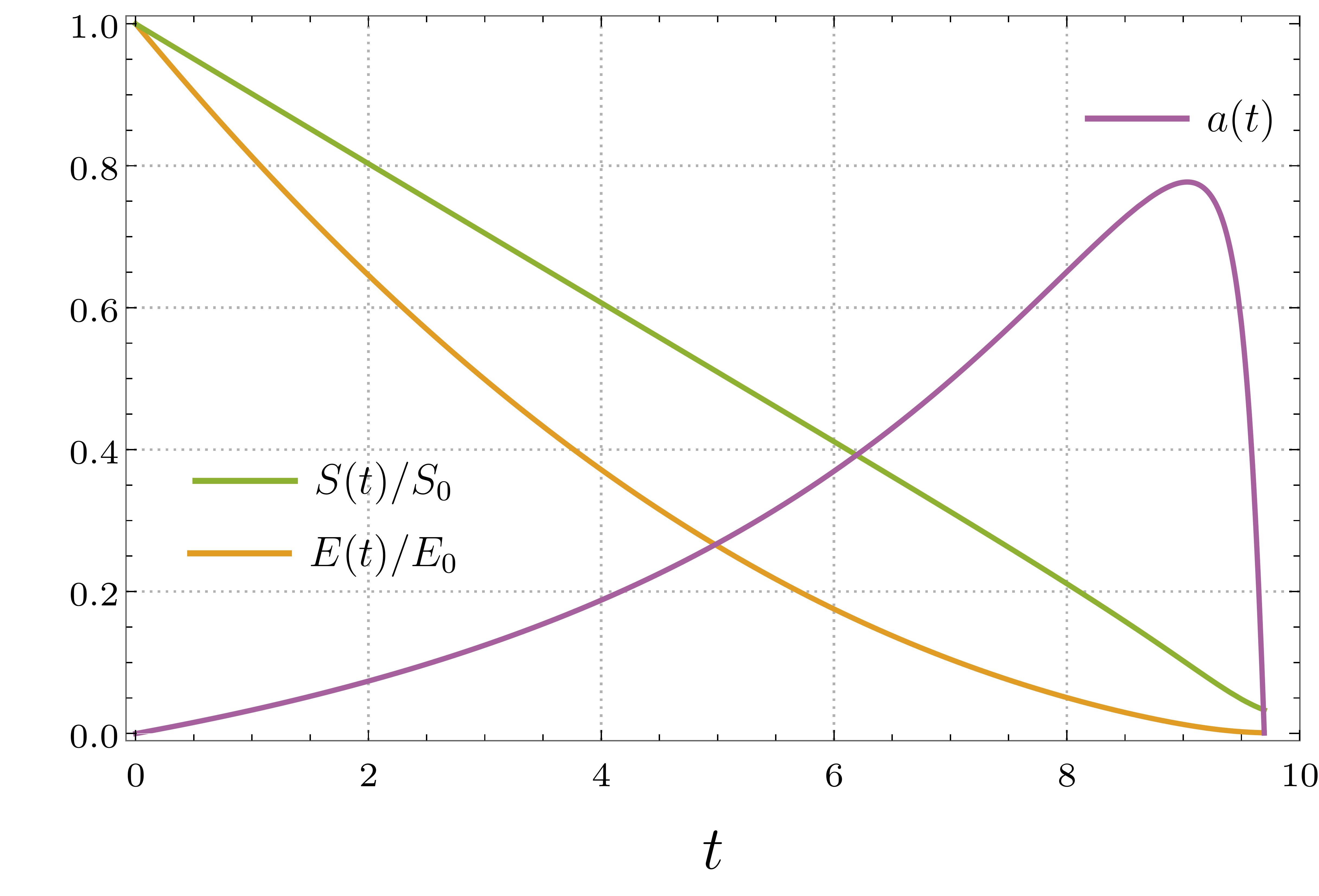}\vspace{-0.05cm}
		\caption{\hspace*{-0.0cm}  The Avramov effect with $\dot{J}(0) = 0.03$.  }\label{figAvr2E}
	\end{subfigure}
	\hspace{0 cm}
	\caption{ Nonlinear generation of rotations (Avramov's effect) of the static BTZ black hole in energy representation. Both BTZ configurations have  initial conditions: $E_0 = 1$, $S_0 = \sqrt{2} \,\pi$, $J_0 = 0$, $\dot{E}_0 = -{16 \zeta (3)}/{\pi ^4}$, $\dot{S}_0 = -{8 \sqrt{2} \zeta (3)}/{\pi ^3}$. \textbf{(a)} $\dot{J}_0 = 0.02=\dot a_0$.
		\textbf{(b)} $\dot{J}_0 = 0.03=\dot a_0$. The final state in both cases is a static configuration ($a_\tau = 0$) with about 0.01\% energy and 3\% entropy left.
	}\label{figAvrE}
\end{figure}

\subsection{Generation of rotations in entropy representation}
The numerical profiles of the optimal parameters $E(t)$, $S(t)$, and $a(t)=\zeta J(t)/E(t)$, describing the evaporation process in the entropy representation, follow  from Eqs. \eqref{EqEenrep} and \eqref{EqJenrep} subject to the initial conditions $J(0)=0$ and $\dot{J}(0)\neq0$. Specific solutions are depicted in Figure \ref{figAvrS}. Once again, small fluctuations in the initial angular momentum rate $\dot{J}(0)\neq 0$ generate an optimal rotation with $a(t)\neq0$. In entropy representation, the resulting optimal rotation corresponds to a relaxation process connecting a static BTZ configurations to an increasingly evaporating near extremal state ($a(t) \approx 1$). Contrary to the energy ensemble case, here complete evaporation of the BTZ black hole is  observed in the final state.

One possible interpretation of this process is the following. An initial fluctuation in the angular momentum rate induces a non-zero rotational state of the BTZ black hole. The subsequent evolution of the specific spin $a(t)$ drives the system toward an extremal configuration, while simultaneously shrinking towards a complete evaporation. In this picture, the Avramov effect in  entropy representation may be understood as a relaxation process that starts with an initially static BTZ configuration, passes through an intermediate rotating phase, and ultimately ends in complete evaporation.

\begin{figure}[H]
	\centering \hspace{-2cm}
	\begin{subfigure}{0.4\textwidth}
		\includegraphics[width=8.3cm,height=5.7cm]{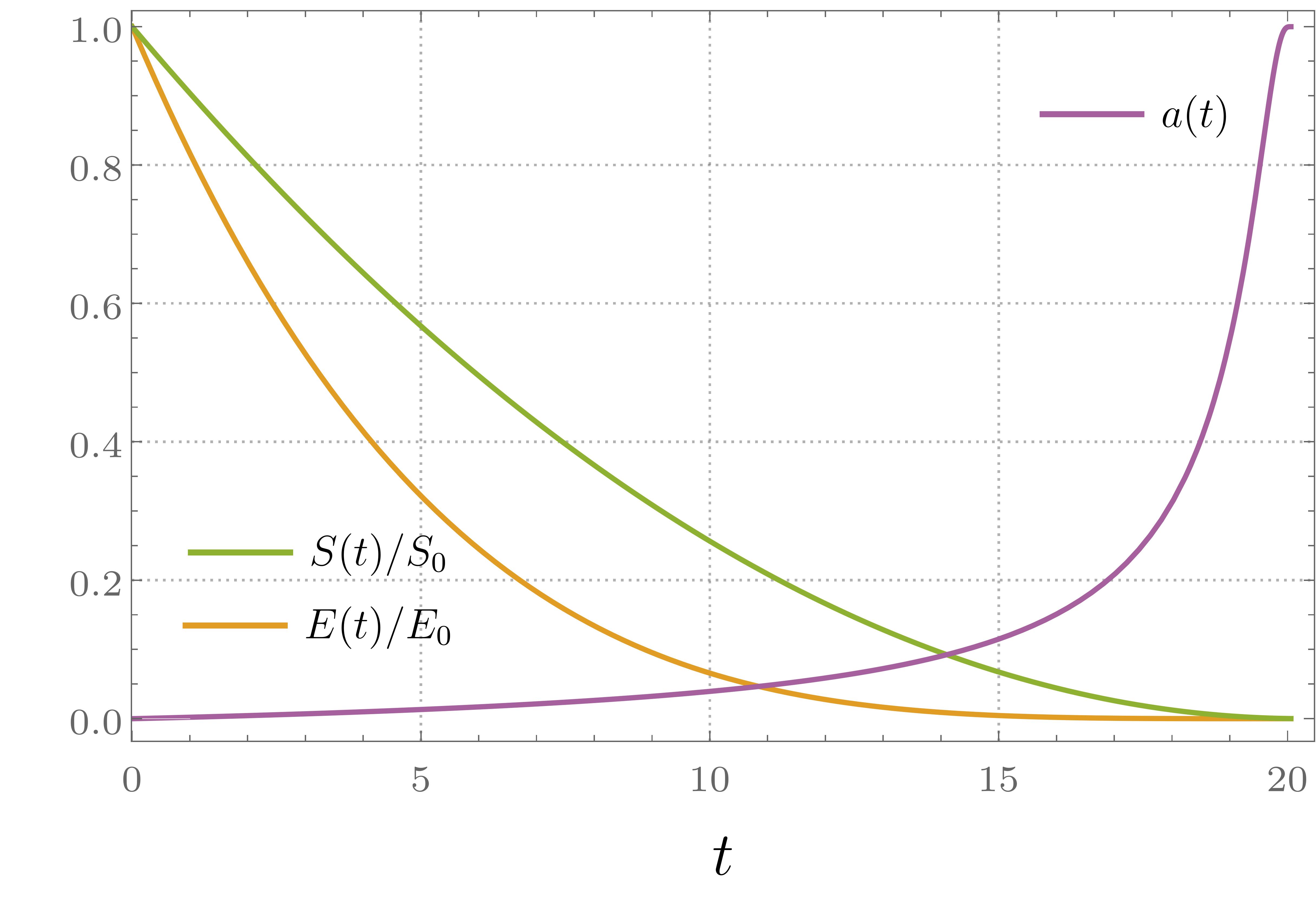}
		\caption{\hspace*{-0.0cm} The Avramov effect with $\dot{J}(0) = 0.002$.}\label{figAvr1S}
	\end{subfigure}
	\hspace{1.5 cm}
	\begin{subfigure}{0.4\textwidth}
		\includegraphics[width=8.3cm,height=5.7cm]{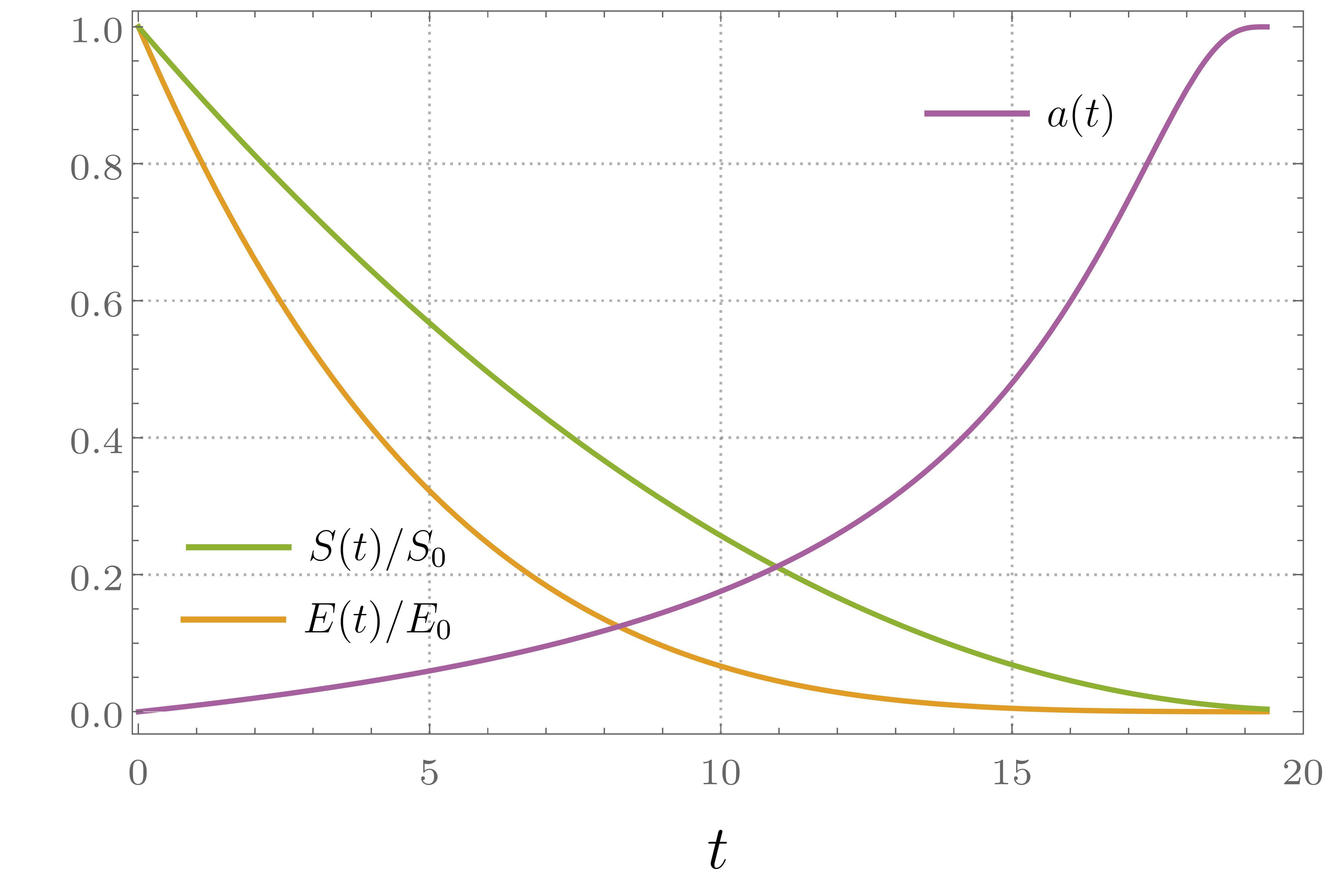}\vspace{-0.05cm}
		\caption{\hspace*{-0.0cm} The Avramov effect with $\dot{J}(0) = 0.009$.}\label{figAvr2S}
	\end{subfigure}
	\hspace{0 cm}
	\caption{Nonlinear generation of rotations (Avramov's effect) of the static BTZ black hole in entropy representation. Both BTZ configurations have  initial conditions: $E(0) = 1$, $S(0) = \sqrt{2} \,\pi$, $J(0) = 0$, $\dot{E}(0) = -{16 \zeta (3)}/{\pi ^4}$, $\dot{S}(0) = -{8 \sqrt{2} \zeta (3)}/{\pi ^3}$. \textbf{(a)} $\dot{J}(0) = 0.002=\dot a_0$.
		\textbf{(b)} $\dot{J}(0) = 0.009=\dot a_0$. At the end the BTZ black hole evaporates completely.
	}\label{figAvrS}
\end{figure}
}

{ 
	\section{Hawking evaporation of the BTZ black hole}\label{secHawkEvapModels}
	In order to benchmark our TGO model against nonoptimal ones, we analyze the Hawking evaporation of the BTZ black hole using the Stefan--Boltzmann power law.
	
	\subsection{Hawking evaporation of the static BTZ black hole}\label{secHESBTZBHa}
	
	In $(d+1)$ dimensions the Stefan--Boltzmann law is given by:
	\begin{equation}
		P= \varepsilon A_{d-1}\sigma_{d+1} T^{d+1},
	\end{equation}
	where \(\varepsilon\) denotes the effective emissivity (graybody factor), $A_{d-1}$ is the surface of the body, $T$ is the temperature, and $\sigma_{d+1}$ is the Stefan--Boltzmann constant:
	\begin{equation}
		\sigma_{d+1}
		=\frac{g_s\Gamma (d+1)\zeta(d+1) }{2^d\pi^{\frac{d+1}{2}}  \Gamma \left(\frac{d+1}{2}\right)} \frac{k^{\,d+1}}{\hbar^{\,d}\,c^{\,d-1}}
		.
	\end{equation}
	Here $g_s$ represents the internal degrees of freedom\footnote{ The parameter $g_s$ represents the effective discrete degeneracy --- the number of independent quantum states available to a single particle at a given energy level:  
		\[ g_s = \sum g_{\text{bosons}} + (1-2^{-d}) \sum g_{\text{fermions}}.
		\]
		The differing numerical weights for fermions and bosons arise from their distinct quantum statistical distributions.}, and $\zeta(d+1)$ is the Riemann zeta-function.
	
	In (2+1) dimensions, the Stefan--Boltzmann law becomes:  
	\begin{equation}
		P= \varepsilon A_{1}\sigma_{3} T^{3}, \quad  \sigma_3 = \frac{g_s\zeta(3) k^3}{\pi^2 \hbar^2 c}.
	\end{equation}
	Since, for an evaporating BTZ black hole we have $P=-dE/dt$ and $A_1=2\pi r_+$, one can write:
	\begin{equation}\label{SBPowerL}
		\frac{dE}{dt}=-\varepsilon \sigma_{3} (2\pi r_+) T^3.
	\end{equation}
	The energy and temperature of the static BTZ are related to the radius of the event horizon by:
	\begin{equation}
		T=\frac{\hbar c}{k} \frac{r_+}{2\pi \ell^2}, \quad E=\frac{c^4 r_+^2}{8G_3\ell^2}.
	\end{equation}
	Using these relations we can write:
	\begin{equation}
		\frac{dE(t)}{dt}=-\alpha E^2(t), \quad \alpha =\frac{16 \hbar G_3^2\varepsilon g_s \zeta(3)}{\pi^4 c^6 \ell^2}.
	\end{equation}
	The solution to this equation determines the Hawking profile of the   energy:
	\begin{equation}\label{eqEnergyHstatic}
		E_H(t)=\frac{E_0}{1+\alpha E_0 t}, \quad E_{0}=E_H(0), \quad \dot E_{H0}=\dot E_H(0)= -\alpha E_0^2.
	\end{equation}
	Since $S=\sqrt{2\lambda E}/\zeta$, we can also write the Hawking profile of the bare entropy on the horizon: 
	\begin{equation}\label{eqEntropyH}
		S_H(t)=\frac{1}{\zeta} \sqrt{ \frac{2\lambda E_0}{1+\alpha E_0 t}}, \quad S_{0}=\frac{\sqrt{2\lambda E_0}}{\zeta}, \quad \dot S_{H0}=\dot S_H(0)=-\frac{\alpha }{\zeta}\sqrt{\frac{\lambda E_0^3}{2}}.
	\end{equation}
	Although the initial rates of change are not required to obtain the solutions, we include them in order to facilitate a later comparison with the optimal profiles.

	Unlike black holes in $(3+1)$ dimensions, which become hotter and ultimately undergo runaway evaporation as they lose mass, a static BTZ black hole in $(2+1)$ dimensions exhibits the opposite behavior: its temperature decreases as its mass diminishes. As a result, rather than evolving toward an explosive endpoint, the BTZ black hole radiates progressively more slowly as it evaporates, approaching a cold, stable configuration or a thermal equilibrium with its surroundings. This is consistent with its positive specific heat, $C = S/3 > 0$. Notably, this means that the Hawking evaporation time becomes infinite. In contrast, the optimal evaporation time of the static BTZ black hole will remain finite, as shown in Sections \ref{secStaticBTZEnergyrep} and \ref{OptimalEvap}.

	\subsection{Hawking evaporation of the rotating BTZ black hole}
	
	Let us consider the Stefan-Boltzmann power law \eqref{SBPowerL} for the rotating BTZ black hole. If we insert $A_1=2\pi r_+$, $r_+$ from \eqref{rp}, and $T$ from \eqref{TOmega}, we find:
	\begin{equation}\label{eqHBTZSBlaw}
		\dot E(t)=-f(a)E^2(t),\quad f(a)=\frac{2\alpha  \sqrt{\big(1-a^2(t)\big)^3} }{ 1+\sqrt{1-a^2(t)}},\quad  a(t)=\frac{\zeta J(t)}{E(t)}.
	\end{equation}
	
	These equations cannot be solved without introducing additional constraints, which can be imposed by adopting the following model proposed by Page \cite{page1976thermal, page2013jcap, Nian:2019buz, Arevalo:2024kmo}:
	\begin{equation}
		\gamma_X=-E^n(t) \frac{d \ln X(t)}{dt},
	\end{equation}
	where $\gamma_X$ is some constant and $X(t)$ represents the parameter that needs to be
	considered for time evolution. The idea of this model is to match $\dot E\propto E^2$ relation in the Stefan-Boltzmann power law \eqref{eqHBTZSBlaw}, thus one should take $n=-1$.  Applying this model to the energy $E(t)$ and the angular momentum $J(t)$ we find:
	\begin{equation}\label{eq_EJ_model}
		\frac{\dot E(t)}{ E^2(t)}=-\gamma_E,\quad \frac{\dot J(t)}{J(t) E(t)} =-\gamma_J,
	\end{equation}
	where $\gamma_{E,J}$ are constants. By comparing the first equation with (\ref{eqHBTZSBlaw}), we obtain $\gamma_E = f(a)$, leading to an algebraic equation for $a(t)$. This implies that during the Hawking evaporation $a(t)$ is time-independent constant, denoted simply by $a=\zeta J_0/E_0$, which allows both equations in \eqref{eq_EJ_model}, together with \eqref{eqHBTZSBlaw}, to be solved. As a result, the Hawking profiles for the energy and the angular momentum of BTZ black hole yield:
	\begin{equation}\label{eqEJprofsKerH}
		E_H(t)= \frac{E_0}{1+ \gamma_E E_0 t},\quad J_H(t)= J_0 \left(1+ \gamma_E E_0 t\right)^{-\frac{\gamma_J}{\gamma_E}}.
	\end{equation}
	Taking the ratio of these two profiles one finds:
	\begin{equation}
		a=\frac{\zeta J(t)}{E(t)} = 	\frac{\zeta  J_0}{E_0}  {\left(1+ \gamma_E E_0 t \right)^{1-\frac{\gamma _J}{\gamma _E}}} \equiv \frac{\zeta  J_0}{E_0},
	\end{equation}
	where for the above equation to be true one has to impose:
	\begin{equation}
		\gamma_J=\gamma_E=\gamma=f(a),
	\end{equation}
	thus the Hawking evaporation profiles become:
	\begin{align}\label{}
		&E_H(t)= \frac{E_0}{1+ \gamma E_0 t},\quad J_H(t)= \frac{J_0}{1+ \gamma E_0 t},  \\[5pt]
		&S_H(t) = \frac{\sqrt{\lambda}}{\zeta }  \bigg( \frac{E_0}{1+ \gamma E_0 t} + \sqrt{\frac{E_0^2}{(1+ \gamma E_0 t){}^2}-\frac{\zeta ^2 J_0^2}{(1+ \gamma J_0 t )^2}} \bigg)^{\!1/2}.
	\end{align}
	Here, we also included the profile of the entropy $S_H(t)$ using \eqref{eqESenergyRep}. It is simple to derive the initial rates for the Hawking evaporation process at $t=0$:
	\begin{equation}
		\dot E_H(0) =-\gamma E_0^2, \quad \dot J_H(0) =-\gamma J_0^2, \quad \dot S_H(0)=-\frac{\gamma  \sqrt{\lambda } \Big(E_0^2 \big(E_0+\sqrt{E_0^2-\zeta ^2 J_0^2}\,\big)-\zeta ^2 J_0^3 \Big)}{2 \zeta  \sqrt{E_0^2-\zeta ^2 J_0^2} \sqrt{E_0+\sqrt{E_0^2-\zeta ^2 J_0^2}}} ,
	\end{equation}
	which will be useful for comparison to the optimal evaporation profiles of the BTZ black hole.

\section{Optimal vs Hawking evaporation}\label{OptimalvsHawking}

It is worth emphasizing that the optimal evaporation process evolves along paths of ``least resistance'' and is therefore not constrained to satisfy the Stefan--Boltzmann blackbody power law in $(2+1)$-dimensions. As a result, Hawking evaporation and optimal evaporation scenarios can differ substantially (see e.g. \cite{Avramov:2025tlh}). Note also that the Hawking evaporation model is independent from the chosen thermodynamic representation.

\subsection{Optimal vs. Hawking  in energy representation: static BTZ}\label{secOvsHenergystatic}

Let us recall the Hawking energy and entropy profiles from \eqref{eqEnergyHstatic} and \eqref{eqEntropyH}:
\begin{align}
	E_H(t)=\frac{E_0}{1+\alpha E_0 t}, \quad |\dot E_{H0}|= \alpha E_0^2, \quad 
	S_H(t)=\frac{1}{\zeta} \sqrt{ \frac{2\lambda E_0}{1+\alpha E_0 t}}, \quad |\dot S_{H0}|=\frac{\alpha }{\zeta}\sqrt{\frac{\lambda E_0^3}{2}},
\end{align}
and compare them to the corresponding optimal profiles (Fig. \ref{figOtoHEnergy}):
\begin{equation}
	E(t)= E_0 \left(1-\frac{|\dot E_0|}{2E_0}t \right)^{\!2}, \quad 
	S(t)=S_0-|\dot S_0| t.
\end{equation}
Since the initial rates $|\dot E_0|$ and $|\dot S_0|$ in the optimal scenario are free parameters, one may choose them to match the Hawking values $|\dot E_{H0}|$ and $|\dot S_{H0}|$. This identification then yields an estimate for the optimal evaporation time:
\begin{equation}\label{eqTevapenergyRepStatic}
	t_{\text{evap}} = \frac{2}{\alpha E_0}.
\end{equation}
The result indicates that, within the optimal framework, larger black holes evaporate more rapidly. It should be stressed, however, that this conclusion holds only when the optimal evolution is initialized with rates equal to those of the Hawking process.

\begin{figure}[H]
	\centering
	\includegraphics[scale=0.47]{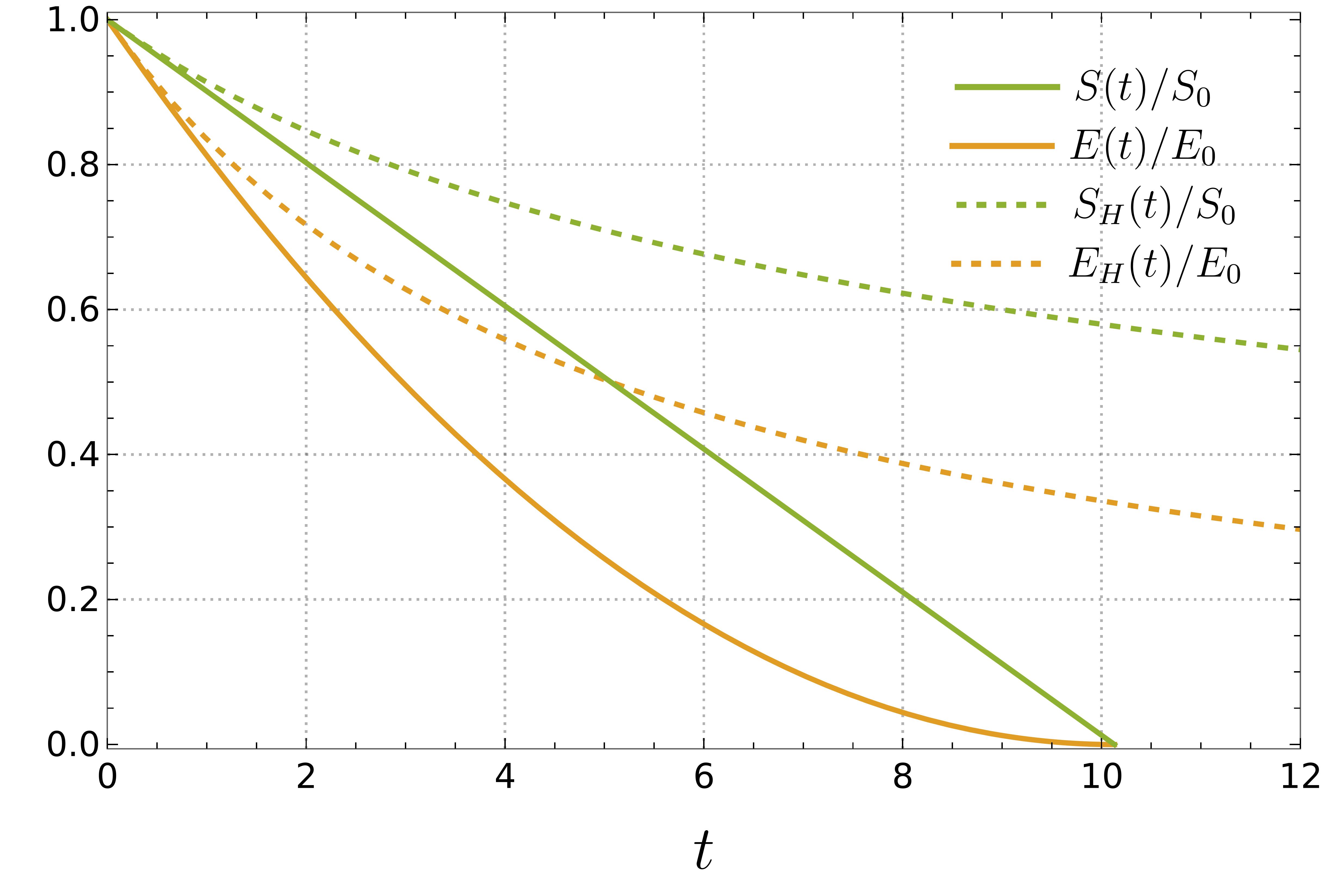}
	\caption{Comparison between optimal energy (solid orange), Hawking energy (dashed orange), optimal entropy (solid green) and Hawking entropy (dashed green) profiles in energy representation for a static BTZ. In Plank units: $\lambda=\pi^2,$ $\zeta=1,$ $\alpha\approx 0.197,$ $E_0=1,$ $S_0=\sqrt{2}\, \pi,$ $|\dot E_0|=|\dot E_{H0}|\approx 0.197,$ $|\dot S_0|=|\dot S_{H0}|\approx 0.439$. The optimal evaporation time is finite, $\tau_{\text{evap}}\approx 10$, in contrast to the Hawking case, where the black hole evaporation proceeds indefinitely.
	}
	\label{figOtoHEnergy}
\end{figure}

\subsection{Optimal vs. Hawking in entropy representation: static BTZ}\label{secOvsHentropystatic}

Let us recall the Hawking energy and entropy profiles from \eqref{eqEnergyHstatic} and \eqref{eqEntropyH}:
\begin{align}
	E_H(t)=\frac{E_0}{1+\alpha E_0 t}, \quad |\dot E_{H0}|= \alpha E_0^2, \quad 
	S_H(t)=\frac{1}{\zeta} \sqrt{ \frac{2\lambda E_0}{1+\alpha E_0 t}}, \quad |\dot S_{H0}|=\frac{\alpha }{\zeta}\sqrt{\frac{\lambda E_0^3}{2}},
\end{align}
and compare them to the optimal profiles (Fig. \ref{figOtoHEntropy}):
\begin{equation}
	E(t) = E_0 \left(1 - \frac{|\dot E_0| }{4 E_0}\, t\right)^{\!4}, \quad S(t) =S_0 \Big(1 - \frac{|\dot S_0|}{2S_0}t \Big)^2.
\end{equation}
Assuming equal rates $|\dot E_0| = |\dot E_{H0}|$ and  $|\dot S_0| = |\dot S_{H0}|$, we can find the optimal evaporation time:
\begin{equation}
	\tau_{\text{evap}} =  \frac {4}{\alpha E_0}.
\end{equation}
Once again, we note that BTZ black holes evaporate for a finite time when considering optimal processes. This time is twice longer than  $\tau_{\text{evap}}$ \eqref{eqTevapenergyRepStatic} from energy representation. 

\begin{figure}[H]
	\centering
	\includegraphics[scale=0.45]{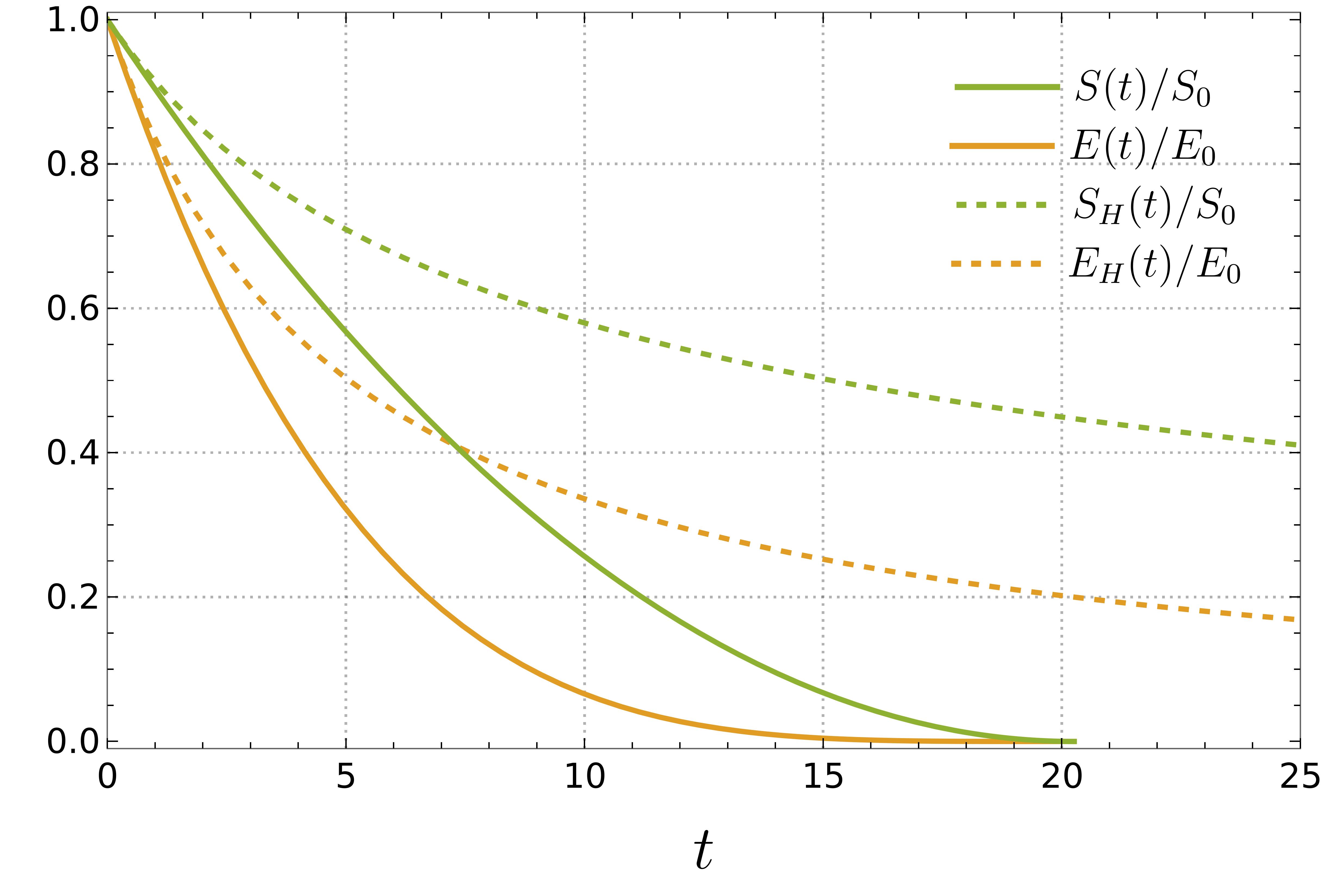}
	\caption{Comparison between optimal energy (solid orange), Hawking energy (dashed orange), optimal entropy (solid green) and Hawking entropy (dashed green) profiles in entropy representation for a static BTZ. Here we adopt Plank units, where: $\lambda=\pi^2,$ $\zeta=1,$ $\alpha\approx 0.197,$ $E_0=1,$ $S_0=\sqrt{2}\, \pi,$ $|\dot E_0|=|\dot E_{H0}|\approx 0.197,$ $|\dot S_0|=|\dot S_{H0}|\approx 0.439$. The optimal evaporation time is $\tau_{\text{evap}}\approx 20$, which is twice longer than $\tau_{\text{evap}}$ from energy representation.
	}
	\label{figOtoHEntropy}
\end{figure}

\subsection{Optimal vs. Hawking in energy representation: rotating BTZ}\label{secOvsHenergyrot}

The rotating case is drastically different from  the static case. In the energy representation, we no longer observe complete evaporation. Instead, the optimal process corresponds to a relaxation toward a static BTZ configuration. Figures \ref{figOHE0.4} and \ref{figOHE0.95} illustrate two representative cases in Planck units ($\lambda=\pi^2$ and $\zeta=1$) with low ($a_0 = 0.4$) and high ($a_0 = 0.95$) initial spins. The solid curves denote the optimal evaporation profiles, while the dashed curves correspond to the Hawking evaporation profiles.

\begin{figure}[H]
	\centering \hspace{-2cm}
	\begin{subfigure}{0.4\textwidth}
		\includegraphics[width=8.3cm,height=5.7cm]{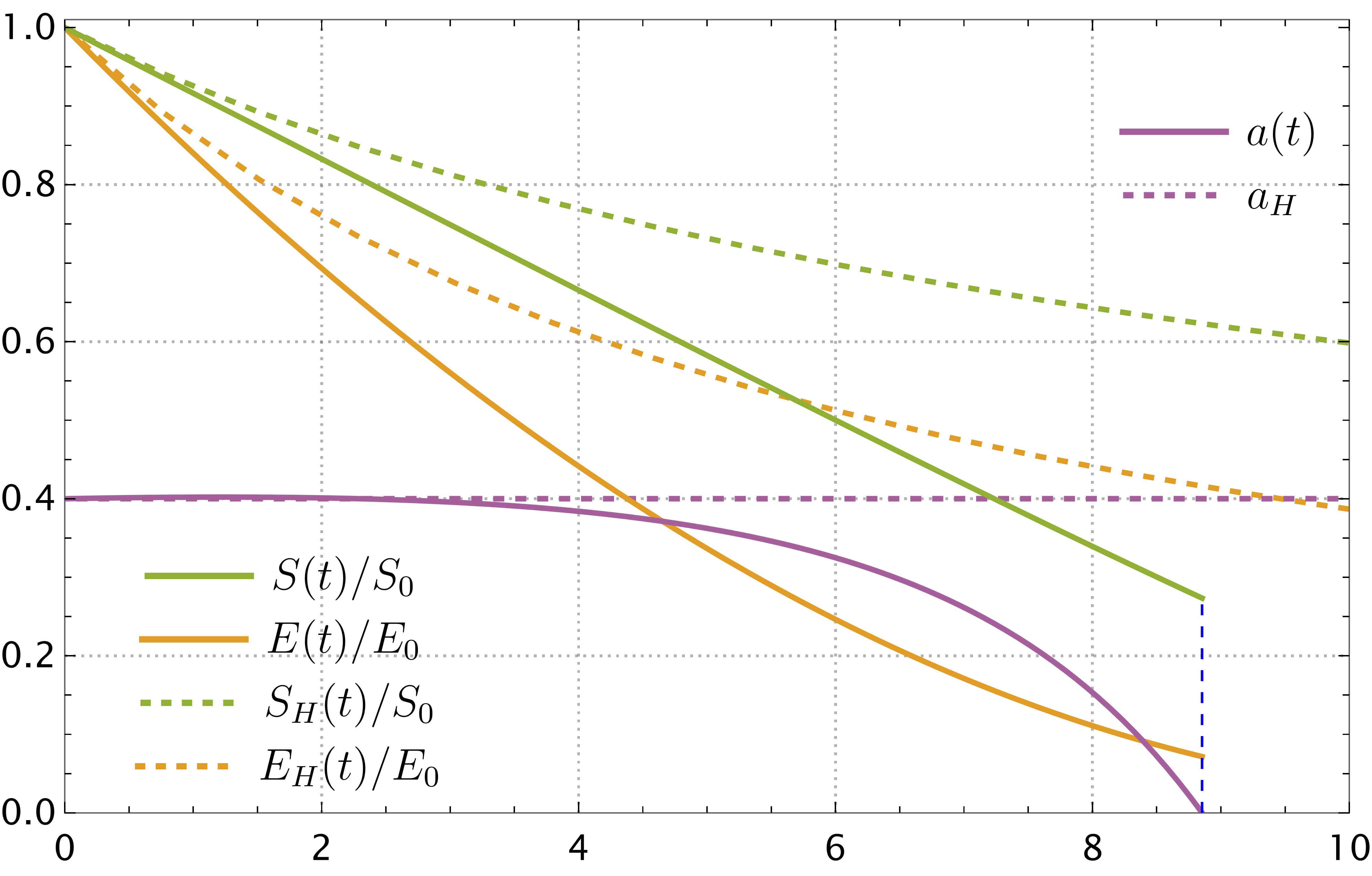}
		\caption{\hspace*{-0.0cm} $a_0=0.4$.}\label{figOHE0.4}
	\end{subfigure}
	\hspace{1.5 cm}
	\begin{subfigure}{0.4\textwidth}
		\includegraphics[width=8.3cm,height=5.7cm]{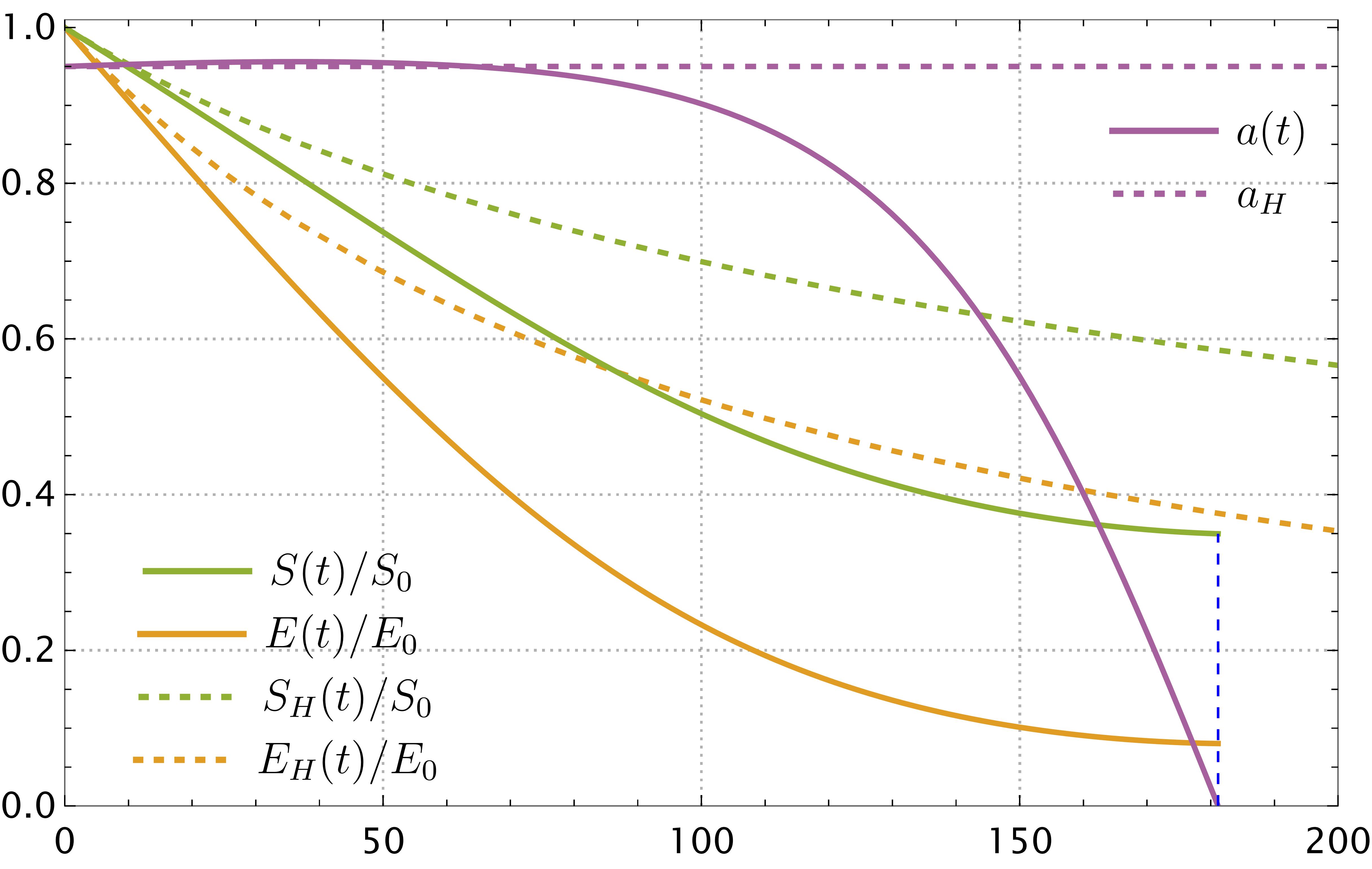}\vspace{-0.05cm}
		\caption{\hspace*{-0.0cm} $a_0=0.95$.}\label{figOHE0.95}
	\end{subfigure}
	\hspace{0 cm}
	\caption{ Comparison of the optimal energy (solid orange), Hawking energy (dashed orange), optimal entropy (solid green), Hawking entropy (dashed green), optimal spin (solid purple), and Hawking spin (dashed purple) profiles for two different initial spin parameters in the energy representation. 
	}\label{figOHE}
\end{figure}

The evaporation processes shown in Figure \ref{figOHE0.4} have initial parameters: $a_0=0.4=J_0$, $E_0=1$, $S_0=4.349$, and $\gamma=0.159$. The corresponding initial rates are $\dot a_0=0.003$, $\dot J_0=\dot J_{H0}=-0.063$, $\dot E_0=\dot E_{H0}=-0.159$, and $\dot S_0=\dot S_{H0}=-0.364$. The optimal process has a finite duration of $\tau\approx 9$ and terminates once the BTZ black hole reaches a static configuration, $a(\tau)=0$, leaving approximately $7$\% of the initial energy and 27\% of the initial entropy.

The evaporation processes shown in Figure \ref{figOHE0.95} have initial parameters: $a_0=0.95=J_0$, $E_0=1$, $S_0=3.60$, and $\gamma=0.009$. The initial rates are $\dot a_0=0.0003$, $\dot J_0=\dot J_{H0}=-0.009$, $\dot E_0=\dot E_{H0}=-0.009$, and $\dot S_0=\dot S_{H0}=-0.018$. In this case, the optimal process lasts significantly longer, with $\tau\approx 180$, and again ends in a static configuration, $a(\tau)=0$, while retaining roughly 8\% of the initial energy and 35\% of the initial entropy.

\subsection{Optimal vs. Hawking in entropy representation: rotating BTZ}\label{secOvsHentropyrot}

In entropy representation we observe complete evaporation. Figures \ref{figOHESn0.4} and \ref{figOHESn0.95} illustrate two representative cases in Planck units ($\lambda=\pi^2$ and $\zeta=1$) with low ($a_0 = 0.4$) and high ($a_0 = 0.95$) initial spins. The solid curves denote the optimal evaporation profiles, while the dashed curves correspond to the Hawking evaporation profiles. A noticeable behavior of the optimal spin profile $a(t)$ in both cases is that it behaves effectively as a constant\footnote{Numerically we have $\dot a_0\sim 10^{-60}$.} during the process until the black hole evaporates.

\begin{figure}[H]
	\centering \hspace{-2cm}
	\begin{subfigure}{0.4\textwidth}
		\includegraphics[width=8.3cm,height=5.7cm]{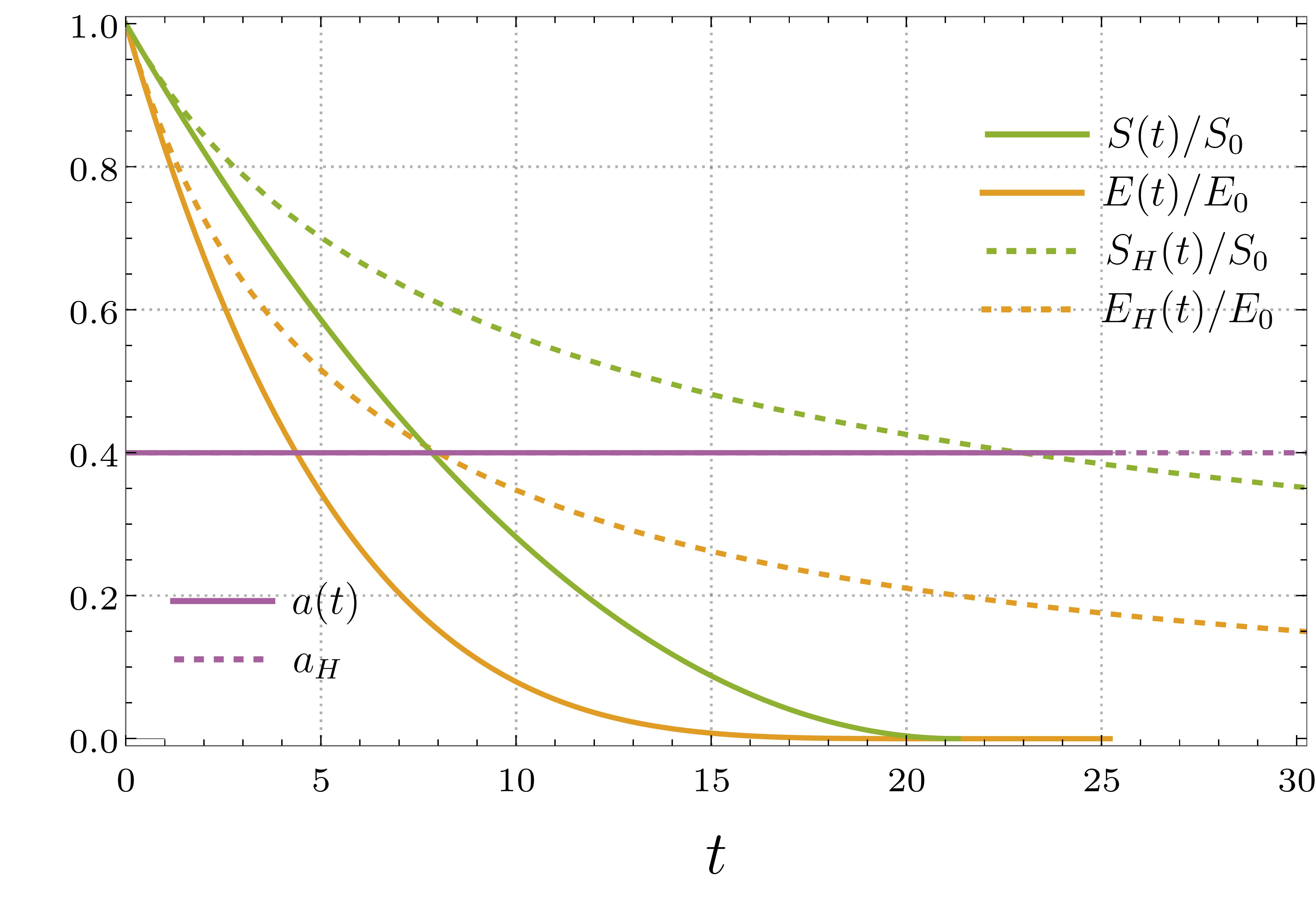}
		\caption{\hspace*{-0.0cm} $\dot a_0=0.4$.}\label{figOHESn0.4}
	\end{subfigure}
	\hspace{1.5 cm}
	\begin{subfigure}{0.4\textwidth}
		\includegraphics[width=8.3cm,height=5.7cm]{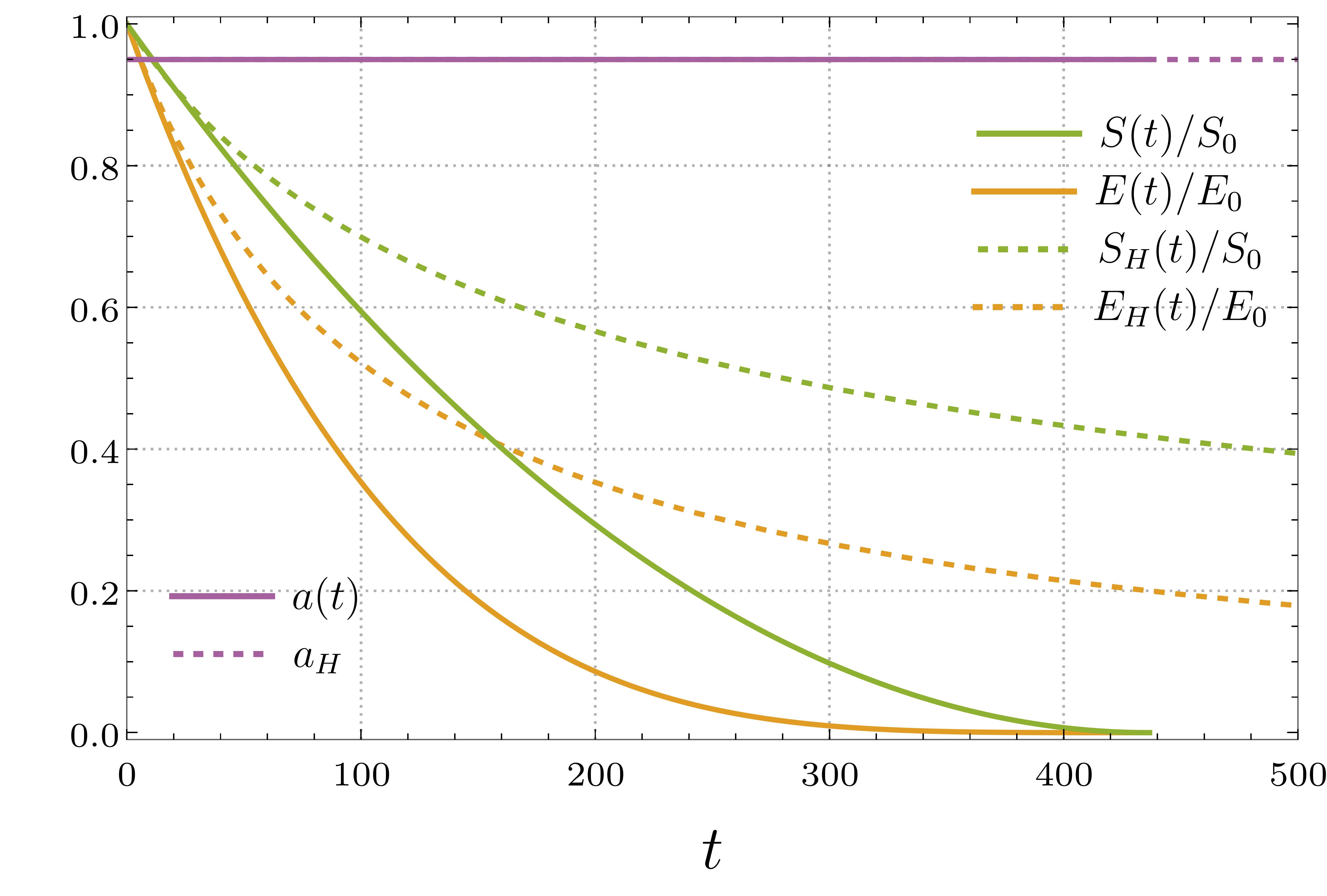}\vspace{-0.05cm}
		\caption{\hspace*{-0.0cm}$\dot a_0=0.95$. }\label{figOHESn0.95}
	\end{subfigure}
	\hspace{0 cm}
	\caption{ Comparison of the optimal energy (solid orange), Hawking energy (dashed orange), optimal entropy (solid green), Hawking entropy (dashed green), optimal spin (solid purple), and Hawking spin (dashed purple) profiles for two different initial spin parameters in the entropy representation. 
	}\label{figOHEn}
\end{figure}

The evaporation processes shown in Figure \ref{figOHESn0.4}  have initial parameters: $a_0=0.4=J_0$, $E_0=1$, $S_0=4.349$, and $\gamma=0.159$. The corresponding initial rates are $\dot a_0=0$, $\dot J_0=\dot J_{H0}=-0.063$, $\dot E_0=\dot E_{H0}=-0.159$, and $\dot S_0=\dot S_{H0}=-0.364$. The optimal process has a finite duration of $\tau\approx 25$ and terminates when the BTZ black hole undergoes complete evaporation.

The evaporation processes shown in Figure \ref{figOHESn0.95}  have initial parameters: $a_0=0.95=J_0$, $E_0=1$, $S_0=3.60$, and $\gamma=0.009$. The initial rates are $\dot a_0=0$, $\dot J_0=\dot J_{H0}=-0.009$, $\dot E_0=\dot E_{H0}=-0.009$, and $\dot S_0=\dot S_{H0}=-0.016$. In this case, the optimal process lasts significantly longer, with $\tau\approx 430$, and ends when the BTZ black hole reaches a state of complete evaporation.
}

{ 
	\section{Intrinsic thermodynamic efficiency of the process}\label{IntrThdEffPr}
	
	We are also interested in quantifying the partitioning of the total black hole energy $E$ into rotational $E_{\rm rot}$, irreducible $E_{\rm irr}$ and emitted/lost $\Delta E$ energy during the process.  The irreducible energy follows from the first law of thermodynamics in energy representation where one has \cite{Yazadjiev:2009}:
	\begin{equation}
		\delta E>\Omega 	\delta J=\frac{\zeta ^2 J}{E+\sqrt{E^2-\zeta ^2 J^2}} 	\delta J.
	\end{equation}
Here $\Omega$ is defined in (\ref{TOmega}). It is easy to rewrite this inequality such as:
	\begin{equation}
		\delta\bigg(\frac{E+\sqrt{E^2-\zeta ^2 J^2}}{2}\bigg)>0.
	\end{equation}
The quantity inside the brackets is the irreducible energy of the BTZ black hole:
\begin{equation}
		E_{irr}=\frac{E+\sqrt{E^2-\zeta ^2 J^2}}{2}=\frac{E}{2} \big(1 + \sqrt{1-a^2}\big).
\end{equation}
where in the last expression we inserted $J=a E/\zeta$. By definition the rotational energy is:
	\begin{align}
		E_{\rm rot}=E-E_{\rm irr}  = \frac{E}{2} \big(1 - \sqrt{1-a^2}\big).
	\end{align}
	The rotational energy reaches its theoretical maximum at the extremal limit $a \to 1$, where $E_{\rm rot} = 0.5 \, E$. This implies that for a rotating BTZ black hole, up to 50\% of the total mass-energy is stored as rotational energy and is, in principle, available for extraction via Penrose-like processes\footnote{Compare to the 29\% of the Kerr black hole.}. Assuming an optimal evolution, this relation remains valid at every instance of $t$:
	\begin{equation}
		E_{\rm rot}(t)  = \frac{E(t)}{2} \big(1 - \sqrt{1-a^2(t)}\big).
	\end{equation}
	
	To rigorously track the redistribution of the initial energy $E_0$ throughout the evolution of the process, we have to look at the energy balance. At any given moment $t$, $E_0$ is partitioned into three distinct physical components: the instantaneous rotational energy\footnote{The kinetic energy associated with the black hole's angular momentum.} $E_{\rm rot}(t)$, the irreducible energy\footnote{The internal energy associated with the horizon area $r_+$, which represents the minimum energy state for a fixed entropy.} $E_{\rm irr}(t)$, and the cumulative dissipated energy\footnote{The total energy flux radiated from the black hole (e.g., via Hawking radiation or gravitational wave emissions) up to time $t$.} $\Delta E(t)$. Given that the total instantaneous mass-energy of the black hole is $E(t) = E_{\rm rot}(t) + E_{\rm irr}(t)$, the energy conservation law relative to the initial state $E_0$ is expressed as:
	\begin{equation}
		\Delta E(t) = E_0 - E(t).
	\end{equation}
	Hence, the global energy dynamic of the BTZ black hole is determined only by the transition between the initial and final states of the system. 
	
	To quantify the fraction of the initial energy $E_0$ currently stored in the rotational part, we define the dimensionless ratio:
	\begin{equation}
		\epsilon_{\rm rot}(t) = \frac{E_{\rm rot}(t)}{E_0} = \frac{E(t)}{2E_0} \big(1 - \sqrt{1-a^2(t)}\big).
	\end{equation}
	This allows for a precise determination of the intrinsic thermodynamic (conversion) efficiency  $\eta(t)=\epsilon_{\rm rot}(t)$ of the process, which is the ratio of the extractable rotational energy to the initial energy input at any time $t$.  This quantity represents the theoretical upper bound of the energy fraction available for extraction (via the Penrose process), stored within the spin of the BTZ black hole\footnote{We can think of it as the ``battery capacity'' of the black hole itself.}. Figures \ref{figOHEetaHigh} and \ref{figOHEeta} depict this optimal efficiency as compared to the efficiency of the Hawking evaporation process at different initial angular momentum rates. 
	
Notably, when the same initial optimal rates as in the Hawking case are imposed, the overall optimal efficiency (solid curves) is lower than the corresponding Hawking evaporation efficiency (dashed curves), as shown in Fig. \ref{figOHEetaHigh}. This behaviour is expected, since optimal processes complete the evaporation in a finite time, whereas Hawking evaporation of the BTZ black hole persists indefinitely. Nevertheless, for sufficiently small negative rates, the optimal processes exhibit a higher intrinsic efficiency over finite time intervals, as illustrated in Fig. \ref{figOHEeta}.
	
	\begin{figure}[H]
		\centering \hspace{-2cm}
		\begin{subfigure}{0.4\textwidth}
			\includegraphics[width=8.3cm,height=5.7cm]{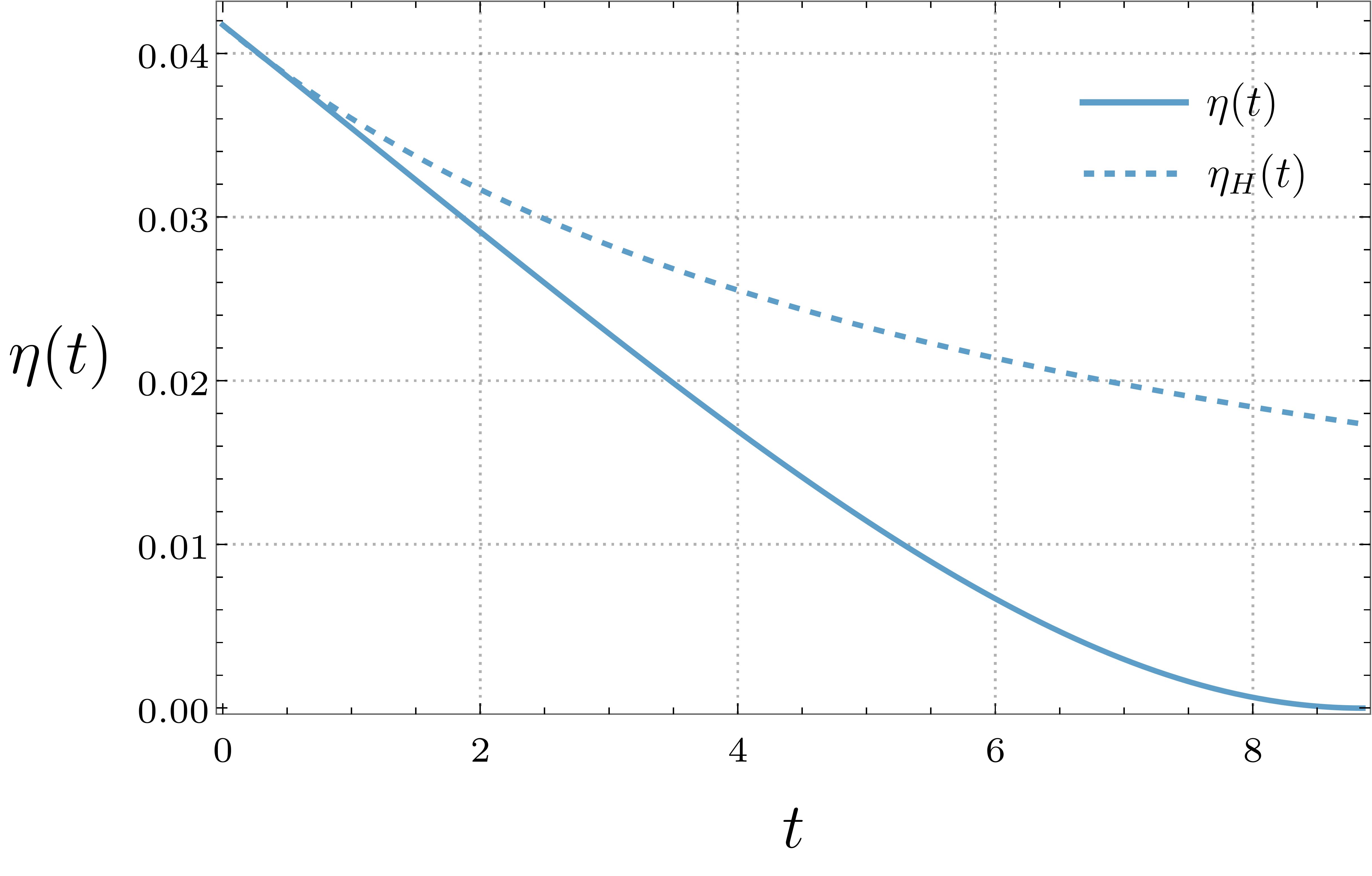}
			\caption{\hspace*{-0.0cm} Efficiency at initial rate $\dot J_0=\dot J_{H0}=-0.063$ in energy representation for the process depicted in Fig. \ref{figOHE0.4}.  }\label{figOHE0.4eta}
		\end{subfigure}
		\hspace{1.5 cm}
		\begin{subfigure}{0.42\textwidth}
			\includegraphics[width=8.3cm,height=5.7cm]{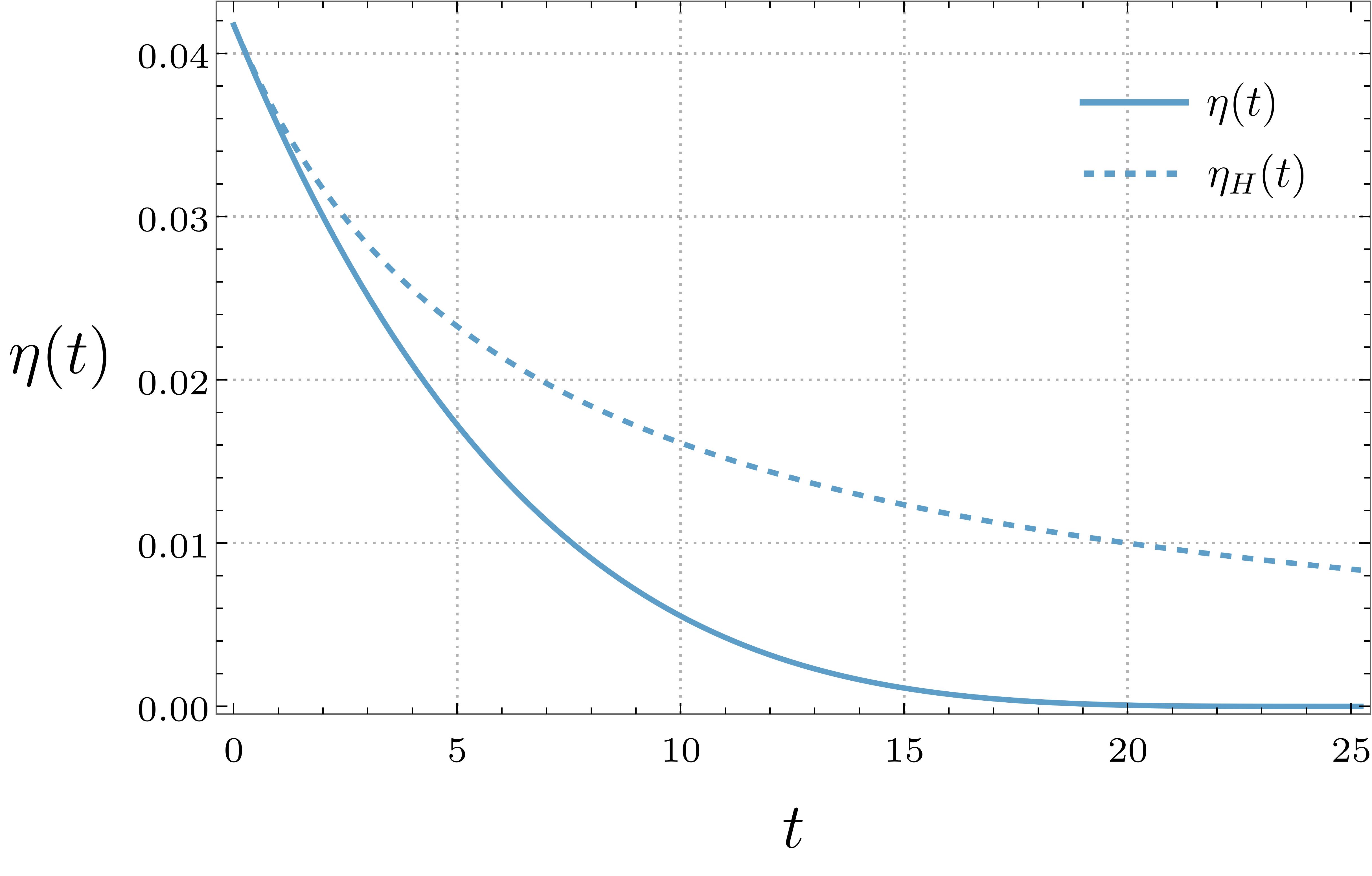}\vspace{-0.05cm}
			\caption{\hspace*{-0.0cm} Efficiency at initial rate $\dot J_0=\dot J_{H0}=-0.063$ in entropy representation for the process depicted in Fig. \ref{figOHESn0.4}. }\label{figOHE0.95eta}
		\end{subfigure}
		\hspace{0 cm}
		\caption{  Efficiency $\eta(t)$ for the optimal (solid) vs. Hawking (dashed) evaporation process in energy and entropy representations with initial angular momentum rates   $\dot J_0=\dot J_{H0}=-0.063$. 
		}\label{figOHEetaHigh}
	\end{figure}

	\begin{figure}[H]
	\centering \hspace{-2cm}
	\begin{subfigure}{0.4\textwidth}
		\includegraphics[width=8.3cm,height=5.7cm]{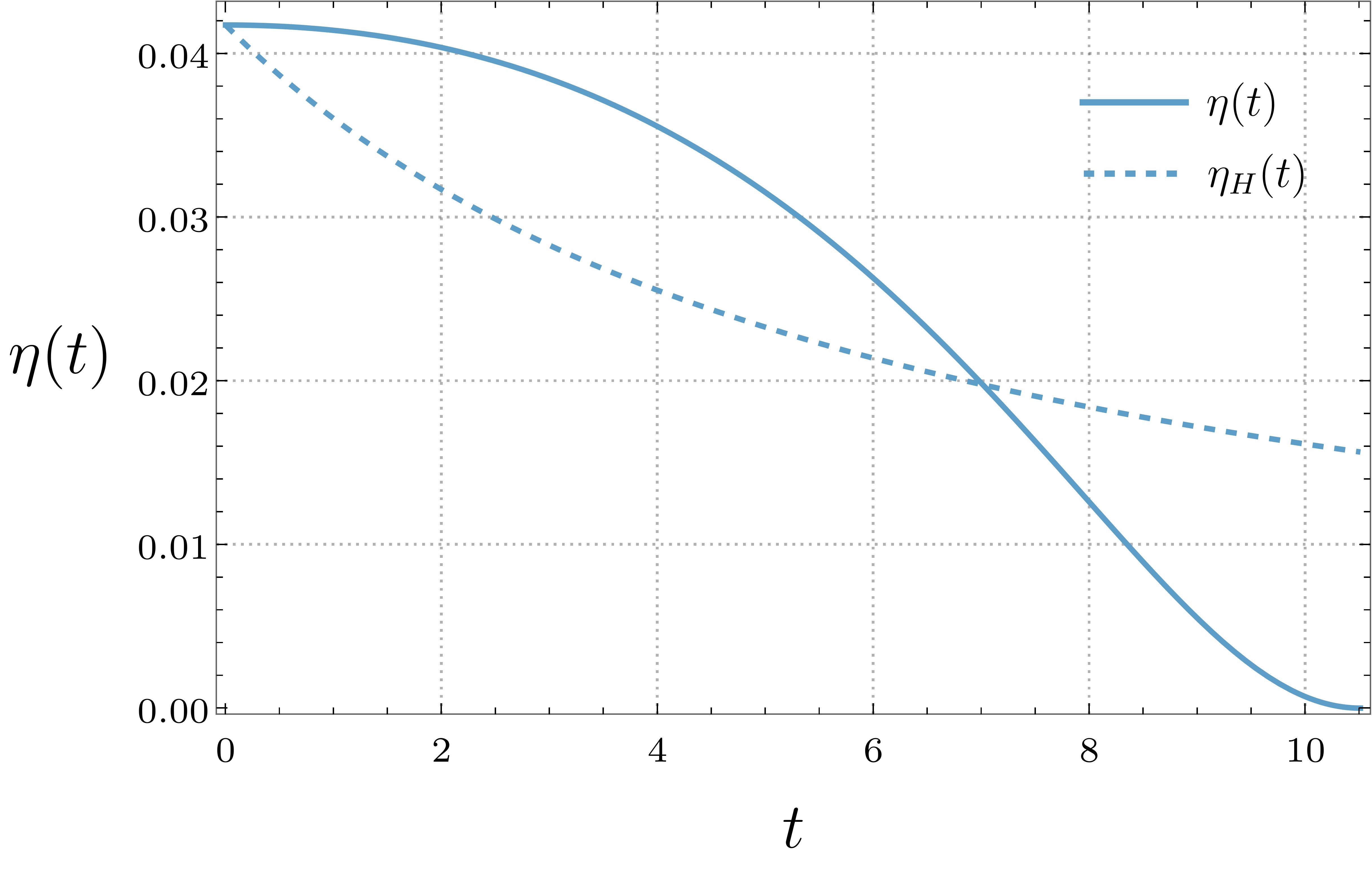}
		\caption{\hspace*{-0.0cm} Efficiency at initial rate $\dot J_0=-0.033$ in energy representation for the process depicted in Fig. \ref{figOHE0.4}. }\label{figOHE0.4etaHigh}
	\end{subfigure}
	\hspace{1.5 cm}
	\begin{subfigure}{0.4\textwidth}
		\includegraphics[width=8.3cm,height=5.7cm]{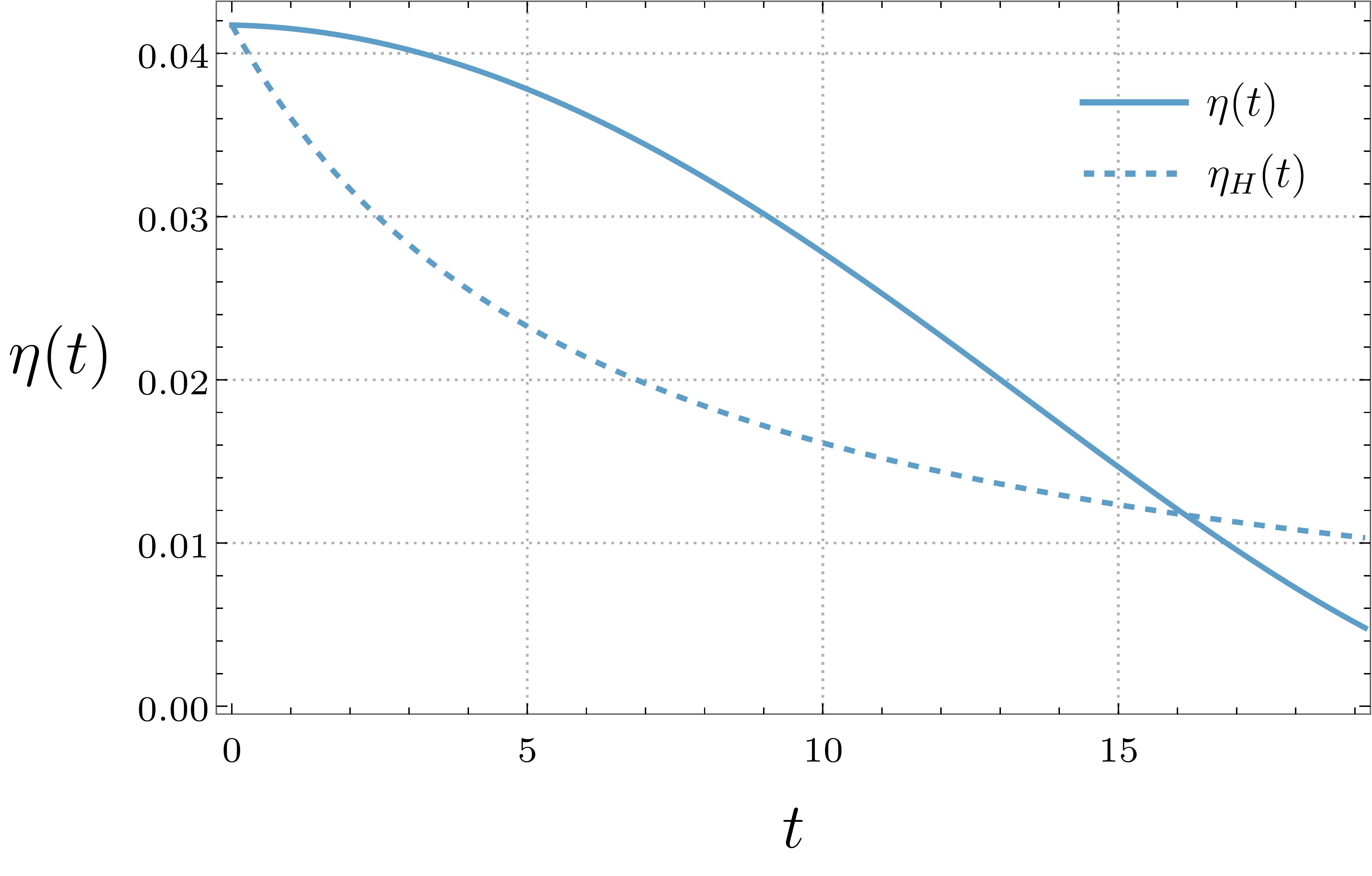}\vspace{-0.05cm}
		\caption{\hspace*{-0.0cm} Efficiency at initial rate $\dot J_0=-0.033$ in entropy representation for the process depicted in Fig. \ref{figOHESn0.4}. }\label{figOHE0.95etaHigh}
	\end{subfigure}
	\hspace{0 cm}
	\caption{  Efficiency $\eta(t)$ for the optimal (solid) vs. Hawking (dashed) evaporation process in energy and entropy representations with initial angular momentum rates   $\dot J_0=\dot J_{H0}=-0.033$. 
	}\label{figOHEeta}
\end{figure}

}

{ 
\section{Conclusion}\label{secConcl}

In this work, we investigated the finite-time optimal thermodynamics of the BTZ black hole in both the energy and entropy ensembles using the thermogeometric optimization framework developed in \cite{Avramov:2025tlh}.

By applying the differential Sylvester criterion \eqref{Sylvester}, we demonstrated that the BTZ black hole is thermodynamically stable. This conclusion is further supported by the positive-definite behavior of the relevant local heat capacities \eqref{eqCOmega}–\eqref{eqCE}. These results also indicate the absence of Davies-type phase transitions in the BTZ state space, aside from the extremal limit.

In the energy representation, we employed the Weinhold thermodynamic metric \eqref{Weinhold}, which possesses a nonvanishing thermodynamic curvature \eqref{eqCurvEn}. Within this framework, we showed that the curvature ($R$) remains positive (Fig.~\ref{figTDRicciErep}), signifying that the corresponding information geometry is elliptic. This geometric feature is reflected also by the positive definiteness of the thermodynamic length for all optimal processes considered.

For the static BTZ black hole in energy representation we obtained analytic expressions for both the thermodynamic length \eqref{ThermoLenghtBTZnorodEn} and the relaxation time \eqref{eqRelTE} associated with the optimal evaporation process. Both quantities were found to scale proportionally with the entropy change $\Delta S = S_0 - S_\tau$ generated along the trajectory. Consequently, larger black holes exhibit lower process probabilities and correspondingly longer evaporation times.

For the static BTZ black hole in the entropy representation, the evaporation thermodynamic length \eqref{eqTDlengthentr}, and consequently the relaxation time \eqref{eqTDtimeentr},  scale with the fourth root of the black hole energy suggesting once again that larger black holes evaporate more slowly.

For the rotating BTZ black hole in energy representation we performed a numerical analysis of the optimal processes induced by specific parameter variations. In all cases examined within the energy representation the system evolves toward a static (non-rotating) configuration with well-defined energy and entropy, preventing complete evaporation. Additionally, we observed that at most two optimal geodesics can connect any pair of admissible macrostates (see Fig.~\ref{figGeodSJ}), a feature permitted only in spaces of positive curvature. A summary of results for a range of initial conditions is given in Table~\ref{tabicerep}.

Contrary to the situation observed in the energy representation, the entropy ensemble for the rotating BTZ black hole allows a richer set of final states, even though its information metric is flat. Our analysis shows that certain optimal trajectories asymptotically drive the system toward near-extremal configurations, consistent with the third law of thermodynamics, which forbids reaching extremality in finite classical time. Other trajectories approach a fixed, nonvanishing specific spin away from extremality, while a third class terminates in a static configuration within finite time. These results are summarized in Table~\ref{tabicsrep}.

When converting to SI units, a consistent interpretation of the thermodynamic length appears either as the minimal energy required for a given process or as the minimal entropy produced along an optimal trajectory. This facilitated the determination of the probability strength factor (the metric scale) $\epsilon$, governing evaporation. In the energy representation we obtained $\epsilon = 1/2$, whereas in the entropy representation we found $\epsilon = -1/4$. In both cases, these values ensure a physically meaningful probability  and guarantee that thermodynamic length remains positive-definite along geodesic paths.

Consequently, we compare the optimal approach with a nonoptimal but analytically tractable blackbody Hawking evaporation model for both static and rotating BTZ black holes. A key result is that optimal evaporation occurs over a finite time interval, whereas the Hawking model yields an effectively infinite evaporation time. In contrast to $(3+1)$-dimensional black holes, which heat up and undergo runaway evaporation as mass decreases, the $(2+1)$-dimensional BTZ black hole exhibits the opposite behaviour: its temperature decreases as its mass is reduced. As a result, instead of evolving toward a runaway endpoint, the BTZ black hole radiates progressively more slowly under Hawking evaporation, asymptotically approaching a cold, stable configuration or thermal equilibrium with its environment. 

For a consistent comparison between optimal and Hawking processes, we impose identical initial rates for both evolutions. Importantly, the Hawking evaporation model is independent of the chosen thermodynamic representation, in contrast to the optimal dynamics. In both the energy and entropy representations, the optimal process leads to faster evaporation than Hawking radiation. However, in the rotating case, we find that in the energy representation complete optimal evaporation is not achieved, whereas in the entropy ensemble full evaporation occurs within a finite time.

We further analyse the dynamical partitioning of the total mass-energy $E$ of a rotating BTZ black hole into irreducible, rotational, and dissipative components. In particular, we show that up to 50\% of the total mass-energy can be stored as extractable rotational energy in the extremal limit ($a \to 1$), indicating a larger capacity for Penrose-type energy extraction than in the $(3+1)$-dimensional Kerr case. Introducing the intrinsic conversion efficiency $\eta(t)$, which is the ratio of the extractable rotational energy to the initial energy input at any time $t$, we demonstrate that although Hawking evaporation persists indefinitely, optimal finite-time processes achieve higher efficiency over finite intervals, provided the angular momentum rates are sufficiently small.

Finally, motivated by the nonlinear structure of the optimal processes, we examine whether a nonvanishing initial angular-momentum rate can drive the BTZ system from an initially static configuration, $a(0)=0$, to a rotating state with nonzero specific spin. This phenomenon was first reported by Dr. Vasil Avramov for the Kerr black hole in his 2026 PhD dissertation \cite{Avramov:2026}, and is therefore referred to as the Avramov effect. Our analysis shows that this effect arises in both thermodynamic representations, albeit with distinct quantitative and qualitative features in the resulting evolution.

It would be interesting to extend this optimization analysis to more general or fundamentally distinct black hole solutions in (2+1) dimensions, including warped black holes \cite{Anninos:2008fx, Tonni:2010gb}, Lifshitz solutions \cite{Ayon-Beato:2009rgu, Sarioglu:2018rhl, Herrera-Aguilar:2021top}, regular black holes \cite{Bueno:2021krl, Karakasis:2021ttn, Sajadi:2023ybm}, noncommutative geometries \cite{Rahaman:2013gw, Hamil:2025nmx}, Chern–Simons black holes \cite{Moussa:1996gm, Moussa:2008sj, Moussa:2015iea}, Gauss–Bonnet-like theories \cite{Hennigar:2020drx}, and other modified-gravity or exotic solutions \cite{Papajcik:2022zve, Karakasis:2022fep, Karakasis:2023ljt, Jusufi:2023fpo}.
}

\section*{Acknowledgments}

We are very grateful to H. Dimov, V. Avramov, S. Yazadjiev, D. Marvakov and P. Ivanov for their useful comments and discussions. G. S. thankfully acknowledges the support by the Sofia University grant 80-10-52/2025. M. R.,  R. R. and T. V. were fully supported by the Bulgarian National Science Fund, grant KP-06-N88/3.  The authors express their sincere gratitude to CERN, SEENET-MTP, ICTP, EPS, the Simons Foundation, and the International Center for Mathematical Sciences (ICMS) in Sofia for their support in funding and organizing various annual scientific events.

\appendix
\section{Planck parameters in (2+1)-dimensions related to $G_3$}\label{appPlancktoSI}
In Planck units ($\hbar=c=G_3=k=1$) every quantity is dimensionless and can be expressed as how many times a given quantity enters in its corresponding SI unit quantity. For example if $L$ is dimensionless length, then its SI version is restored by $L_{(\text{SI})}=L L_\text{P}^{(3)}$, where $L_\text{P}^{(3)}$ is the (2+1)-dimensional Planck length related to $G_3$. In this case, the replacement rule is actually $L\to L_{(\text{SI})}/L_\text{P}^{(3)}$. The (2+1)-dimensional Planck system of units, related to $G_3$, can be extracted by a simple dimensionless analysis from:
\begin{equation}
    P=\hbar^x G_3^y c^z k^u=[M^{u+x-y} L^{2 u+2 (x+y)+z} T^{-2 u-x-2 y-z} K^{-u} ].
\end{equation}
The relevant (2+1)-dimensional Planck units are given by:
\begin{equation}
    L_\text{P}^{(3)}=\frac{\hbar G_3}{c^3}, \,\,\,M_\text{P}^{(3)}=\frac{c^2}{G_3}, \,\,\,E_\text{P}^{(3)}=\frac{c^4}{G_3}, 
    \,\,\, t_\text{P}^{(3)}=\frac{\hbar G_3}{c^4},\,\,\, T_\text{P}^{(3)}=\frac{c^4}{G_3 k},\,\,\,S_\text{P}^{(3)}=k,\,\,\,J_\text{P}^{(3)}=\hbar,
\end{equation}
where $L_\text{P}^{(3)}$ is length, $M_\text{P}^{(3)}$ is mass, $E_\text{P}^{(3)}$ is energy, $t_\text{P}^{(3)}$ is time,  $T_\text{P}^{(3)}$ is temperature, $S_\text{P}^{(3)}$ is entropy, and $J_\text{P}^{(3)}$ is angular momentum. The BTZ thermodynamics in Planck units is defined by:
\begin{align}
    \lambda \to \pi ^2,\quad \zeta \to \frac{1}{\ell },
\end{align}
where the AdS radius $\ell$ is now a dimensionless parameter. As an example, let us restore the SI units of the evaporation thermodynamic length from (\ref{eqLTDevapEnergyRep}). In Planck units it is given by:
\begin{equation}
\mathcal{L}=\frac{S_0}{\pi \ell},
\end{equation}
where $\mathcal{L}$, $S_0$ and $\ell$ are dimensionless. Since $\mathcal{L}^2_{(\text{SI})}$ represents energy, its SI unit is  Joule. Accordingly, $S_0^{(\text{SI})}$ has units of Joules/Kelvin, and $\ell_{(\text{SI})}$ has units of length. Thus, the following replacement rules restore the appropriate factors of the fundamental constants:
\begin{equation}
\mathcal{L}^2\to\frac{\mathcal{L}^2_{(\text{SI})}}{E_\text{P}^{(3)}},\quad S_0\to \frac{S_0^{(\text{SI})}}{k},\quad \ell\to\frac{\ell_{(\text{SI})}}{L_\text{P}^{(3)}}.
\end{equation}
At the end, one may omit the (SI) notation from the quantities.

\section{The BTZ solution in Planck and other units}

Solving \eqref{eqBTZTD} for the horizon radii $r_\pm$ in terms of the energy $E$ and angular momentum $J$, one obtains:
\begin{equation}\label{rp}
r_+=\frac{2 \ell}{c^2}  \sqrt{\left(1+\sqrt{1-a^2}\right) G_3 E},\quad r_-=\frac{2 G_3 a E \ell }{c^2 \sqrt{\left(1+\sqrt{1-a^2}\right)G_3 E}},\quad a=\frac{c J}{E \ell },
\end{equation}
where $a \in [0,1)$ is the dimensionless specific spin parameter of the rotating BTZ black hole.
Using these expressions, the BTZ metric (\ref{eqBTZMetric}) can be written in the form:
\begin{equation}
ds^2= \left(\frac{8 G_3 E }{c^4}-\frac{r^2}{\ell ^2}\right) c^2 dt^2+\frac{dr^2}{\dfrac{r^2}{\ell ^2}+\dfrac{ 16 \ell ^2 G_3^2 a^2 E^2 }{c^8 r^2}-\dfrac{8 G_3 E}{c^4}}-\frac{ 8 a E G_3 \ell }{c^3} dt {d\varphi}+r^2 d\varphi^2,
\end{equation}

In Planck units $c=G_3=1$ one has $E=M$ and:
\begin{equation}
 ds^2=\left(8M-\frac{r^2}{\ell ^2}\right)  dt^2 + \frac{dr^2}{\frac{16 J^2}{r^2}+\frac{r^2}{\ell ^2} -8M}-8 J dt d\varphi+r^2 d\varphi^2.
\end{equation}

Note that by setting $c=8G_3=1$  one recovers the original form of the BTZ metric \cite{Banados:1992wn, Carlip:1995qv}:
\begin{equation}\label{eqBTZMetricTradF}
 ds^2=\left(M-\frac{r^2}{\ell ^2}\right)  dt^2 + \frac{dr^2}{\frac{J^2}{4 r^2}+\frac{r^2}{\ell ^2} -M}-J dt d\varphi+r^2 d\varphi^2.
\end{equation}

\bibliographystyle{utphys}
\bibliography{References}

\end{document}